\documentclass[twocolumn,aps,prb,eqsecnum,citeautoscript,floatfix,notitlepage]{revtex4-2}

\usepackage{braket,slashed}
\usepackage{mathtools,amsfonts,amsmath}
\usepackage{silence}
\numberwithin{equation}{section}

\usepackage{hyperref}
\hypersetup{
  colorlinks=true,
  linkcolor=blue,
  urlcolor=blue,
  citecolor=blue,
  pdfborder={0 0 0},
}

\usepackage{eso-pic}
\usepackage{empheq, tcolorbox, listings}
\usepackage{tikz}
\usetikzlibrary{tikzmark}
\usepackage{float}
\usepackage{wrapfig,framed}
\usepackage{enumitem}
\usepackage{natbib}
\usepackage{bm}
\usepackage{ulem}
\usepackage{mathrsfs}
\usepackage{euscript}

\usepackage[export]{adjustbox}
\usepackage[caption = false]{subfig}

\usepackage[margin=0.75in]{geometry}
\usepackage[utf8]{inputenc}
\usepackage[title]{appendix}
\newcommand{\equref}[1]{Eq. ~(\ref{#1})}
\newcommand{\figref}[1]{Fig. \ref{#1}}
\newcommand{\subfigref}[1]{Fig. \ref{#1}}
\newcommand{\subfigrangref}[2]{Figs. \ref{#1}-\ref{#2}}

\def\LSCO{La$_{2-x}$Sr$_x$CuO$_4$}
\def\LBCO{La$_{2-x}$Ba$_x$CuO$_4$}

\def\BSCCO{Bi$_2$Sr$_2$CaCu$_2$O$_{8+\delta}$}

\def\LESCO{La$_{1.8-z}$Eu$_{0.2}$Sr$_x$CuO$_4$}

\def\C60{A$_x$C$_{60}$}

\def\HgCu3{HgCa$_2$Cu$_3$O$_{8+y}$}
\def\HgCu4{HgBa$_2$Ca$_3$Cu$_4$O$_{10+y}$}
\def\TlCu{Tl$_2$Ba$_2$CuO$_{6+\delta}$}
\def\TlCu3{Tl$_2$Ba$_2$Ca$_2$Cu$_3$O$_{10+y}$}
\def\TlCu4{Tl$_2$Ba$_2$Ca$_3$Cu$_4$O$_{12+y}$}

\def\BiCu3{Bi$_2$Sr$_2$Ca$_{2}$Cu$_3$O$_y$}

\def\8LSCO{La$_{1.88}$Sr$_{.12}$CuO$_4$}
\def\110LNSCO{La$_{1.5}$Nd$_{0.4}$Sr$_{0.1}$CuO$_{4}$}
\def\stage4LCO{La$_{2}$CuO$_{4+\delta}$}
\def\Y248{YBa$_2$Cu$_4$O$_8$}

\def\NbSe2{NbSe$_2$}
\def\TaSe2{TaSe$_2$}
\def\TiSe2{TiSe$_2$}
\def\NaCoOH2O{Na$_{0.3}$CoO$_{2y}$H$_2$O}
\def\MgB2{MgB${}_2$}

\def\URu2Si2{URu$_2$Si$_2$}

\def\Ba122{Ba(Fe$_{1-x}$Co$_x$)$_2$As$_2$}

\def\htr{high temperature superconductor}

\begin{document}
\title{Electronic structure of topological defects in the pair density wave superconductor}
\author{Marcus Rosales and Eduardo Fradkin}
\affiliation{Department of Physics and Institute for Condensed Matter Theory, University of Illinois Urbana-Champaign, 1110 West Green Street, Urbana, Illinois 61801, USA}
\begin{abstract}{Pair density waves (PDWs) are a inhomogeneous superconducting states whose Cooper pairs possess a finite momentum resulting in a oscillatory gap in space, even in the absence of an external magnetic field. 
There is growing evidence for the existence of PDW superconducting order in many strongly correlated materials, particularly in the cuprate superconductors and in several other different types of systems.
A feature of the PDW state is that inherently it has a  CDW as a composite order associated with it. Here we study the structure of the electronic topological defects of the PDW, paying special attention to the half-vortex  and its electronic structure that can be detected in STM experiments. We discuss tell-tale  signatures of the defects in violations of inversion symmetry,  in the excitation spectrum and their spectral functions in the presence of topological defects. We discuss the ``Fermi surface'' topology of Bogoliubov quasiparticle of the PDW phases, and we briefly discuss the role of quasiparticle interference.}
\end{abstract}
\date{\today}
\maketitle

\section{Introduction\label{Intro}} 

One of the central problems in condensed matter physics is understanding the phases of strongly correlated systems such as high-T$_c$ superconductors. Experimental and theoretical research during the past decade has clearly shown that a characteristic feature of these systems is that their complex phase diagrams have phases with different types of charge, spin and, superconducting orders which are intertwined rather than competing with each other \cite{Fradkin2015}.
The prototype quantum materials with intertwined orders are the cuprate superconductors, which possess a very rich phase diagram hosting antiferromagnetic order, high-temperature $d$-wave superconductivity, CDW order, nematic order, and, at least in the lanthanum family of cuprates, incommensurate SDW order. Experiments done during the past decade have provided increasing evidence that, in addition to the $d$-wave superconducting order, in these systems a type of superconducting order, known as a pair density wave, may be at play \cite{Agterberg-2020}.
In addition, and in contrast to conventional superconductors where the superconducting (SC) state is born from a normal Fermi liquid,
the  ``normal'' (high-temperature)  state of all the cuprates is a strange metal, a metal without well-defined electronic quasiparticle.

\begin{figure}[hbt]
\vspace{-5pt}
\hspace{-5pt}
\includegraphics[width=0.5\textwidth]{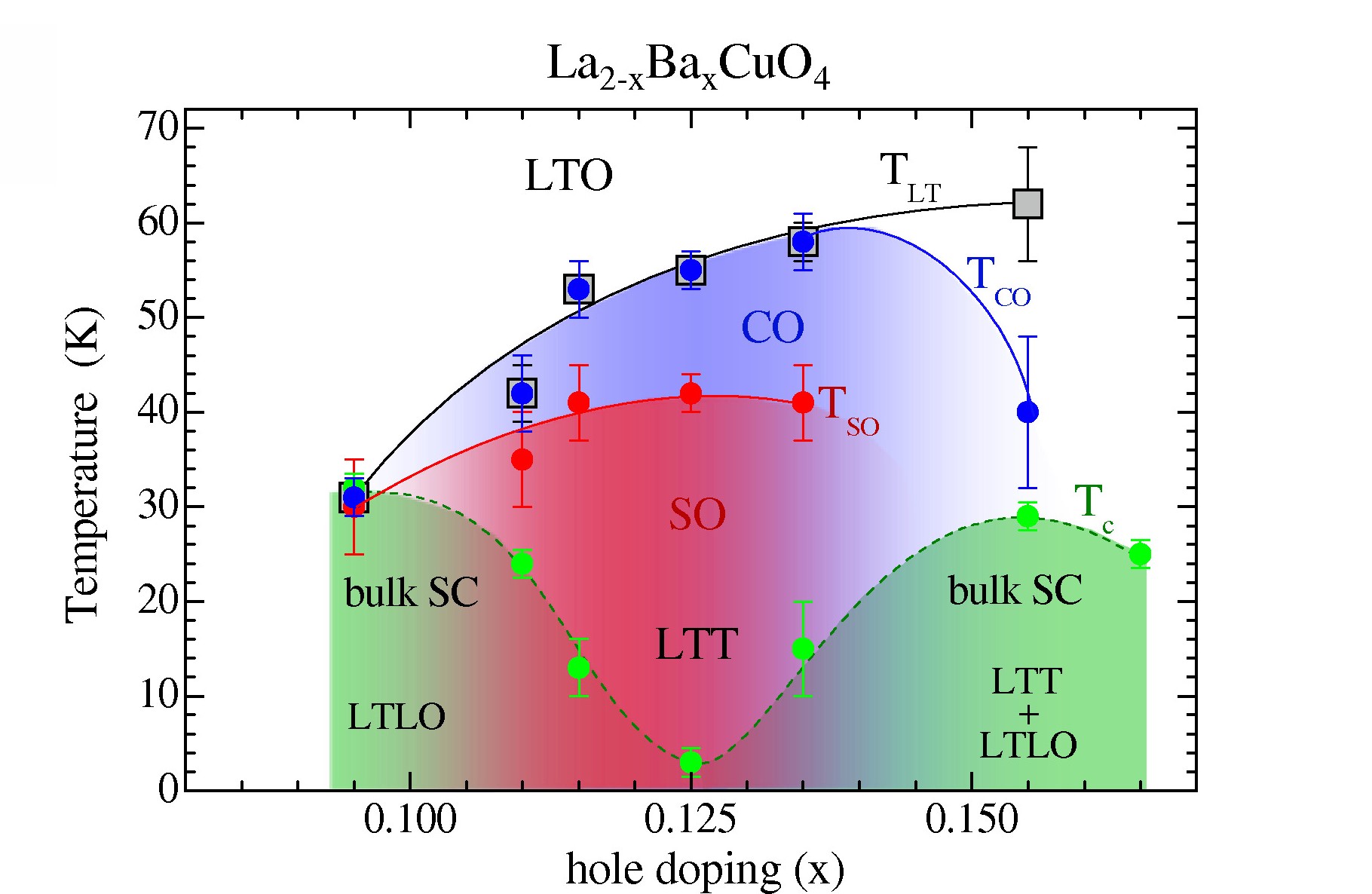}
\hspace{-5pt}
\caption{(Online Color)  Experimental data of LBCO indicating various phases for temperature, $T$, vs hole-doping, $x$. Various orders exist simultaneously under the superconducting dome, which may indicate the onset of the PDW phase. Data from Ref. \cite{Hucker-2011}.} 
\label{figure_1}
\end{figure}

Of particular interest is the cuprate material La$_{2-x}$Ba$_x$CuO$_2$ (LBCO). This cuprate superconductor, the original high T$_c$ material, has the interesting phase diagram  provided in \figref{figure_1} \cite{Hucker-2011}. Instead of a single SC dome, as most other cuprates have, {\LBCO} has a pronounced anomaly at  $x=1/8$ hole doping where the transition temperature to the $d$-wave SC state is suppressed dramatically from 35K to about 4K where the Meissner state is observed. 
In this regime a remarkable set of phase transitions are observed \cite{Li-2007}: static charge density-wave (CDW)  and spin-density-wave (SDW) orders onset at ~52K and ~42K, respectively. Below the onset of the SDW order the $c$-axis resistivity, $\rho_c$, increases with decreasing temperature while the $ab$-plane resistivity, $\rho_{ab}$, decreases rapidly. Superconducting phase fluctuations in the $ab$-planes onset at about 35K, and at approximately 16 K a two-dimensional Berezhinskii-Kosterlitz-Thouless (BKT) transition to a two-dimensional SC state is observed. On the other hand, the resistive transition where the $c$-axis resistivity vanishes happens only at 10K, and the full Meissner $d$-wave SC state is reached only at  $T_c \sim 4$K. This ``dynamical layer decoupling'' behavior is also observed in {\LBCO} in the presence of a $c$-axis magnetic field away from $x=1/8$ \cite{Zhong-2018} and in underdoped {\LSCO} in magnetic fields $B \sim~8$T \cite{Schafgans-2010} where a magnetic-field induced SDW was observed long ago \cite{Lake-2002}. Similar behaviors have been found in LSCO doped with Zn \cite{Lozano-2021} and with iron \cite{Huang-2021}.

The remarkable dynamical layer decoupling observed at the 1/8 anomaly of {\LBCO} implies that the interlayer Josephson effect is suppressed in this regime. Berg and coworkers \cite{Berg-2007}  proposed that that the complex behavior of {\LBCO} at $x=1/8$ was evidence for the presence of the pair-density-wave made evident by the lattice structure of the low-temperature-tetragonal (LTT) crystal phase of {\LBCO} \cite{Axe-89,Axe-94}.

Evidence for PDW order also exists in cuprate superconductors that do not have the LTT crystal structure such as {\BSCCO}. In this case much of the evidence was found in STM experiments \cite{Fujita-2012} and Josephson tunneling spectroscopy \cite{Du-2020}. STM experiments in the vortex halo of {\BSCCO} have revealed tell-tale evidence for PDW order in that regime \cite{Edkins-2018,Wang-2018}.
Recent experiments have provided evidence for PDW order to be present also in the superconducting states of several strongly correlated materials including the heavy fermion material UTe$_2$ \cite{Aishwarya-2023,Gu-2023}, the iron superconductor Fe(Se,Te) \cite{Liu-2023}, the pnictide EuRbFe$_4$As$_4$ \cite{Zhao-2023}, and kagome superconductors such as CsV$_3$Sb$_5$ \cite{Zhao-2021,Chen-2021}. In this study we will focus on the pristine PDW with no uniform component present. In a future publication we will address how the uniform component affects various plots seen in this paper. 

The PDW is a superconducting state in which Cooper pairs with finite center of mass momentum 
$\bm Q$ condense. 
In such a state the local pairing amplitude $\Delta(\bm r)$ is periodic function of position whose period is $2\pi/|\bm Q|$. 
The order parameter of the PDW has the same symmetry as  the Larkin-Ovchinnikov (LO) state \cite{Larkin-1965}. 
However, the PDW differs from the LO state in several important ways: 
(1) the LO state is created by a magnetic field through the Zeeman 
coupling to the spins and it is spin-polarized, and (2) as a result, it has a broken time-reversal invariance. In 
contrast, the PDW arises \textit{in the absence of a magnetic field}. Another superconducting state with 
a finite-momentum Cooper pair is the Fulde-Ferrell (FF) state \cite{FF}, which, like the LO state, was also envisioned as arising in the presence of a Zeeman coupling to an external magnetic field. 
Much like the  LO state, the FF state breaks time reversal, but, in addition, it  breaks
inversion symmetry. The amplitude of the FF state is constant in space.  
Finally the ordering wave vectors of the LO and the FF states are tuned by the strength of the magnetic field, and,
consequently, the periodicity of both states is much larger than the lattice constant. 

In the cuprates, and in other candidate materials for PDW superconductors, the periodicity is a few lattice spacings. 
For all these reasons the ``classic'' FF and LO states are not suitable to explain the observed phenomenology 
of the cuprates and other materials. 

The phenomenology of the PDW state \cite{Berg-2007,Berg-2009,Berg-2009a,Lee-2014,Agterberg-2020} 
(and the phenomenology of all high-temperature superconductors) strongly suggests 
that, with some possible exceptions, a PDW is necessarily a strong coupling state, which cannot be explained in 
terms of the conventional BCS theory of superconductivity \cite{Bardeen-1957,Schrieffer-1964}. 
Nevertheless, BCS-type approaches have been 
developed to explain the PDW \cite{Loder-2010,Loder-2011,Wardth-2017,Wardth-2018} which require that the interactions be 
large compared to the kinetic energy of the holes. In this regime BCS theory is not reliable. Numerical simulations of 
$t-J$ and extended Hubbard models have shown evidence that PDW ground is at least a strong competitor to be ground 
state ~\cite{Himeda-2002,Corboz-2014,Ponsioen-2023,Huang-2022,Jiang-2023}. Quasi-one-dimensional models have also shown 
that their phase diagrams contain PDW phases \cite{Soto-Garrido-2015}. Under special circumstances weak coupling models 
do predict the existence of PDW phases, but typically they require some other strong coupling physics to take place 
first, such as a Pomeranchuk instability in the triplet channel \cite{Soto-Garrido-2014}. We should note that a recent 
study \cite{Shaffer-2023} predicts the occurrence of a PDW state in transition metal dichalcogenide materials. To the best of our 
knowledge the only microscopic model which is unambiguously known to have a (large) PDW phase is the one-dimensional 
Kondo-Heisenberg chain \cite{Berg-2010,Jaefari-2012}.

The purpose of this paper is to investigate physics of the PDW state by studying the electronic structure of the topological defects of this state: the half-vortex, the double dislocation and the Abrikosov vortex. The study that we undertake here is relevant to the understanding of the features of these defects revealed by STM experiments. Here we adopt a phenomenological description of the PDW and we will not concern ourselves with the 
possible physical mechanism(s) associated with this superconducting state. 
For concreteness we will consider a system with a square lattice (kagome and honeycomb lattices 
have also been considered). 
On a square lattice the PDW state may be unidirectional, which breaks spontaneously both translation symmetry and the 
point group symmetry of the square lattice, or bidirectional, which is invariant under the point group symmetry but 
breaks translation symmetry along two directions. Let $\Delta(\bm r)$ be the local  amplitude for a spin singlet 
superconductor (where $\bm r$ is a lattice site; in the case of a local $d$-wave state $\bm r$ is the superconducting 
amplitude on a bond between two nearest-neighbor sites $\bm r$ and $\bm r'$).
We will consider the simpler case of an unidirectional spin-singlet PDW with a period close to $8a_0$, which is 
appropriate for the lanthanum cuprates \cite{Agterberg-2020}. The local pairing amplitude can be expanded in Fourier 
components
\begin{equation}
    \Delta(\bm r)
    = \Delta_0(\bm r)+
    \Delta_{\bm Q}
    (\bm r)
    e^{i\bm Q\cdot \bm r}
    +
    \Delta_{-\bm Q}
    (\bm r)
    e^{-i\bm Q\cdot \bm r},
    \label{PDW}
\end{equation} 
where $\Delta_0(\bm r)$ is the uniform component. Here we denoted by $\Delta_{\pm \bm Q}(\bm r)$ the two 
PDW components with wave vectors $\pm \bm Q$.
 We will not include higher harmonics in the order parameter occurring at $n\mathbf{Q}$ since in the ordered state 
 these orders are slaved to the fundamental and hence are not independent dynamical degrees of freedom \cite{Berg-2009b}.

The PDW equilibrium state is a phase in which the uniform component vanishes, $\langle \Delta_0 \rangle=0$, 
and the two Fourier components have the same expectation value, 
$\langle \Delta_{\bm Q} \rangle=\langle \Delta_{-\bm Q} \rangle=\Delta_{\rm PDW}$. 
If $\langle \Delta_0 \rangle \neq 0$ and $\Delta_{\rm PDW} \neq 0$ the resulting state is a striped superconductor. 
Most proposed PDW states (outside the lanthanum cuprates) are actually striped superconductors. 
Finally, in the FF state $\langle \Delta_{\bm Q} \rangle$ or (exclusive) $\langle \Delta_{-\bm Q} \rangle$ 
are not zero. 

In this paper we will consider only a unidirectional PDW state and hence assume that $\langle \Delta_0 \rangle=0$.
The central results of this paper are concerned with experimental signatures associated with the induced 
CDW ``daughter'' states of the PDW parent state \cite{Berg-2009,Agterberg-2008}. 
The induced CDW order with wave vector $\bm K = 2\bm Q$ whose order parameter field is 
$\rho_{\bm 2\bm Q}(\bm r) \sim \Delta_{\bm Q} \Delta_{-\bm Q}^*$ 
[and similarly for $\rho_{-2 \bm Q}(\bm r)$] is described in more detail in Sec. \ref{LG-section}, including the 
topological defects of the PDW state we are interested in. 
The unidirectional PDW has two complex order parameters, $\Delta_{\pm \bm Q}(\bm r)$, 
which means two amplitude fields and two phase fields. This means that the order parameter of the PDW transforms under 
a $U(1) \times U(1)$ global symmetry, where the first factor is the conventional global $U(1)$ gauge invariance of 
a superconductor and the second factor represents the invariance under continuous translations of the 
{\it incommensurate} PDW state.
In Refs. \cite{Berg-2009,Agterberg-2008} it is shown that as a result of the topology of the target space of the 
order parameter(s) of the PDW, this state has three distinct topological defects: 
a superconducting Abrikosov vortex, a half-vortex, and a double dislocation.

The electronic structure of the PDW has Bogoliubov quasiparticle with (Bogoliubov) Fermi surfaces which define pockets of 
quasiparticle states \cite{Berg-2009,Baruch-2008}. The stability of Bogoliubov Fermi surfaces has been 
established in several studies \cite{Agterberg-2018,Berg-2008-QPI}. 
In the PDW state the half-vortex has an energy cost which diverges logarithmically with sample size. 
In contrast, in the presence of a non-vanishing uniform SC component, $\langle \Delta_0\rangle\neq 0$, 
the energy cost is instead 
{\it linearly divergent} and, hence, half-vortices and anti-half-vortices are {\it confined} into pairs. 
Experimental evidence for half-vortices pairs has been found in STM experiments in {\BSCCO} 
by Du and coworkers \cite{Du-2020}.



In this work we describe the half-vortex as a state in which one of the two order parameters of the PDW, say $\Delta_{\bm Q}(\bm r)$, has a vortex while the other order parameter, $\Delta_{-\bm Q}(\bm r)$, does not. 
Since {\it one} of the SC order parameters vanishes at the 
defect location while the other does not, at the defect core we have an FF type state which breaks inversion symmetry. 
In this sense, the half-vortex has an FF halo. This behavior is analogous to the halo associated with the 
Abrikosov vortex in a system 
in which uniform SC and PDW orders compete \cite{Edkins-2018,Wang-2018}. 
Our construction of the half-vortex follows the same strategy used by Wang and coworkers \cite{Wang-2018} 
for the Abrikosov vortex halo of an uniform superconductor with a PDW as a subleading order. In contrast, in the case of a Abrikosov vortex \textit{both} PDW order parameters wind with the same topological charge, whereas in the double dislocation they wind with opposite topological charges.

We then investigate the electronic structure of the PDW in the presence of topological defects by embedding the 
resulting configuration into the Bogoliubov-de Gennes Hamiltonian. Here we use a non-interacting band structure suitable 
for a superconductor in a copper oxide plane. Using this effective Hamiltonian with the defect background we calculated 
the local density of states  and a function of bias for a model of an STM measurement at a point contact with a normal 
metal. We also compare the results on the half-vortex with the same calculation done for a full Abrikosov vortex and for 
a double dislocation of the CDW.

The paper is organized as follows. In Sec. \ref{LG-section} we introduce the Landau-Ginzburg equations to be used in 
our analysis. 
In this section the notion of intertwined order is reviewed and we discuss the composite order parameters of interest in 
Sec. \ref{pdw-order-parameters}. An overview of the relationship between experiment and induced order is briefly 
discussed there as well. 
In Sec. \ref{topological-defects} we review the topological defects of the PDW superconducting state. 
Section \ref{NLSM} is devoted to the construction of a static half-vortex. Here we discuss how the profiles of the 
components of the PDW order parameters are modeled in the case of a half-vortex. 
In Sec. \ref{Hamiltonian} the effective mean field Hamiltonian for the PDW with static topological defects  used in 
our simulation is introduced as well as the Green functions and the related spectral function. Here we present results 
for  the local density of states (LDOS), used to acquire plots for the numerical solutions. In Sec. \ref{Hamiltonian} 
the numerical parameters and the explicit form of the order parameters used in the simulation are discussed. 
In Sec. \ref{LDOS} the plots of the charge density wave order of the PDW in the presence of defects are shown, and various other 
aspects of the vortex cores are explored and discussed. In Sec. \ref{spec-section} we discuss the spectral functions of 
the PDW order and display the dispersion relation for a PDW with a half-vortex. Finally, in Sec. \ref{summary} we  
discuss some implications of our results and  summarize the most salient results. 
Several appendixes are devoted to technical details. Details of the Bogoliubov transformation are given in 
Appendix \ref{BV} and the setup for the numerical diagonalization of the Bogoliubov-de Gennes Hamiltonian is sketched in 
Appendix \ref{numerical-diagonalization}. In Appendix \ref{zero-temp-Green}, we give details of the retarded Green 
function at zero temperature. In Appendix \ref{DOS_Sup} we compare the tunneling DOS for a PDW,  an FF state and for the half-vortex of the PDW. In Appendix \ref{spec_appendix}  we present data on spectral functions for PDW states with $s$- and $d$- wave form factors.

\section{Landau-Ginzburg Theory\label{LG-section}}

In this section we review the Landau-Ginzburg (LG) theory for a unidirectional PDW state \cite{Berg-2009}, which will be used to describe the profiles of our order parameters in the presence of static topological defects.  In a a later section we will calculate the local density of states (LDOS) in the presence of topological defects that we will describe in this section. Because our defects are static, we will be neglecting fluctuations, so our analysis here will be done at the level of mean field theory. Furthermore, we also work deep in the PDW phase where the order parameters have well defined local amplitudes, which is justified at low temperatures.

\subsection{Free energy of the PDW state}
\label{free-energy-pdw}

\par The free energy of the LG theory is chosen to have the form:
    \begin{equation}
    \begin{aligned}
        \mathcal{F}[\Delta_0,\Delta_{\mathbf{Q}},\Delta_{-\mathbf{Q}}]
        =
        \mathcal{F}_{\text{sc}}[\Delta_0]
        &+
        \mathcal{F}_{\text{pdw}}[\Delta_{\mathbf{Q}},\Delta_{-\mathbf{Q}}\big].   
    \end{aligned}
        \label{Landau-Ginzburg}
    \end{equation}
    The first term in \equref{Landau-Ginzburg} describes uniform superconductivity, which takes the familiar form 
    \begin{equation}        
        \mathcal{F}_{\text{sc}}
        =
        \frac{1}{2m^*}\Big|\big(-i\nabla
        +2e\mathbf{A}\big)\Delta_0\Big|^2
        +\frac{a}{2}|\Delta_0|^2
        +\frac{b}{2}|\Delta_0|^4.
        \label{LG-SC}
    \end{equation}

In this paper our main results correspond to a defect-free PDW in the absence of an uniform component: $\Delta_0=0$. We will mention the effects of a non-zero uniform component when relevant, but in the majority of this paper we neglect it.
Furthermore, we will assume that the PDW is unidirectional, characterized by a single ordering wave vector $\bm Q$. 
We will also ignore the effects of disorder. This is important since, unlike a conventional uniform superconductor, 
a PDW can couple to local charge disorder through the interaction of the induced CDW associated with the PDW. 
In fact, in the presence of disorder the distinction between unidirectional and bidirectional orders is lost as these 
PDW components get mixed with each other \cite{Robertson-2006}.

The second term in \equref{Landau-Ginzburg} describes the free energy of the unidirectional PDW. In Eq. \eqref{PDW} we presented the expansion of the  local pairing amplitude $\Delta(\bm r)$ in its Fourier components $\Delta_{\pm \bm Q}(\bm r)$, where $\Delta_{\pm \bm Q}(\bm r)$ are the order parameters of the unidirectional PDW state with ordering wave vector $\bm Q$. Since the local pairing amplitude $\Delta(\bm r)$ is a complex field, the two PDW components $\Delta_{\pm \bm Q}$ are not the complex conjugate of each other (as they would have been for a CDW) but are two independent complex fields. The PDW free energy has the form \cite{Berg-2009,Berg-2009a,Agterberg-2008,Fradkin2015}:

 \begin{widetext}
    \begin{align}
        \mathcal{F}_{\text{pdw}}[\Delta_{\mathbf{Q}},\Delta_{-\mathbf{Q}}\big]
    =&
        \kappa\Big(|\nabla \Delta_{\mathbf{Q}} |^2+|\nabla \Delta_{-\mathbf{Q}} |^2\Big)+r \Big(| \Delta_{\mathbf{Q}} |^2+
        | \Delta_{-\mathbf{Q}} |^2\Big)
        +
        u \Big(| \Delta_{\mathbf{Q}} |^2
    +
        |\Delta_{-\mathbf{Q}} |^2\Big)^2
        +\gamma |\Delta_{\bm Q}|^2|\Delta_{-\bm Q}|^2,
        \label{LG-PDW}
    \end{align} 
   \end{widetext}
   where $r=T-T_c^{\rm pdw}$, and $T_c^{\rm pdw}$ is the (mean-field) critical temperature for the PDW superconductor. 
   We will assume that the coupling constant $\gamma<0$ as needed for an attractive interaction needed for a PDW, 
   an LO-type state. Instead, a repulsive value of the coupling constant, $\gamma >0$, favors an FF type state.

The PDW state is described by two independent \textit{complex} order parameters fields $\Delta_{\pm \bm Q}(\bm r)$. Since they are complex fields they can be decomposed into amplitude and phase fields, $\Delta_{\pm \bm Q}(\bm r)=|\Delta_{\pm \bm Q}(\bm r)|\, \exp(i \theta_{\pm \bm Q}(\bm r))$, respectively.
For general values of the coupling constants $u$ and $\gamma$ the free energy for the PDW order parameters, 
Eq. \eqref{LG-PDW},  is invariant under the $U(1) \times U(1)$ global symmetries 
$\theta_{\pm \bm Q}(\bm r) \to \theta_{\pm{\bm Q}}(\bm r)+\vartheta_{\pm \bm Q}$, where $\vartheta_{\pm{\bm Q}}$ are 
two independent transformations of the phases of the complex fields $\Delta_{\pm \bm Q}(\bm r)$. 
In the special case in which $\gamma=0$ this global symmetry is enhanced from $U(1) \times U(1)$ to $U(2)$.

In many superconductors of interest, such as cuprates {\BSCCO} and {\LBCO} in the Meissner state, a uniform superconducting order parameter $\Delta_0$ is present. This results in the following additional ``lock-in''  terms that couple the three superconducting order parameters and it must be added to the free energy of Eq. \eqref{Landau-Ginzburg}:
\begin{equation}
    \mathcal{F}_{I}=\beta_1 |\Delta_0|^2(|\Delta_{\bm Q}|^2+|\Delta_{-\bm Q}|^2)+\beta_2 (\Delta_0^*)^2 \Delta_{\bm Q} \Delta_{-\bm Q}+ {\rm c.c.}
    \label{LG-I}
\end{equation}
The first term is the usual biquadratic term which is attractive (repulsive) for $\alpha<0$ ($\alpha>0$). 
The second term breaks the $U(1) \times U(1)$ global symmetry down to the global $U(1)$ symmetry of 
the uniform superconductor and locks (mod $\pi$) the phase fields of the PDW order parameters 
$\Delta_{\pm \bm Q}(\bm r)$ to the phase field of the uniform superconducting order parameter $\Delta_0(\bm r)\equiv|\Delta_0(\bm r)| \exp(i \theta_0(\bm r))$, 
which transforms under global gauge transformations in the usual way, $\theta_0(\bm r) \to \theta_0(\bm r)+\vartheta_0$.

\subsection{Order parameters of the PDW state}
\label{pdw-order-parameters}

With the above SC orders we can construct the following set of composite order parameters \cite{Berg-2009,Berg-2009a}:
\begin{align}
    \rho_{\bm Q}(\bm r)
    &=
    \Delta_{0}(\bm r) \Delta_{\bm Q}^*(\bm r)
    \label{1Q},\\
    \rho_{2\bm Q}(\bm r)
    &=
    \Delta_{\bm Q}(\bm r) \Delta_{-\bm Q}^*(\bm r)
    \label{2Q},\\
    \Delta_{\text{4e}}
    ({\bm r})
    &=
    \Delta_{\mathbf{Q}}({\bm r})
     \Delta_{-\mathbf{Q}}({\bm r}).
    \label{4e}
\end{align}
The  two order parameters of Eqs.\eqref{1Q} and \eqref{2Q} are interpreted as the ${\bm Q}$ and $2{\bm Q}$ components of a CDW associated with the PDW SC order. The order parameter of Eq.\eqref{4e} represents  an uniform charge $4e$ superconductor.

Following the analysis of Berg \textit{et al.} \cite{Berg-2009} we decompose the phase fields of the PDW  order parameters as
\begin{equation}
\theta_{\pm \bm Q}(\bm r)=\theta_+(\bm r)\pm \theta_-(\bm r).
\label{theta-pm}
\end{equation}
 Under the global $U(1) \times U(1)$ gauge transformations defined above the order parameters transform as follows
\begin{align}
\Delta_{\pm \bm Q}(\bm r)&\to \exp(i (\vartheta_+\pm \vartheta_-)) \Delta_{\pm \bm Q}(\bm r),\\
\Delta_0(\bm r) &\to \exp(i \vartheta_0) \Delta_0(\bm r),\\
\rho_{\bm Q}(\bm r) &\to \exp(i(\vartheta_0 -\vartheta_+)) \exp(i \vartheta_-) \rho_{\bm Q}(\bm r),\\
\rho_{2\bm{Q}}(\bm r) &\to \exp(i 2\vartheta_-) \rho_{2\bm Q}(\bm r),\\
\Delta_{4e}(\bm r) &\to \exp(i 2\vartheta_+) \Delta_{4e}(\bm r),
\label{transformations}
\end{align}
where we defined the global gauge transformations $\vartheta_\pm=(\vartheta_{\bm Q} \pm \vartheta_{-\bm Q})/2$.

Under a global \textit{electromagnetic} gauge transformation all three superconducting order parameters must transform as charge $2e$ complex fields and, consequently, $\vartheta_0=\vartheta_+$.
With this identification the CDW order parameter $\rho_{\bm Q}$ is manifestly 
invariant under global gauge transformations.  Similarly, the order parameter $\Delta_{4e}$ transforms under global 
gauge transformations as a charge $4e$ field.

On the other hand, the composite order parameter field $\rho_{\pm2{\bm Q}}({\bm r})$ has 
the same transformation as that of the order parameter 
for an  {\it incommensurate}  CDW under an arbitrary global {\it translation}.
The slowly varying relative phase of the two PDW order parameters is identified with the Goldstone mode 
of the spontaneously broken translation invariance of the PDW state.
In an incommensurate (unidirectional) CDW state with wave vector $\bm K$,  
the local charge density $\rho(\bm r)$ has the Fourier expansion
\begin{equation}
 \rho(\bm r)=\bar \rho + \rho_{\bm K}(\bm r) \exp(i \bm K \cdot \bm r) 
 +    \rho_{-\bm K}(\bm r) \exp(-i \bm K \cdot \bm r)+\ldots,
 \label{eq:density-expansion}
\end{equation}
where $\rho_{\bm K}=\rho_{-\bm K}^*$ since $\rho(\bm r)$ is real and invariant under global 
gauge transformation, and where the ellipsis denotes higher harmonics of the density wave. 
Thus, the PDW has an associated  charge density 
modulation with wave vector $\bm K=2\bm Q$ \cite{Berg-2007}. An arbitrary relative phase transformation by $\theta$ 
(mod $2\pi$) is then equivalent to a displacement of the charge density profile by $2\theta/|\bm K|$. 
In the case of a CDW which is commensurate with the underlying lattice with period $p$ lattice spacings, $pa_0$, 
the CDW wave vector is $|\bm K|=2\pi/pa_0$. 
In this case the allowed transformations of the relative phase take discrete $p$ values. In this case
the $U(1)$ symmetry group of translations reduces to the discrete (cyclic) group $\mathbb{Z}_p$. 
In this case the PDW is locked to the lattice, 
it has  $p$ equivalent ground states, and the Goldstone mode of translations is gapped.

  
Alternatively, the incommensurate CDW may be a present as a preexisting order with 
wave vector $\bm K$. 
Such a CDW cannot  couple to an (also incommensurate) PDW unless the CDW ordering wave vector $\bm K$ 
and the PDW ordering wave vector $\bm Q$ satisfy the {\it mutual commensurabilty condition} $\bm K=2\bm Q$. 
This interaction is described by an additional trilinear term in the free energy of the form 
\begin{equation}
\mathcal{F}_{PDW-CDW}=g \,\rho_{\bm K} \Delta_{\bm Q} \Delta^*_{-\bm Q}+{\rm c.c.},
\label{mutual}
\end{equation}
 where $g$ is a coupling constant \cite{Berg-2009}. 
Translation invariance then requires that the mutual commensurability condition is satisfied. 
The same requirement exists for a coupling between a spin density wave $SDW$. This effect  is seen in {\LBCO} 
at the charge-ordering transition \cite{Miao2019} and in {\LESCO} \cite{Lee2022} at low temperatures. 
The existence of this interaction yields some interesting physics not allowed for an uniform SC state: the PDW is sensitive 
to charged impurities due to their coupling to the CDW. 
Thus, unlike  uniform superconducting order, which is weakened only by disorder, 
even small amounts of charge disorder destroys true long-range incommensurate PDW order.

\subsection{Topological defects of the PDW state}
\label{topological-defects}

We will now discuss the topological defects of the PDW phase. For now we will set $\Delta_0=0$, and we will be brief as many details exist in the literature \cite{Berg-2009,Agterberg-2008}. 
The phase fields, which we denote by $\theta_{\pm \bm Q}(\bm r)$, are periodic and defined mod $2\pi$. 
Hence, the topological singularities of the phase fields $\theta_{\pm \bm Q}(\bm r)$ have integer-valued 
winding numbers, $m_{\pm \bm Q}$. This implies that $\theta_\pm(\bm r)$, the average and relative phase fields 
$\theta_\pm(\bm r)=(\theta_{\bm Q}\pm \theta_{-\bm Q})/2$, be defined mod $\pi$. 
We  denote the topological charge of the average phase $\theta_+(\bm r)$ by the vortex charge $q_v$ 
and the topological charge 
of the relative $\theta_-(\bm r)$, the dislocation topological charge of the CDW, by $q_d$. They are given by
\begin{equation}
q_v=\frac{1}{2} (m_{\bm Q}+m_{-\bm Q}), \qquad q_d=\frac{1}{2} (m_{\bm Q}-m_{-\bm Q}).   
\label{top-charges}
\end{equation}
We will label the topological defects by the combinations $(q_v, q_d)$, 
the vorticity and dislocation charges, or equivalently, $(m_{\bm Q}, m_{-\bm Q})$. 
In the simplest cases they are (1) the \textit{superconducting (Abrikosov) vortex} with topological charges 
$(q_v, q_d)=(\pm 1, 0)$ [or, equivalently, $(m_{\bm Q}, m_{-\bm Q})=(\pm 1, \pm 1)$], 
(2) the \textit{half-vortex} (bound to a single CDW dislocation) with topological charges 
$(q_v, q_d)=(\pm 1/2, \pm 1/2)$ [or, equivalently, $(m_{\bm Q}, m_{-\bm Q})=(\pm 1, 0)$], 
and (3) the CDW \textit{double dislocation} with topological charges $(q_v, q_d)=(0 , \mp 1)$ 
[or, equivalently, $(m_{\bm Q}, m_{-\bm Q})=(\pm 1, \mp 1)$]. These identifications imply that a conventional superconducting vortex is equivalent to  {\it both} PDW order parameters $\Delta_{\pm \bm Q}$ having a vortex. The half-vortex is equivalent to a vortex in $\Delta_{\bm Q}$ but {\it not} in $\Delta_{-\bm Q}$ (and vice versa), and it has a single-dislocation, as required by Eq.\eqref{2Q}. Finally, a double dislocation is a vortex in $\Delta_{\bm Q}$ and an anti-vortex in $\Delta_{-\bm Q}$ (and vice versa).

In the PDW state the energy of all three types of topological defects is logarithmically 
divergent, leading to the rich phase diagram of Ref. \cite{Berg-2009}. 
However, if the superconducting state also has a uniform component, $\Delta_0\neq 0$, 
the lock-in term of Eq. \eqref{LG-I} predicts a \textit{linearly divergent} energy cost resulting in a confined neutral pair of half-vortices which cannot be excited thermally. 
In this case only the superconducting vortices and the double dislocations have logarithmic energy and 
govern the phase diagram.

In this paper we will focus primarily on the properties of the half-vortex which we will 
regard as a \textit{static} topological defect of an ordered PDW state which we will take to be of the 
Larkin-Ovchinnikov (LO) type \cite{Larkin-1965}. Hence, we will assume that at long distances the PDW amplitudes are 
equal and constant in space, $|\Delta_{\bm Q}|=|\Delta_{-\bm Q}|$. 
However, at the core of the half-vortex one of these two amplitudes, say $\Delta_{\bm Q}$, 
must vanish while the other amplitude does not. 
As a result, the core of the half-vortex is in a Fulde-Ferrell (FF) state \cite{FF}, and
 inversion symmetry is broken at the core of the half-vortex since
 $|\Delta_{\bm Q}|\neq |\Delta_{-\bm Q}|$. On the other hand, none of these considerations apply to either the Abrikosov vortex or to the double dislocation.

As in all superconductors, the Abrikosov vortex of the PDW arises in the presence of a magnetic field. On the other hand, the half-vortex and the double dislocation can appear due to the interaction of the PDW state with sufficiently strong charged impurities. 
This is possible since the CDW order parameter $\rho_{2\bm Q}(\bm r)$ of the PDW has a linear coupling to 
charged impurities potentials, whereas gauge invariance requires that the superconducting order parameters interact only quadratically through $|\Delta_{\pm \bm Q}(\bm r)|^2$.

The properties of all three topological defects will be discussed in Sec. \ref{Hamiltonian}, 
where we specify the form of the order parameters used in the numerics. 
The associated charge density wave patterns of each defect will be compared in Sec. \ref{LDOS} as well 
as their spectral functions in Sec. \ref{spec-section}.

\par An extremely useful experimental technique for detecting and visualizing  CDW patterns is scanning tunneling microscopy (STM) \cite{Fischer-2007,Hoffman-2002,Howald-2003a,Kivelson-2003}. STM has been used to investigate in great detail the charge order present in the superconducting phase of {\BSCCO} \cite{Lawler-2010,pasupathy-2008,Mesaros-2016} and in the vortex halo \cite{Edkins-2018,Wang-2018}. Relevant to the existence of PDW order is the experimental evidence for static (pinned) half-vortices in the superconducting phase of {\BSCCO} found in STM experiments by Du \textit{et al.} \cite{Du-2020}. These authors argued that the jumps in the PDW SC winding number are located around the charge dislocations. In this regime the  term proportional to $\beta_2$ in Eq. \eqref{LG-I} locks the phase of the PDW order parameter to the phase of the uniform superconducting state. The result is a confinement of the half-vortices into (half) vortex-anti-vortex pairs. Hence, in the phase in which the uniform order parameter is present,  $\Delta_0\neq 0$, the energy of the half-vortex is linearly divergent and half-vortices cannot exist in isolation. In this regime charged impurity potentials can separate the half-vortices and anti-half-vortices as static defects. 

\par The basic setup for an STM consists of an atomically sharp metallic tip, with a featureless Fermi surface, biased at some voltage $V$ relative to the sample. The voltage difference induces a tunneling current, $I_T(V)$, which is used to map out the electronic structure at the surface of a material. In the regime in which the STM operates, the differential conductance $g(V)=dI_T/dV$ is proportional to the one-particle density of states (DOS) $\rho(\epsilon=eV)$ at an energy $\epsilon=eV$.

\section{The Half-Vortex \label{NLSM}}

In this section we will model the profile of our half-vortex of an unidirectional PDW and we  will assume that there is no uniform SC order. In Sec. \ref{Hamiltonian} we will embed the configuration of the half-vortex in a Bogoliubov-de Gennes Hamiltonian for the $d$-wave superconducting state of a CuO$_2$ plane and investigate its effects on the electronic structure. The Abrikosov vortex and the double dislocation will also be considered in later sections, so we mention their solutions here as well. We should note that the associated CDW dislocation was given in Ref. \cite{Agterberg-2008}.

We will seek an extremal solution of the Landau-Ginzburg free energy density $\mathcal{F}_{pdw}(\Delta_{\bm Q}(\bm r), \Delta_{-\bm Q}(\bm r))$ for the PDW order given in Eq.\eqref{LG-PDW}. 
In Sec. \ref{LG-section} we showed that a vortex with half magnetic flux quanta can be realized by putting a $2\pi$ phase winding in one of the two superconducting components $\Delta_{\pm \bm Q}$ of the unidirectional PDW order parameter. Thus, we will require the order parameter $\Delta_{\bm Q}(\bm r)$ to have a unit vortex and set the other component  to be real $\Delta_{-\bm Q}(\bm r)=|\Delta_{-\bm Q}(\bm r)|$, and we set its phase $\theta_{-\bm Q}=0$

A half-vortex is an extremal solution of the PDW free energy $\mathcal{F}_{pdw}$ which at  long distances behaves as a vortex in $\Delta_{\bm Q}$:
\begin{align}
\lim_{|\bm r| \to \infty} \Delta_{\bm Q}(\bm r)&=\Delta_{pdw} \, \exp(i \varphi), \nonumber\\ 
\lim_{|\bm r| \to \infty} \Delta_{-\bm Q}(\bm r)&=\Delta_{pdw},
\label{bc-long-distance}
\end{align}
where $\tan\varphi=y/x$, where $\bm r=(x,y)$. Notice that we require the amplitude $\Delta_{pdw}$ asymptotically to be the same for both $\Delta_{\pm \bm Q}$ so that at long distances we have a PDW (LO) state.

On the other hand, since $\Delta_{\bm Q}$ has a vortex, its amplitude must vanish at the origin. Instead, $\Delta_{-\bm Q}$ does not have to vanish at the origin, and it will take some finite value, which we denote as $\Delta_{ff}$. In other words, in the presence of a half-vortex the superconducting order has an FF component. Hence, at short distances the half-vortex must have the behavior
\begin{align}
 \lim_{|\bm r| \to 0} \Delta_{\bm Q}(\bm r)&=0, \nonumber \\
 \lim_{|\bm r| \to 0} \Delta_{-\bm Q}(\bm r)&=\Delta_{ff}.
 \label{bc-short-distance}
\end{align}
\noindent The precise profile of the configuration of the half-vortex depends on the parameters of the PDW free energy $\mathcal{F}_{pdw}$ of Eq. \eqref{LG-PDW}: the stiffness $\kappa$, the PDW critical temperature  $T_c^{pdw}$, and the coupling constants $u$ and $\gamma<0$.  The way the asymptotic values, $\Delta_{pdw}$ and $\Delta_{ff}$, are attained depends on all the parameters of the free energy. There are two significant length scales (which also depend on these parameters): the scale over which the amplitude of $\Delta_{\bm Q}(\bm r)$ decreases from $\Delta_{pdw}$ to zero (the ``core'' of the half-vortex), and the scale over which $\Delta_{-\bm Q}(\bm r)$ interpolates between $\Delta_{pdw}$ and $\Delta_{ff}$ (the FF ``halo'' of the half-vortex). 

In order to obtain an explicit expression for the field configuration of the half-vortex we will use a non-linear sigma model approximation similar to the one used by Wang and coworkers  in their study of the PDW halo of a superconducting vortex \cite{Wang-2018} (see also Ref. \cite{Babaev02pr}). Thus, we define a three-component unit vector field $\bm n (\bm r)$ such that
\begin{equation}
    {\bm n}(\bm r)  =\frac{1}{\Delta} \, (\textrm{Re} \Delta_{\bm Q}(\bm r), \textrm{Im} \Delta_{\bm Q}(\bm r), \Delta_{-\bm Q}(\bm r))
    \label{n-nlsm}
\end{equation}
with $\bm n^2 (\bm r)=1$ everywhere, and where $\Delta$ will be determined below. Here we used that $\Delta_{-\bm Q}$ is a real field.
With these assumptions the free energy density of the PDW, Eq. \eqref{LG-PDW}, becomes [with $\bm n =(n_x, n_y, n_z)$]
\begin{align}
\mathcal{F}_{pdw}&={\kappa} \Delta^2 \left(\nabla {\bm n}\right)^2+(T_c^{pdw}-T)\Delta^2+u \Delta^4\nonumber\\
&- |\gamma|  \Delta^4 n_z^2(n_x^2+n_y^2).
\label{L-pdw-nslm1}
\end{align}
We will set $\Delta$ to be the value $\bar \Delta$ that minimizes the free energy in the uniform PDW phase; that is, using
\begin{equation}
  {\bm n}_{pdw}=\frac{1}{\sqrt{2}} (1, 0, 1)  
\end{equation}  
we find 
\begin{equation}
 \bar \Delta=\sqrt{\frac{2(T_c^{pdw}-T)}{4u-|\gamma|}}.
 \label{Delta-bar}
\end{equation}
The free energy density $\mathcal{F}_{nlsm}$ of the unit vector field $\bm n (\bm r)$ becomes
\begin{equation}
\mathcal{F}_{nlsm}={\bar \kappa} (\nabla \bm n)^2-v n_z^2 (n_x^2+n_y^2)+\textrm{const},
\label{F-nlsm}
\end{equation}
where we used the definitions ${\bar \kappa} \equiv \kappa \bar \Delta^2$ and $v \equiv  |\gamma|  \bar \Delta^4$, with $\bar \Delta$ given in Eq. \eqref{Delta-bar}.

We will now construct the half-vortex of the PDW using the non-linear sigma model (NLSM) of Eq.\eqref{F-nlsm}. We assume that the  ordering wave vector of the PDW is oriented along the $x$ axis and write 
$\mathbf{Q}\equiv Q\mathbf{e}_x$. We will also  assume that the phase field $\varphi(\mathbf{r})$ winds by $2\pi$, and we define its branch cut along the positive $x$ axis. With these assumptions we define the unit vector
\begin{equation}
\mathbf{e}_{\mathbf{r}}=\cos [\varphi(\mathbf{r})] \mathbf{e}_x+\sin [\varphi(\mathbf{r})]\mathbf{e}_y,    
\label{e_r}
\end{equation} 
and write the $O(3)$ NLSM field $\bm n(\bm r)$ in the form
\begin{equation}
\mathbf{n}(\mathbf{r})=
\text{sin}[\alpha(\mathbf{r})]
\mathbf{e}_{\mathbf{r}}
+
\cos[\alpha(\mathbf{r})]\mathbf{e}_{z}.
\label{alpha}
\end{equation} 

 In Sec. \ref{LG-section} we defined the half-vortex as a configuration in which the phase field of $\Delta_{\bm Q}$ winds by $2\pi$ at infinity while the order parameter field $\Delta_{-\bm Q}$ does not wind and is defined to be real. The order parameters of the PDW in the half-vortex state are required to obey the boundary conditions of Eq. \eqref{bc-long-distance} (at long distances) and Eq. \eqref{bc-short-distance} (at short distances). As a result, the half-vortex has an FF type order within the core and asymptotically far from the core is of LO-type. Such a state breaks inversion symmetry in the core of the half-vortex. In terms of the NLSM field $\bm n(\bm r)$ the boundary conditions of Eqs. \eqref{bc-long-distance} and \eqref{bc-short-distance} become 
\begin{align}
\lim_{r\rightarrow 0}{\bm n}(\mathbf{r})&=(0, 0, 1),\label{nlsom-short}\\
\lim_{r \rightarrow \infty}{\bm n}(\bm r)&=\frac{1}{\sqrt{2}} (\cos \varphi(\bm r),\sin \varphi(\bm r), 1),    
\end{align}
where, as before, we defined the phase $\varphi (\bm r)$, to be the azimuthal angle measured from the positive $x$ axis, with $\tan \varphi(\bm r)=y/x$, which winds by $2\pi$ on a large circle. To satisfy these boundary conditions we will require the field $\alpha(\bm r)$ to be isotropic, $\alpha(\bm r)\equiv \alpha(r)$, and to satisfy the boundary conditions
\begin{equation}
\lim_{r\rightarrow 0}\alpha(r)=0,\;\;\; \lim_{r\rightarrow \infty}\alpha(r) = \pi/4.
\label{bc-alpha}
\end{equation}

\begin{figure}[hbt]
\subfloat[\label{figure_2_1}]{
{\includegraphics[width = .5\textwidth]{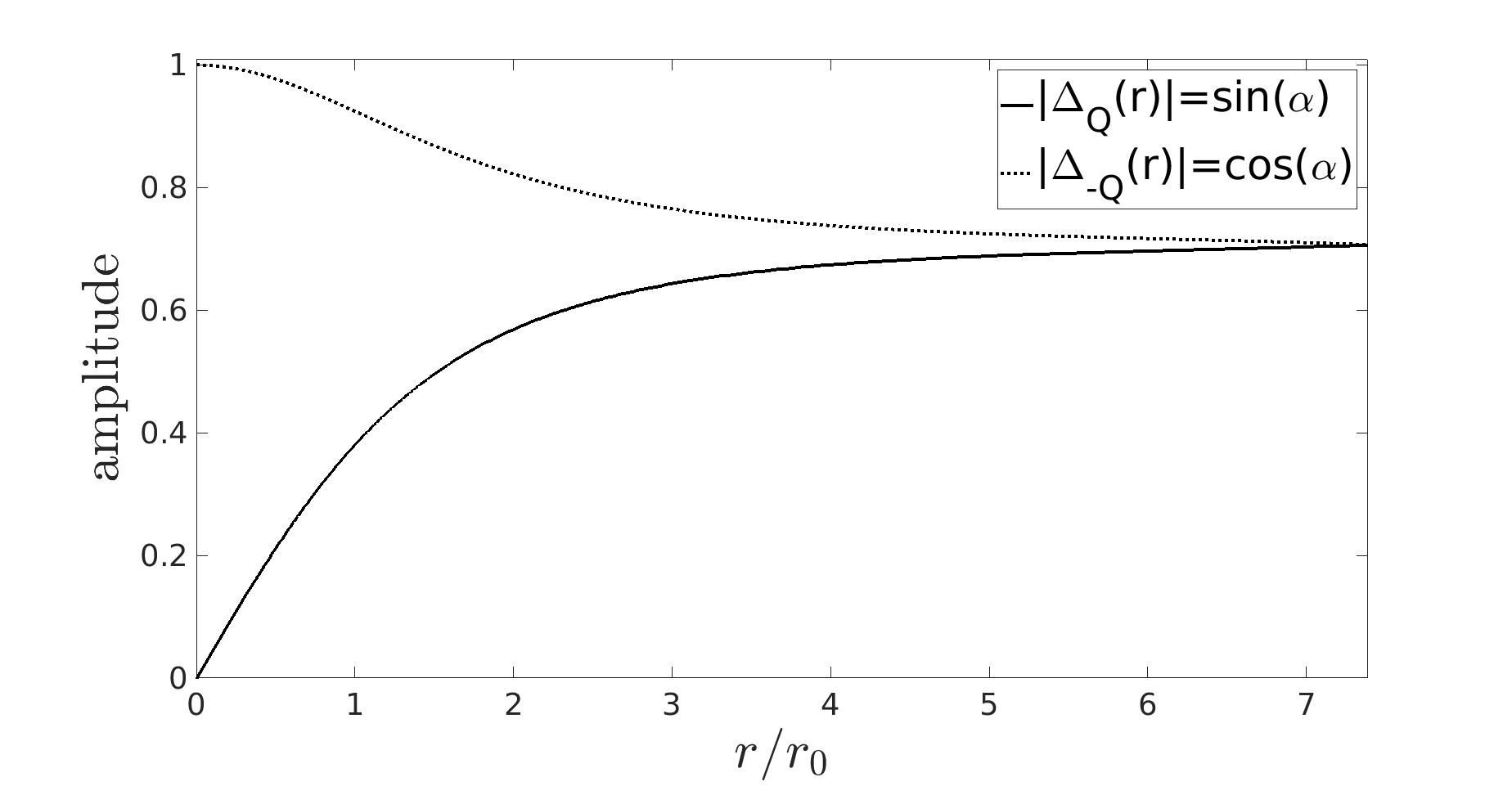}}
}
\caption{Plots of the profiles for the two components of the PDW order parameters, $|\Delta_{\pm \bm Q}(r)|$, measured in units of $\bar\Delta$ given in Eq. \eqref{Delta-bar}. We  define the vortex radius to be the place where $|\Delta_{-\bm Q}|$ has an inflection point, in this plot this is at $r/r_0=1$.}
\label{figure_2}
\end{figure}

After performing some algebra, we can recast the total free energy of the NLSM, Eq.\eqref{F-nlsm} into the following form:
\begin{align}
    F[\mathbf{n}]&=2\pi \bar \kappa \int r dr \Bigg[ 
    \Big( \frac{\partial \alpha}{\partial r}\Big)^2+\frac{\text{sin}^2(\alpha)}{r^2}
    -\frac{v}{\bar \kappa}\text{sin}^2(\alpha)\text{cos}^2(\alpha)
    \Bigg]
    \\
    &=2\pi \bar \kappa \int dt \Bigg[ 
    \Big( \frac{\partial \alpha}{\partial t}\Big)^2+\text{sin}^2(\alpha)
    -e^{2t}\text{sin}^2(\alpha)\text{cos}^2(\alpha)
    \Bigg],
    \label{F-nlsm-alpha}
\end{align}
where in line we defined
$r_0^2=\bar \kappa/v$, and made the change of variables  $ t=\text{ln}(r/r_0)$, where $t\in (-\infty,\infty)$, for $r\in [0, \infty)$. In our numerics the half-vortex radius will be set to to be $r_0/a_0=8n$ where $a_0$ is the lattice spacing and $n$ is an integer, which we vary.  Upon extremizing the free energy $F(\bm n)$ of Eq.\eqref{F-nlsm-alpha} we find that $\alpha(t)$ must obey the ``equation of motion'':
\begin{equation}
  \frac{d^2\alpha}{dt^2}=\frac{1}{2}\Big[
1-\text{cos}(2\alpha(t))e^{2t}
\Big]\text{sin}[2\alpha(t)]
\label{eom-alpha}
\end{equation}
such that the boundary conditions of Eq. \eqref{bc-alpha} now become
\begin{equation}
 \lim_{t \to -\infty}\alpha(t)=0, \qquad   \lim_{t \to \infty}\alpha(t)=\pi/4.
 \label{bc-alpha-t}
\end{equation}

A numerical solution of the equation of motion \eqref{eom-alpha} yields the optimal solution for the half-vortex. Plots of the magnitudes of the PDW components (in units of $\bar{\Delta}$) as a function of distance from the vortex core are provided in \figref{figure_2_1}. As is clear from this figure, as the amplitude $\Delta_{\bm Q}(r)$ {\it decreases} as $r \to 0$, the amplitude of $\Delta_{-{\bm Q}}(r)$ {\it increases} as $r \to 0$. In other words, the core of the half-vortex behaves as a FF state which breaks inversion symmetry. The  parameter $r_0$ can be used to define the radius of the half-vortex and it is set at the inflection of the $\Delta_{-{\bm Q}}(\mathbf{r})$ field (at $r/r_0=1$ in \figref{figure_2}), which is related to the coherence length of the Cooper pairs. We can then see that the degree of inversion symmetry breaking depends on the area of the core of the half-vortex. In the absence of the uniform component $\Delta_0$ the branch cut of the half-vortex is unobservable resulting in a  free energy that is only logarithmically divergent. However, if $\Delta_0\neq 0$ the branch cut becomes observable and behaves as a {\it domain wall}. In this case the energy of the half-vortex becomes linearly divergent.

The profiles of the Abrikosov vortex and the double dislocation are obtained using a similar approach. In the cases of these topological defects both components of the PDW order parameter 
fields $\Delta_{\pm \bm Q}(\bm r)$ have vorticity. In the case of the Abrikosov vortex we consider solutions of the Landau-Ginzburg equations with the the same vorticity and set  
$\Delta_{-{\bm Q}}(\mathbf{r})=\Delta_{{\bm Q}}(\mathbf{r})\equiv \Delta(\bm r)$, where $\Delta(\bm r)$ is a conventional Abrikosov vortex. 
Instead, in  the case of the double dislocation we consider solutions in which the two PDW order parameters have equal and opposite vorticity, 
$\Delta_{\bm Q}(\bm r)=\Delta(\bm r)$ and $\Delta_{-{\bm Q}}(\mathbf{r})=\Delta^*(\mathbf{r})$, where again $\Delta(\bm r)$ is a conventional vortex solution.
The vortex solution has the form 
\begin{equation}
   \Delta(\bm r)=\bar \Delta f(r/r_0) \exp(i \varphi(\bm r)), 
   \label{profile}
\end{equation}
where $r_0$ is the radius of the vortex and $\varphi(\bm r)$ is the azimuthal angle on the plane. 
The profile function $f(r/r_0)$ is calculated numerically and satisfies the boundary conditions $\lim _{r\to 0} f(r/r_0)=0$ and $\lim_{r \to \infty} f(r/r_0)=1$.


\section{PDW Bogoliubov-de Gennes Hamiltonian with Topological defects}\label{Hamiltonian}

In this section we describe the Bogoliubov-de Gennes Hamiltonian  on the square lattice with a cuprate electronic structure with the configurations of the PDW order parameter in the background of the topological defects introduced in Secs. \ref{LG-section} and \ref{NLSM}. We will focus on the effects on the electronic states.

In order to study the effects of the different topological defects of the PDW state in the associated CDW order and in the electronic structure   we consider a model which couples our electronic degrees of freedom to the local amplitude of the superconducting order parameter in the background of each defect, denoted by the pair field $\Delta(\bm r, \bm r')$ in the bonds $(\bm r, \bm r')$ of the square lattice. In what follows we will define $\Delta(\mathbf{r},\mathbf{r}')$ as the embedding to the square lattice of the solutions of the Landau-Ginzburg equations for the pair field of a PDW in the background of the different topological defects. 
\par We consider four configurations of the PDW order parameter, 
(1) the uniform PDW state, (2) the half-vortex, (3) the Abrikosov vortex and (4) the double dislocation. Defined relative to the origin of the $\bm r$ plane, the  configurations of the PDW order parameter take the following generic form
\begin{equation}
    \Delta_i(\mathbf{r},\mathbf{r}')=\bar \Delta \, F(\mathbf{r},\mathbf{r}')f_i(\mathbf{r}).
    \label{PDW_OP_def}
    \end{equation}
Here $\bar \Delta$ is the amplitude of the SC gap given in Eq. \eqref{Delta-bar}, $F(\mathbf{r},\mathbf{r}')$ is the SC form factor, and $f_i(\mathbf{r})$ are the profiles and winding numbers of the four configurations of the PDW order parameters listed above. On a square lattice the form factor $F(\bm r, \bm r')=1$ for an $s$-wave SC state. In a  $d$-wave SC state, which is our focus, the form factor is $F(\bm r, \bm r')= 1 (-1)$ for a bond $(\bm r, \bm r')$ on the $x$ axis ($y$ axis) of the square lattice, and changes sign under a $\pi/2$ rotation. Using the results of section \ref{NLSM} the explicit forms of the functions $f_i(\mathbf{r})$'s  are
    \begin{equation}
    \begin{aligned}
        f_1(\mathbf{r})
        =&
        \cos(\mathbf{Q}\cdot \mathbf{r}),
        \\
        f_2(\mathbf{r})
        =&\frac{1}{2}
       \Big(\sin\big(\alpha(\mathbf{r})\big)
         e^{i\mathbf{Q}\cdot \mathbf{r}+i\varphi(\mathbf{r})}
         +\cos\big(\alpha(\mathbf{r})\big)
         e^{-i\mathbf{Q}\cdot \mathbf{r}}
         \Big),
         \\
        f_3(\mathbf{r})
        =&
       f(r/r_0)
        \cos(\mathbf{Q}\cdot \mathbf{r})
        e^{i\varphi(\mathbf{r})},
         \\
        f_4(\mathbf{r})=&
        f(r/r_0)
         \cos[\mathbf{Q}\cdot \mathbf{r}+\varphi(\mathbf{r})].
         \label{order-parameter-configs}
    \end{aligned}
    \end{equation}
\begin{figure*}
\subfloat[\label{figure_3_1}]
{\includegraphics[width = .33\textwidth]{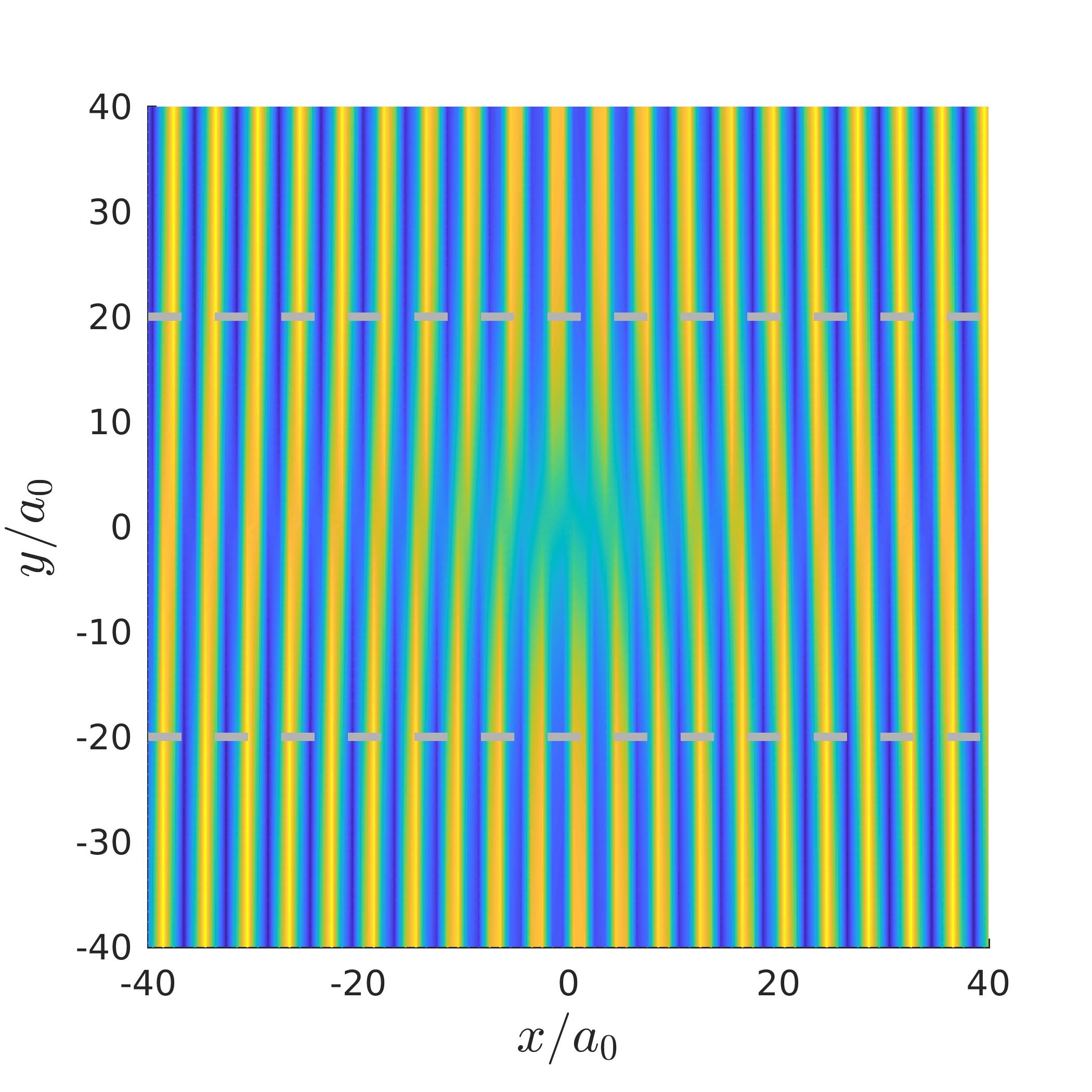}
} 
\subfloat[\label{figure_3_2}]
{\includegraphics[width = .33\textwidth]{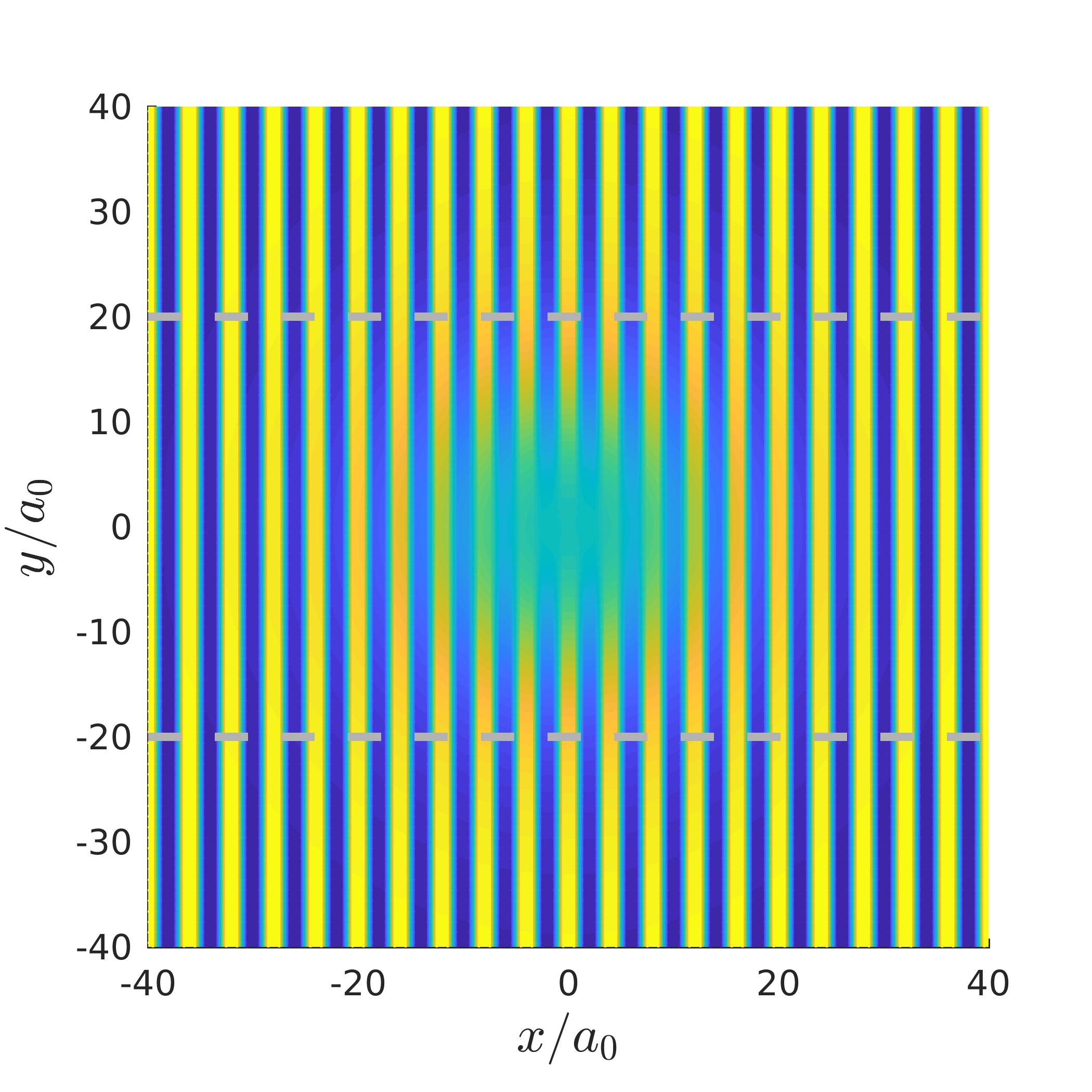}
} 
\subfloat[\label{figure_3_3}]
{\includegraphics[width = .33\textwidth]{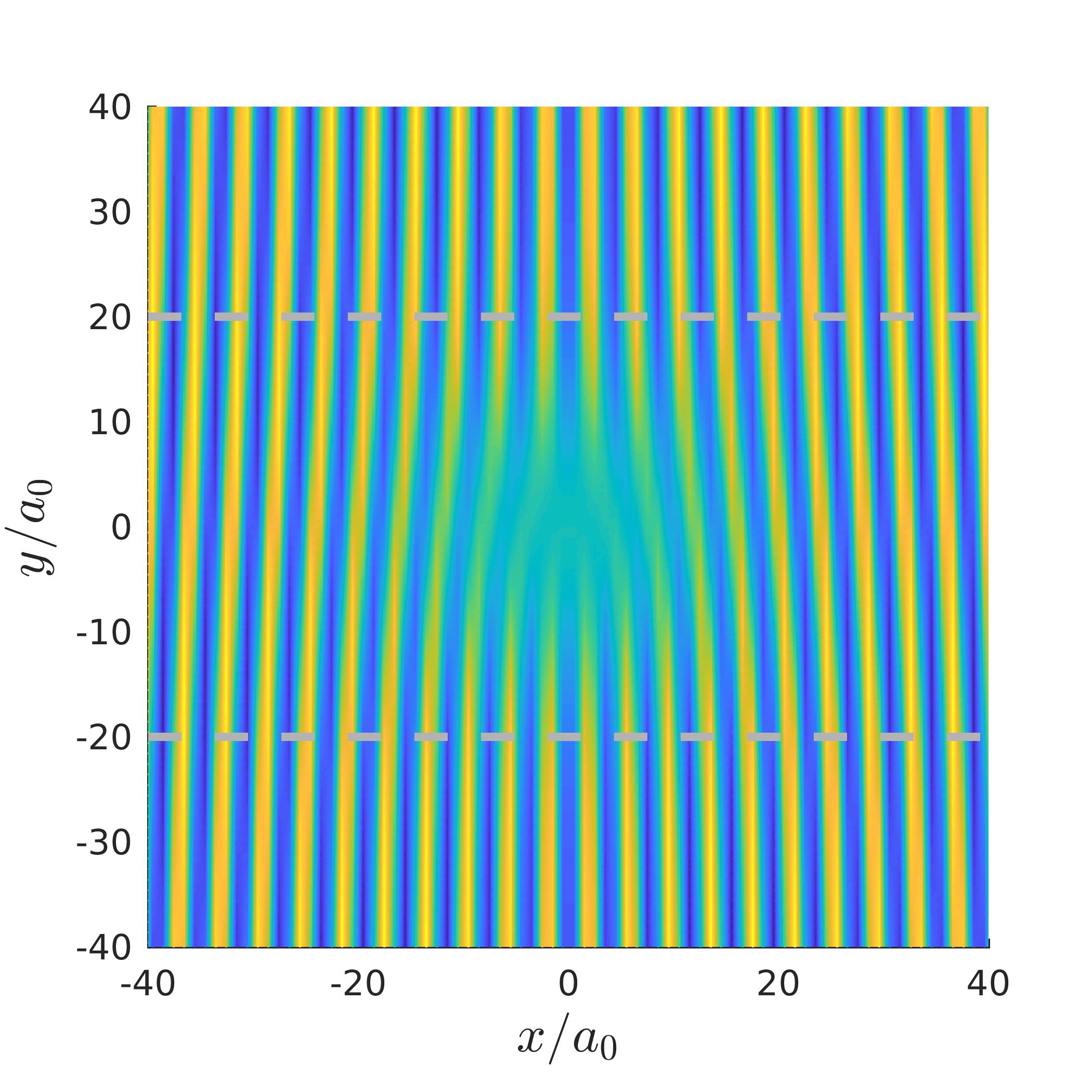}
}
\\
\subfloat[\label{figure_3_4}]
{\includegraphics[width=0.33\textwidth]{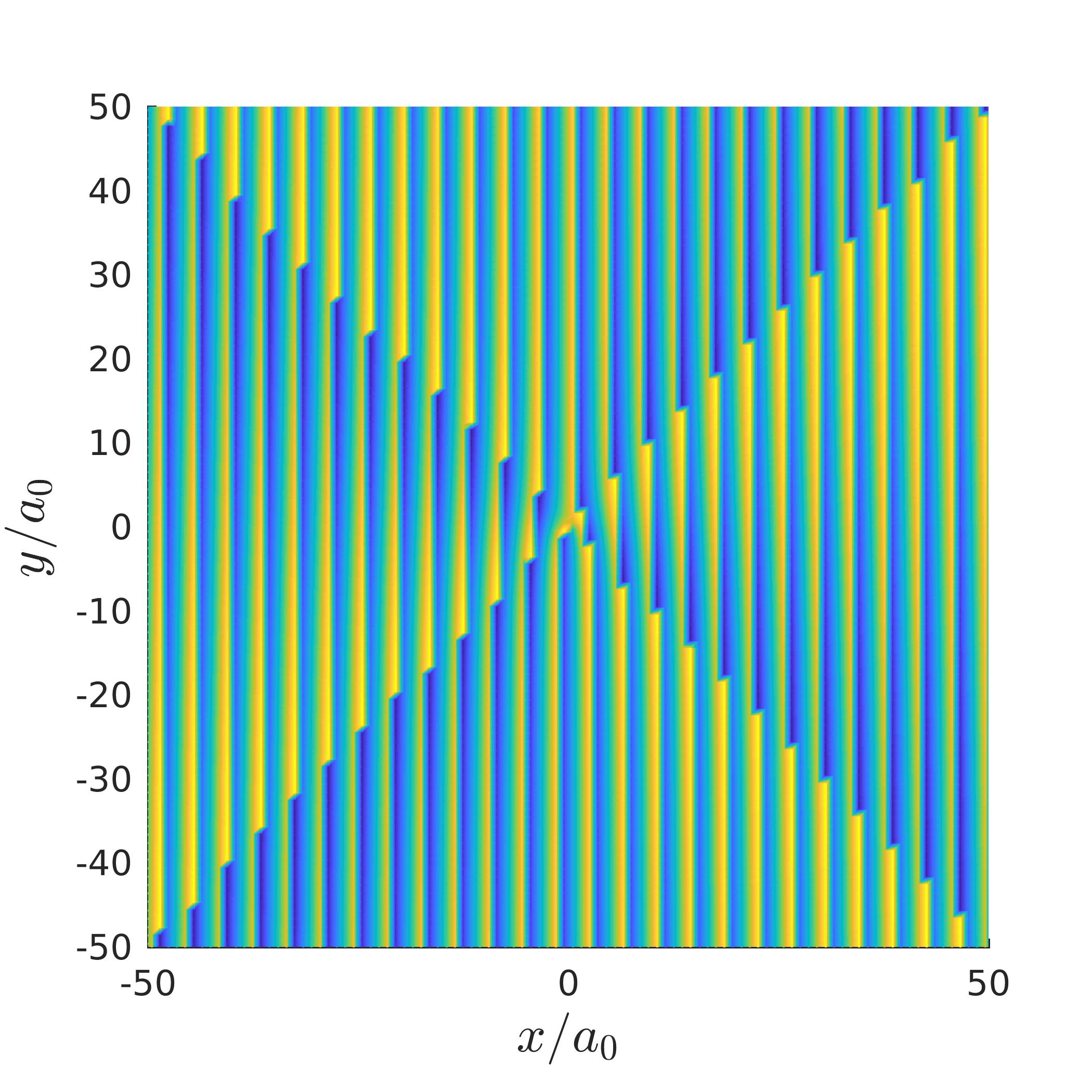}}
\subfloat
[\label{figure_3_5}]
{\includegraphics[width=0.33\textwidth]{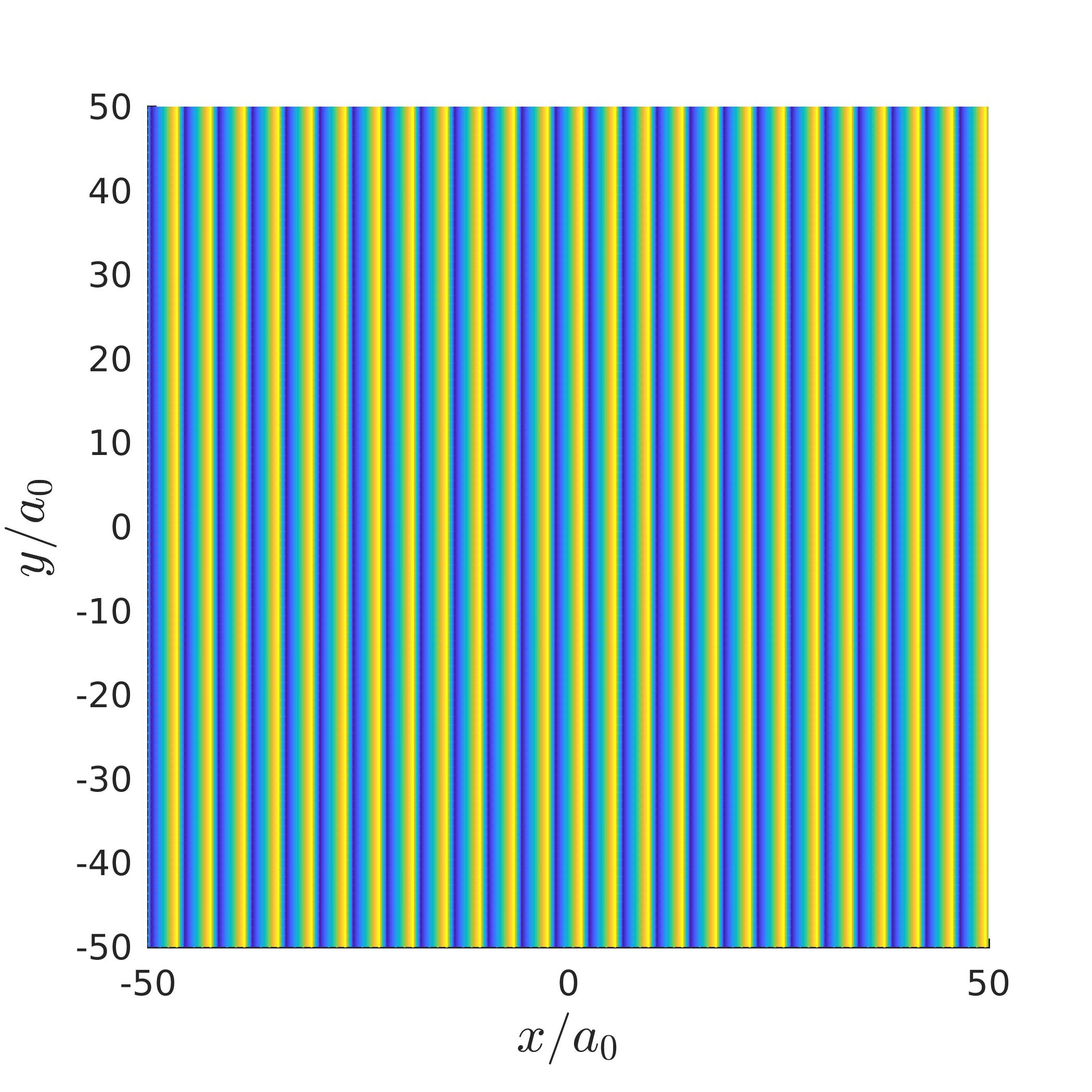}}
\subfloat[\label{figure_3_6}]
{\includegraphics[width=0.33\textwidth]{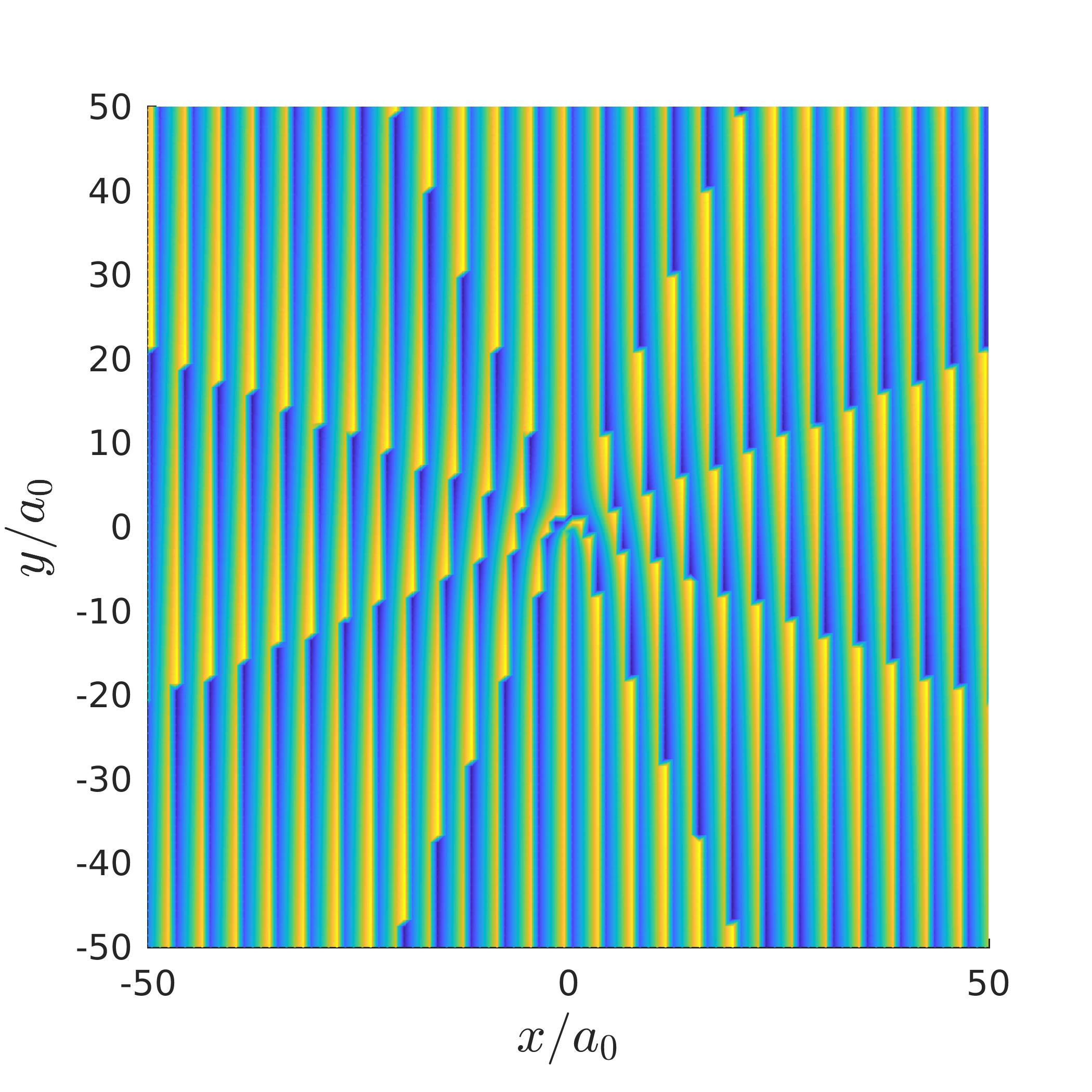}} 
\\
\subfloat[\label{figure_3_7}]
{\includegraphics[width=0.45\textwidth]{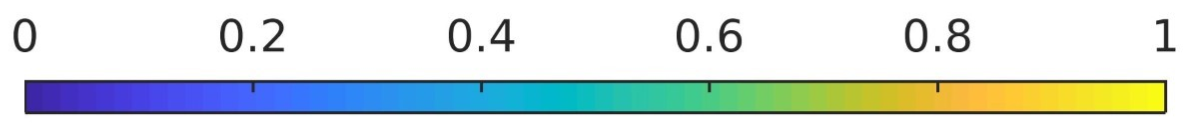}} 
\caption{(Online Color) Plots of the real parts of the  daughter CDWs $\rho_{2\mathbf{Q}}(\mathbf{r})$  given by \equref{2Q} defined in terms of the functions given in \equref{order-parameter-configs}. Here the three topological defects have a vortex radius of $r_0=16a_0$ being: (a) the half-vortex, (b) the Abrikosov vortex, and (c) the double dislocation. The dotted lines shown in light gray are guides to count the CDW peaks to find the associated Burgers vectors, being the difference of the top line and the bottom, As expected, in (a) for the half-vortex (which has a single dislocation) they skip by one, in (b) for the Abrikosov vortex they do not skip, and in (c) for the double dislocation they skip by two. In the second row, plots (d)-(f), we include the corresponding $\text{arg}\big(\rho_{2\mathbf{Q}}\big)$  for (d) the half-vortex, (e) the Abrikosov vortex and (f) the double dislocation. The jumps in phase seen in these panels are $\pi/2$, and they sum up to the expected dislocation charge associated with a given defect.
The color bar provided in (g) pertains to all plots. For (a)-(c) it corresponds to the scale of the defect (that is, we normalized these plots), and for (d)-(f) it represents units of $2\pi$.
}
\label{figure_3}
\end{figure*}
\begin{figure*}
\vspace{0pt}
\hspace{0pt}
\subfloat[\label{figure_4_1}]
{\includegraphics[width=0.33\textwidth]{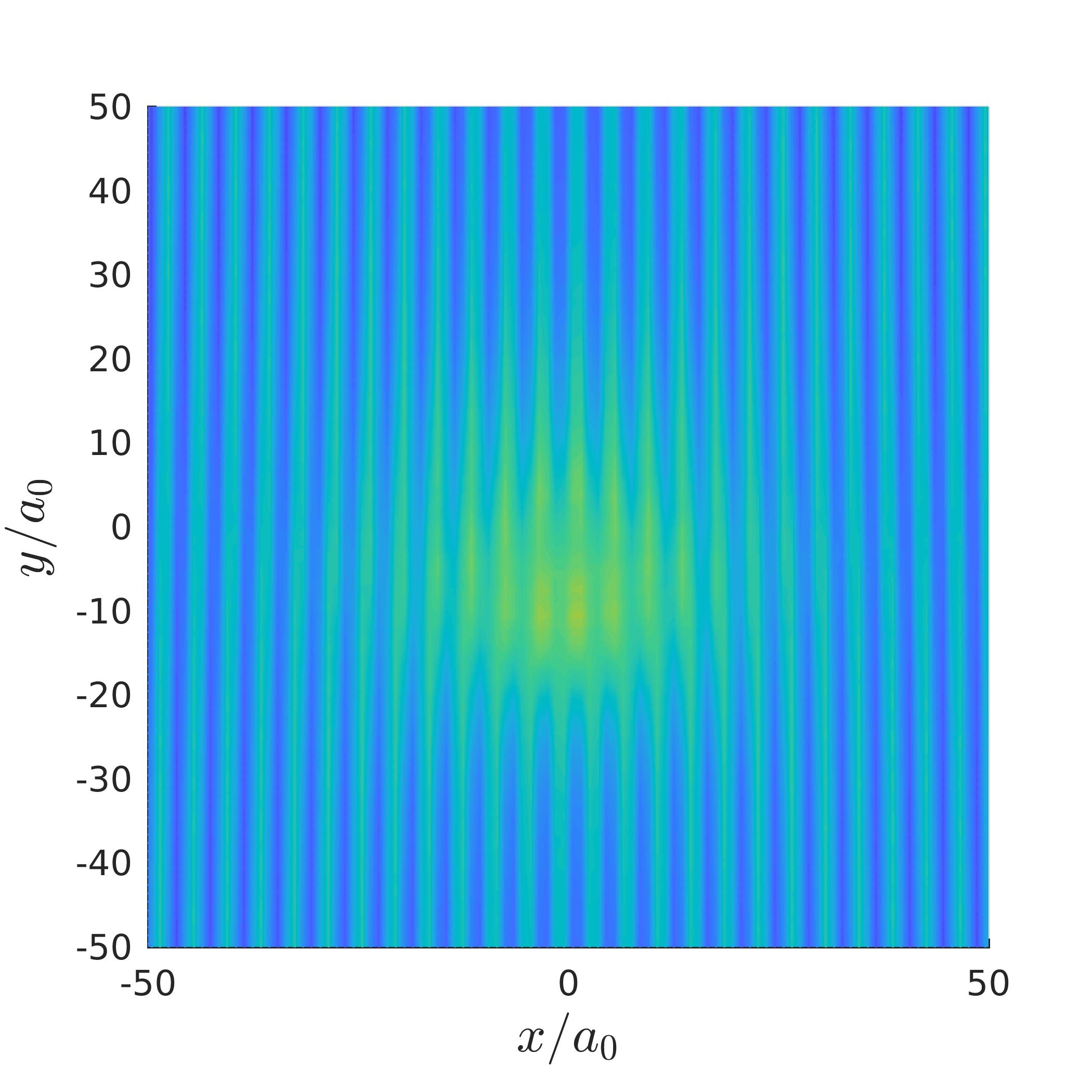}}
\subfloat
[\label{figure_4_2}]
{\includegraphics[width=0.33\textwidth]{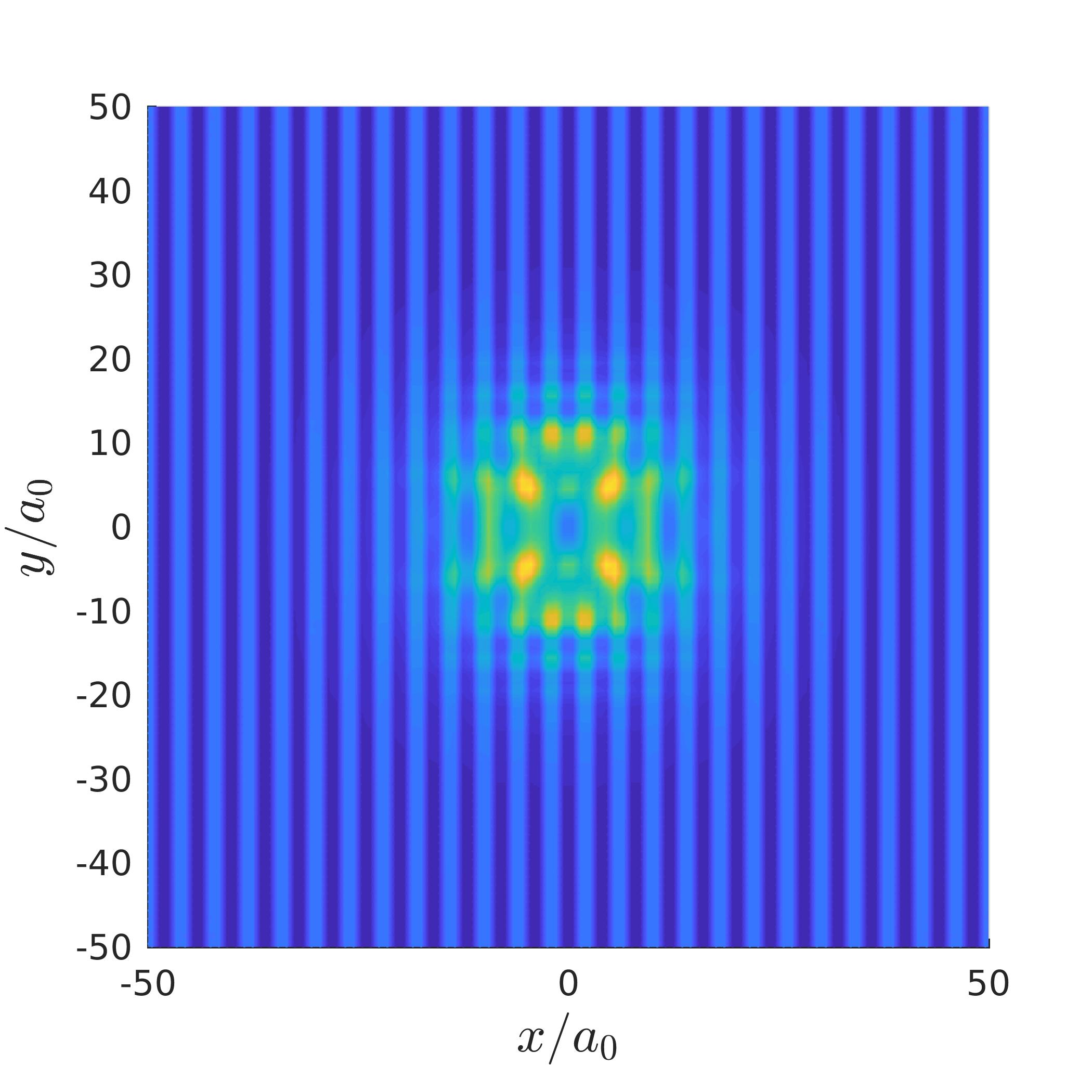}}
\subfloat[\label{figure_4_3}]
{\includegraphics[width=0.33\textwidth]{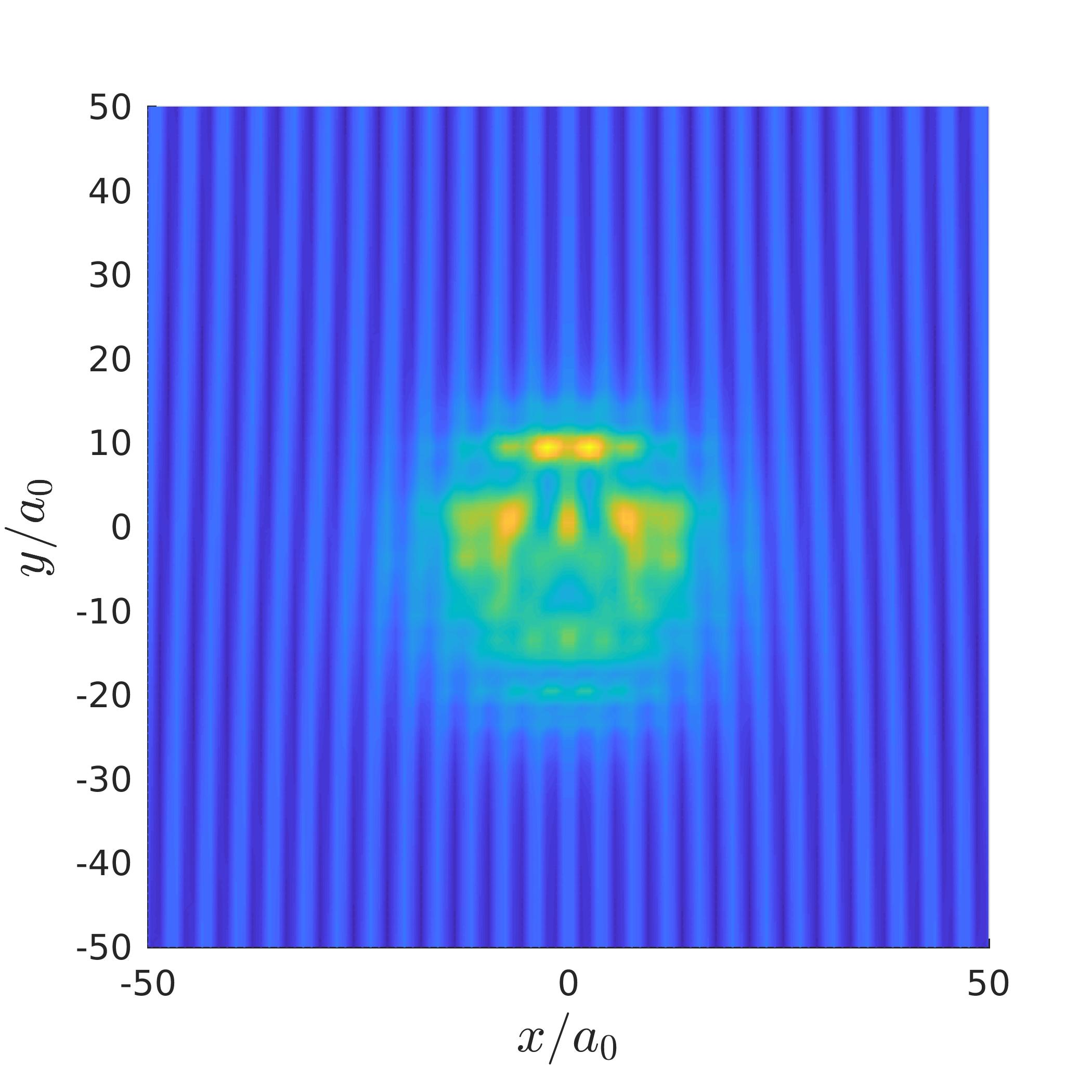}} 
\\
\subfloat[\label{figure_4_4}]
{\includegraphics[width=.5\textwidth]{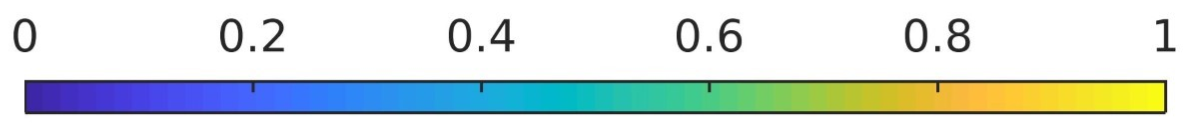}} 
\hspace{0pt}
\caption{(Online Color) The LDOS profiles of a PDW in the presence of the three topological defects given in Eq. \eqref{order-parameter-configs}. Using the states of the BdG Hamiltonian [Eq.\eqref{H_BdG}] the LDOS evaluated at $\omega=0.25\;\bar{\Delta}$ for (a) the half-vortex, (b) the Abrikosov vortex, and (c) the double dislocation. In (d) we provide a normalized color bar for each of the plots. Note that each specific defect is rescaled according to their own maximum value, so that the vortex core shows up clearer.} 
\label{figure_4}
\end{figure*}

{\noindent}Here $\alpha(\mathbf{r})$ is the angle we used to parametrize the NLSM in Eq.\eqref{alpha}, whose numerical solution was found in Sec. \ref{NLSM}. 
The vortex profile function $f(r/r_0)$ is defined in Eq. \eqref{profile}. Finally, the complex phase, $\varphi(\mathbf{r})$, is the azimuthal angle on the plane, and it winds by $2\pi$ in all the expressions in which it appears.
In Figs. \ref{figure_3_1}-\ref{figure_3_3} we show the profile of the composite order parameters $\rho_{2\mathbf{Q}}(\mathbf{r})$ (that is, we take the real part of this expression) in the presence of the three topological defects of our PDW order. The Burgers vector associated with a given charge dislocation can be found by simply counting the difference in the CDW peaks found above and below the vortex cores. The dotted lines are guides used to indicate where to do the counting. The profiles of the associated CDW order in the presence of the defects are shown in Sec. \ref{LDOS}. We also include $\text{arg}(\rho_{2\mathbf{Q}})$ in Figs. \ref{figure_3_4}-\ref{figure_3_6}. Notice the four jumps in phase by $\pi/2$ for the half-vortex [Fig. \ref{figure_3_4}] and the eight jumps for the double dislocation [Fig. \ref{figure_3_6}], while there are none for the full vortex [Fig. \ref{figure_3_5}], reflecting the expected amount of dislocation charge present in each defect.

\subsection{Hamiltonian and Observables}
\label{BdGH}

The Bogoliubov-de Gennes (BdG) Hamiltonian for the lattice model is
        \begin{align}
    \hat{H}_i&= -\sum_{\mathbf{r},\mathbf{r}',\sigma} t(\mathbf{r}-\mathbf{r}')
    \hat{c}^\dagger_{ \mathbf{r}\sigma}\hat{c}_{\mathbf{r}'\sigma}
    +\sum_{\mathbf{r},\mathbf{r}'}\Big(\Delta_i(\mathbf{r},\mathbf{r}')
    \hat{c}^\dagger_{ \mathbf{r}\uparrow}\hat{c}^\dagger_{\mathbf{r}'\downarrow}
    +\text{H.c.}    
    \Big)
    \label{H_BdG}
    \end{align}
for each configuration of the SC amplitudes  $\Delta_i(\bm r, \bm r')$ [see Eqs. (\ref{PDW_OP_def}) and (\ref{order-parameter-configs})].
The normal state band structure we will be using is parameterized with values of hopping amplitudes of a 
tight-binding model on the square lattice  chosen to best fit Angle-Resolved Photoemission Spectroscopy 
(ARPES) experiments in the high-temperature superconductors {\LBCO} and {\BSCCO} \cite{He-2008,Vishik-2010}. 
The explicit parameters used (in units of eV) are: 
$t=0.25,\; t^{'}=-0.031863,\;t^{''}=0.016487,\; t^{'''}=0.0076112$, and $\mu=-0.16235$. We also take the superconducting amplitude $\bar \Delta=60$ meV. 
In all cases we assumed that the superconducting order parameter $\Delta(\bm r, \bm r')$ is a
unidirectional PDW along the $x$ direction with period eight lattice spacings, with a wave vector 
$\bm Q=(\pi/4, 0)$ (in units with $a=1$).

\par Since the Fermi surface of the cuprates is not spherically symmetric, the PDW states along the nodal and anti-nodal directions have different features. The same applies for a putative FF state. Below we  will show the Bogoliubov spectrum in the core of the half-vortex resembles that of a pristine FF state trapped inside. 
A wave vector oriented along the anti-nodal direction results in a fully gapped FF state, whereas for a state oriented in the nodal direction the resulting spectrum has nodes. In the situation of interest the FF state in the core of the half-vortex of the PDW is gapped.

The details of the diagonalization procedure can be found in Appendixes \ref{BV} and \ref{numerical-diagonalization}. In short, we define a Nambu spinor, $\psi_+^T=[\mathbf{c}_{\uparrow},\mathbf{c}^\dagger_{\downarrow}]$, which helps us perform the exact diagonalization (see Appendix \ref{numerical-diagonalization}). These define our quasiparticle operators, $\hat{b}_l$ and $\hat{b}_l^\dagger$, which annihilate the BCS ground state and create single particle excitations with energy $E_l$, respectively \footnote{Here we take as the zero of energy the ground state value, $E_G$.}.
\par As in the case of a uniform superconductor, the excited states are an admixture of electrons and holes. We find our electron creation/annihilation operators are related to linear combinations of our quasiparticle operators: $ \hat{c}_{i\sigma }=
        v^*_{il}
        \hat{b}_l^\dagger
        +
        \sigma
        u_{il}\hat{b}_l  
$. Here repeated indices are summed over, and the coefficients are the real space coherence factors. 
\begin{figure*}
\vspace{0pt}
\hspace{0pt}
\subfloat[\label{figure_5_1}]
{\includegraphics[width=0.325\textwidth]{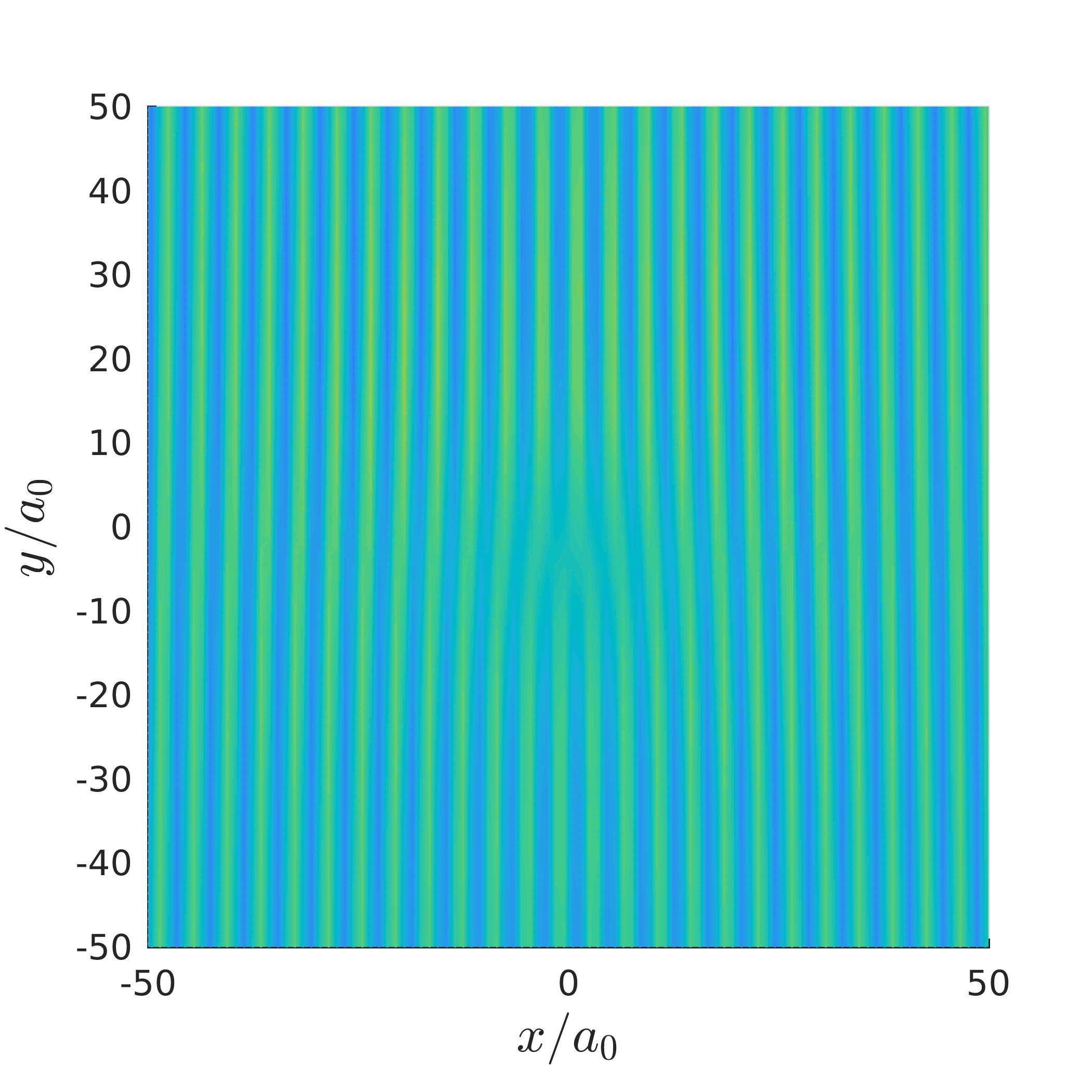}}
\subfloat[\label{figure_5_2}]
{\includegraphics[width=0.325\textwidth]{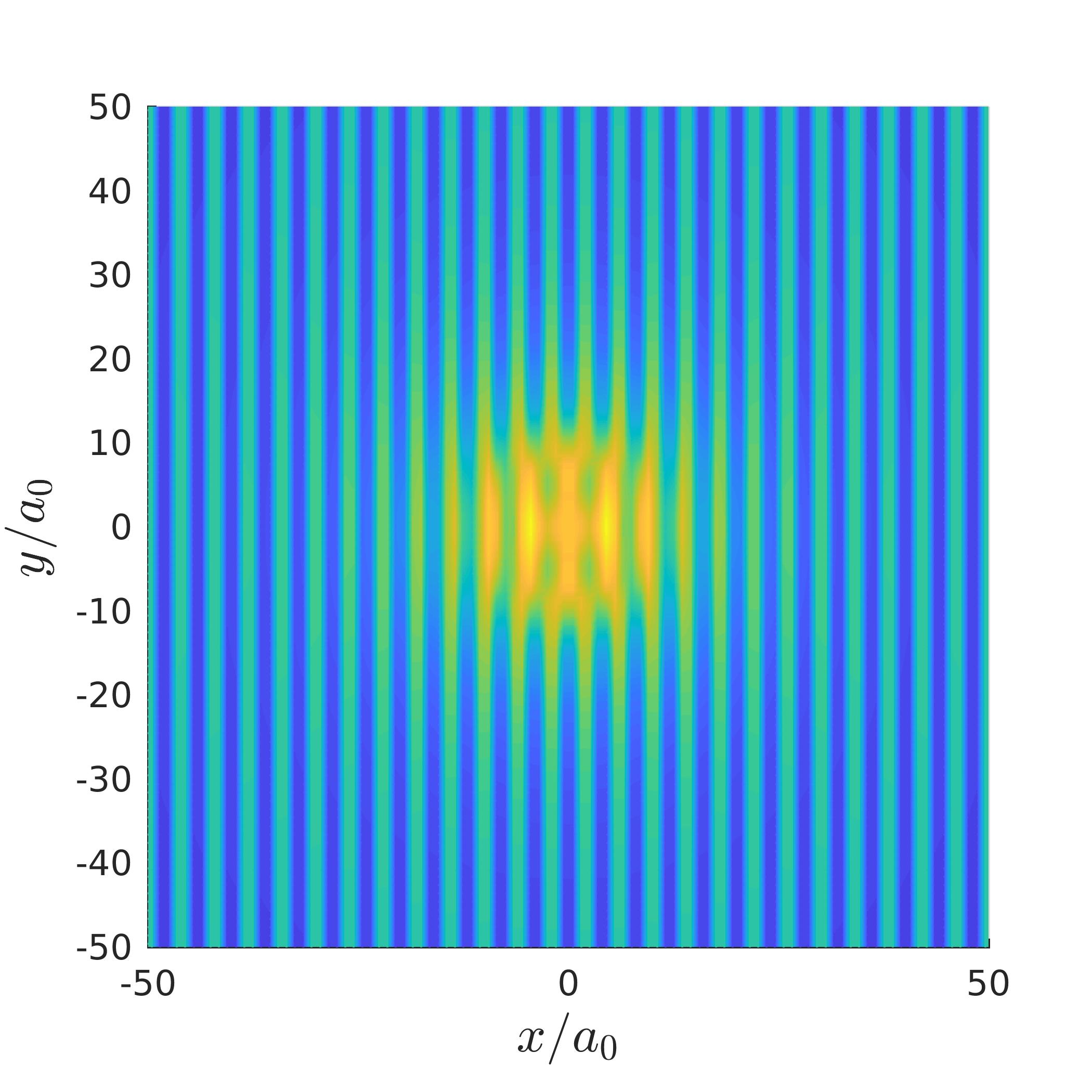}}
\subfloat[\label{figure_5_3}]
{\includegraphics[width=0.325\textwidth]{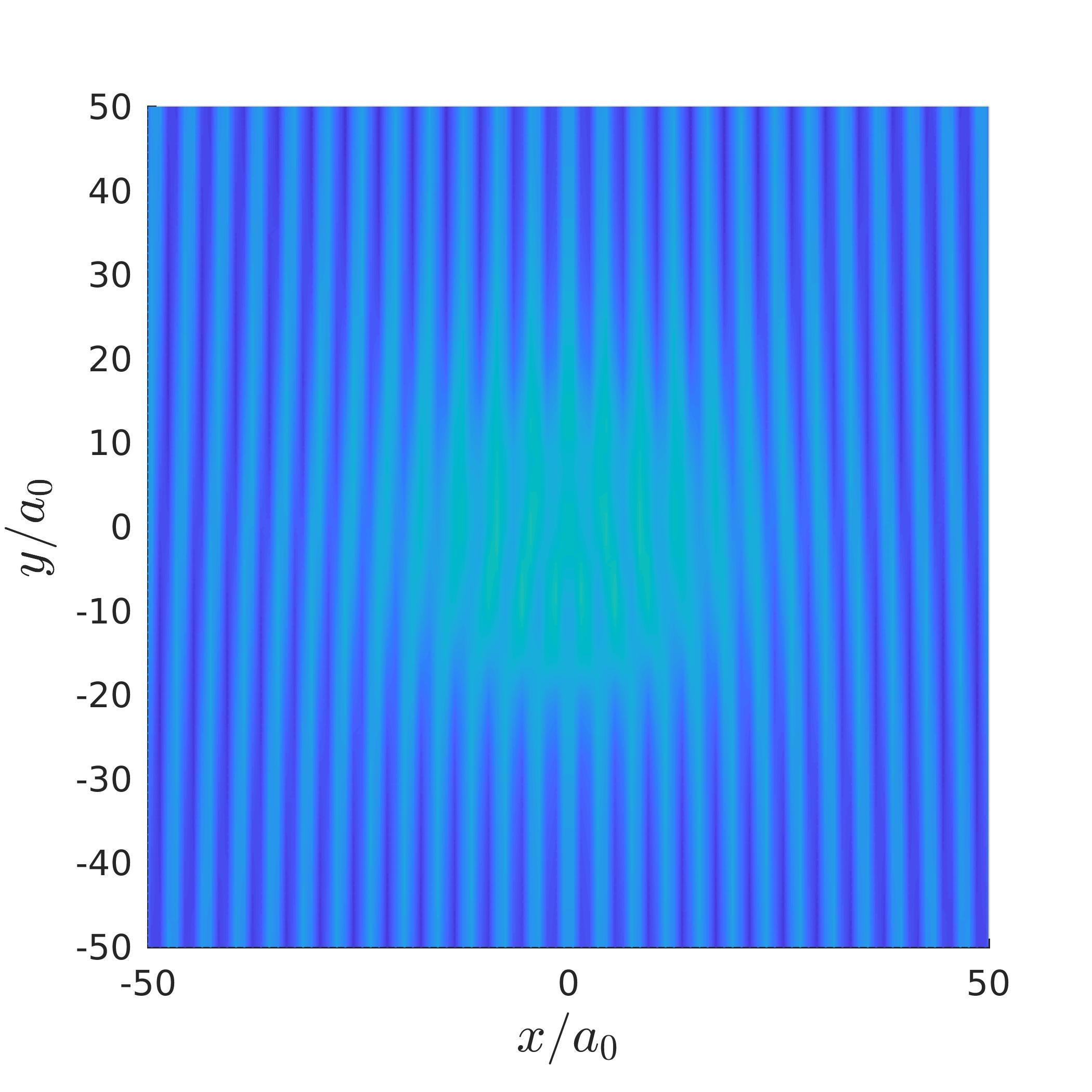}}
\\
\hspace{0pt}
\caption{(Online Color) Static charge density profiles obtained by integrating the LDOS out to a voltage of $1.25\, \bar \Delta$ in the presence of the three topological defects given in Eq.\eqref{order-parameter-configs}. Using the states of the BdG Hamiltonian [Eq. (\eqref{H_BdG})] the LDOS for a PDW defects for (a) the half-vortex, (b) the Abrikosov vortex, and (c) the double dislocation. We take a scale normalized to the specific defect for each of these plots.}
\label{figure_5}
\end{figure*}

In order to compare the spectroscopic properties of our system obtained from the states of the BdG Hamiltonian to experiment we use the zero temperature retarded Green functions and their Fourier transforms (see Appendix \ref{zero-temp-Green}). We will focus on two quantities of experimental interest, the LDOS $L(\bm r, \omega)$  and the spectral function $A(\bm k, \omega)$. In Appendix \ref{zero-temp-Green} we show that these quantities are given by
\begin{align}
 & L(\mathbf{r}_i,\omega)
  =
  -\frac{1}{\pi}\text{Im}\Big(G(\mathbf{r}_i,\mathbf{r}_i,\omega) \Big) \nonumber \\
  =
 & \frac{1}{\pi}\sum_{E_l\geq 0}
  \Bigg(
  \frac{\epsilon}{(\omega-    E_l)^2+\epsilon^2}|u_{il}|^2
  +  
  \frac{\epsilon}{(\omega+ E_l)^2+\epsilon^2}|v_{il}|^2\Bigg),
  \label{LDOS-eq}
  \end{align}
  and
  \begin{align}
  &A(\mathbf{k},\omega)
    =
    -\frac{1}{\pi }\text{Im}\Big(G(\mathbf{k},\mathbf{k},\omega) \Big)\nonumber\\
    &=\frac{1}{\pi }\sum_{E_l\geq 0}
    \epsilon
    \Bigg(
      \frac{|\tilde{u}_{l}(\mathbf{k})|^2}{(\omega- E_l)^2+\epsilon^2}
        +  \frac{|\tilde{v}_{l}(\mathbf{k})|^2}{(\omega+ E_l)^2+\epsilon^2}
    \Bigg)
    \label{Spectral}
\end{align}
\noindent where $\tilde u_l(\bm k)$ and $\tilde v_l(\bm k)$ are the eigenvectors of the BdG equations in momentum space and the energy resolution will be taken to be $\epsilon=2.5$ meV. Our simulations were also conducted on a $400\times 400$ lattice to achieve the desired resolution for our spectral functions and Fourier transforms of the LDOS. We leave the consideration of the anomalous Green functions and its relations to Cooper pair tunneling for a future study.

\subsection{Electronic Structure of the PDW Topological defects}
\label{LDOS}

In this subsection we analyze our numerical results for the LDOS, computed using 
Eq. \eqref{LDOS-eq}, for the configurations of the PDW order parameter with the three topological defects defined 
in Eq. \eqref{order-parameter-configs}. The intertwining of the PDW defects with the induced CDW order will be discussed 
in detail, as well as the structure of the charge distribution induced by these defects. The main focus will be on experimental signatures associated with the CDW pattern induced by the half-vortex and the double dislocation. The superconducting properties of the PDW half-vortex will be discussed in Sec. \ref{tunneling-DOS}.

In \figref{figure_3} we plot the profiles of (\textit{i.e.}, we take the real part of) the resulting CDW order parameter $\rho_{2\bm Q}$ near the three topological defects given in \equref{order-parameter-configs} using the definition of \equref{2Q}. We note that in our numerics we take the form factor $F(\mathbf{r},\mathbf{r}')$ to be defect-free $d$-wave. Since the PDW order breaks the point group symmetry of the lattice, the form factor associated with the unidirectional PDW phase should be an admixture of $s$-wave and $d$-wave \cite{Wang-2018}. However, as was discussed in \cite{Agterberg-09Q}, there are robust features which are essentially the same for both form factors. In Appendix \ref{spec_appendix}, \figref{figure_14}, we present the spectral functions for an order parameters with an $s$-wave form factor, but our primary focus will be on $d$-wave SC.

In \figref{figure_4}  we show the  changes in the LDOS 
of a the PDW state with the three topological defects whose CDW order parameter $\rho_{2\bm Q}$ near the defects are shown in \figref{figure_3}.
The LDOS of these defects are shown in \subfigrangref{figure_4_1}{figure_4_3} for a probing voltage of $0.25\Delta_0$. These were obtained by computing numerically the tunneling density of states of the electronic states obtained from the BdG equations for the three defects. These patterns exhibit a sinusoidal PDW oscillatory component of four lattice spacings, as expected for a CDW with ordering wave vector $\bm Q_{cdw}=2\bm Q$ 
(see Eq.\eqref{2Q}), superposed with various effects arising from the changes induced by the topological defects on the 
eigenstates of the BdG equation.  

The charge density profiles associated with each of these defects reveals some of the most salient 
signatures of the  PDW order. First and foremost, \subfigref{figure_4_1} shows that the half-vortex which can 
indeed be thought of as a dislocation in the CDW order parameter $\rho_{2\bm Q}(\bm r)$ pinned to a 
half-SC-flux-quanta. The predicted forms of the other two topological defects have been discussed in the 
literature ~\cite{Agterberg-2008,Berg-2009b,Agterberg-2020}. 
The double dislocation is shown in \subfigref{figure_4_3}. As was the case for \figref{figure_3}, the Burgers vector associated with these charge dislocations 
can be found by simply counting the difference in the CDW peaks found above and below the vortex cores. The full vortex 
has no dislocation charge [\subfigref{figure_4_2}]. Notice, however, the phase of background density wave pattern of the 
full vortex is shifted by $\pi$ relative to the pattern of the other two defects, which can easily be seen at $x/a_0=0$.

In \figref{figure_5} we plot the integrated LDOS (\textit{i.e.}, the static charge density) for all three defects to a voltage of $1.25\,\bar{\Delta}$, well above the PDW SC gap. 
A comparison of these plots with \figref{figure_3} shows, as expected, that the integrated LDOS yields the CDW pattern (for details of this approach see 
Ref. \cite{Kivelson-2003}). As expected,  in both figures the CDW order parameter is suppressed in the core of the defects where one or both components of the PDW
 order parameters $\Delta_{\pm \bm Q}$ are suppressed.

Next we  notice the additional patterns seen within the core of the defects in Figs. \ref{figure_4_2} and \ref{figure_4_3}, the Abrikosov vortex and the 
double dislocation. {\it Both} PDW  order parameters $\Delta_{\pm \bm Q}$ vanish in the core of the double dislocation and of the Abrikosov vortex; hence, the 
additional electronic structure residing in their cores, revealed by the LDOS, is due to quasiparticle states. Although the quasiparticle states are responsible for the 
additional LDOS, they are not bound to the core of these two types of topological defects. The PDW has pockets of quasiparticle and quasi-holes in momentum space.  This 
interpretation is confirmed by a computation of the Fourier transforms in momentum space of the LDOS at different energies for the Abrikosov vortex.
 Figure \ref{figure_13} in Appendix \ref{LDOS_appendix} shows the quasiparticle spectrum in the presence of the vortex, which confirms that these are  
 propagating states and are not bound to the core of the defect. Thus this structure in the LDOS has to be interpreted as due to quasiparticle interference (QPI) 
 at the defects.
\begin{figure*}[hbt]
\vspace{0pt}
\hspace{0pt}
\subfloat[\label{figure_6_1}]{\includegraphics[width=0.26\textwidth]{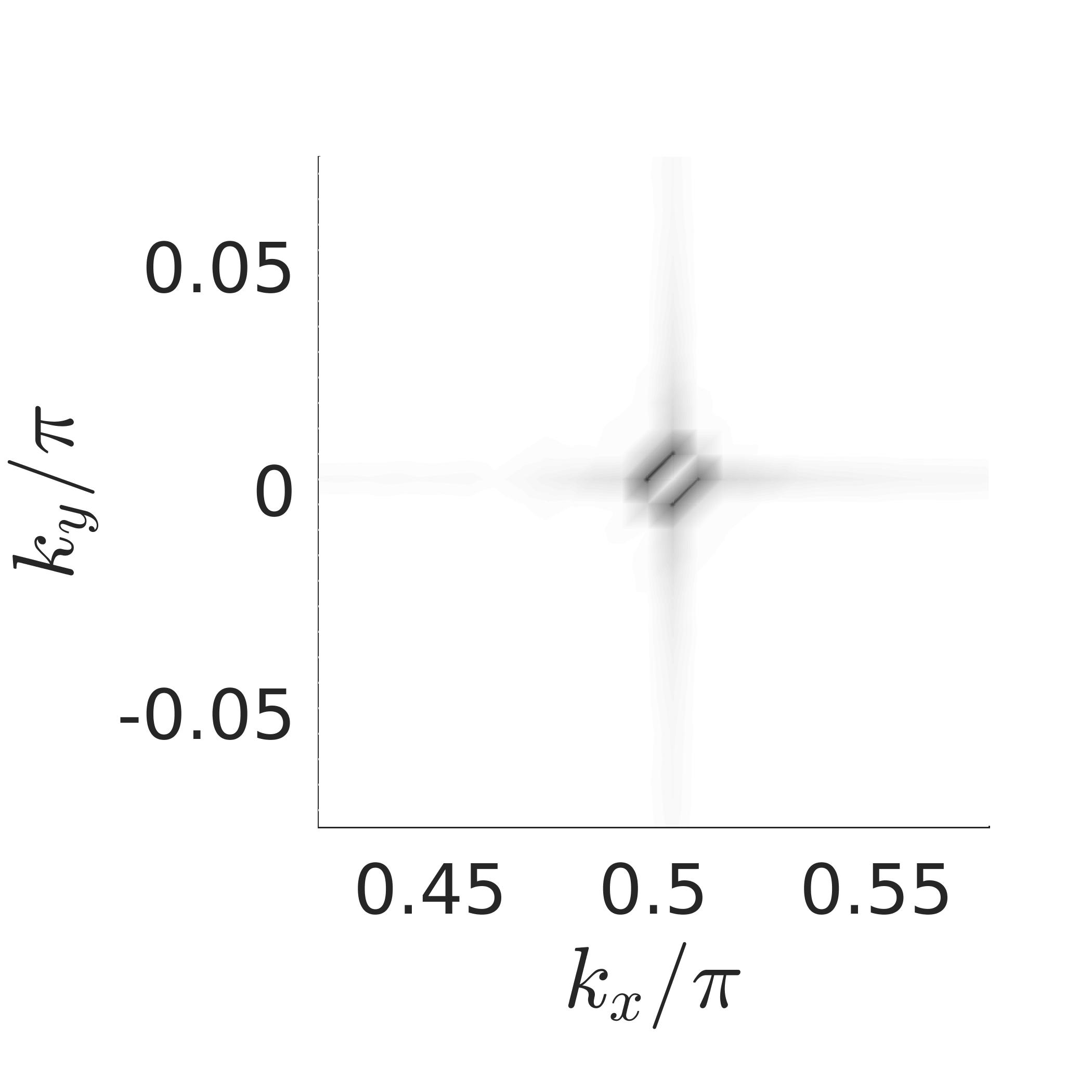}}
\subfloat[\label{figure_6_2}]{\includegraphics[width=0.26\textwidth]{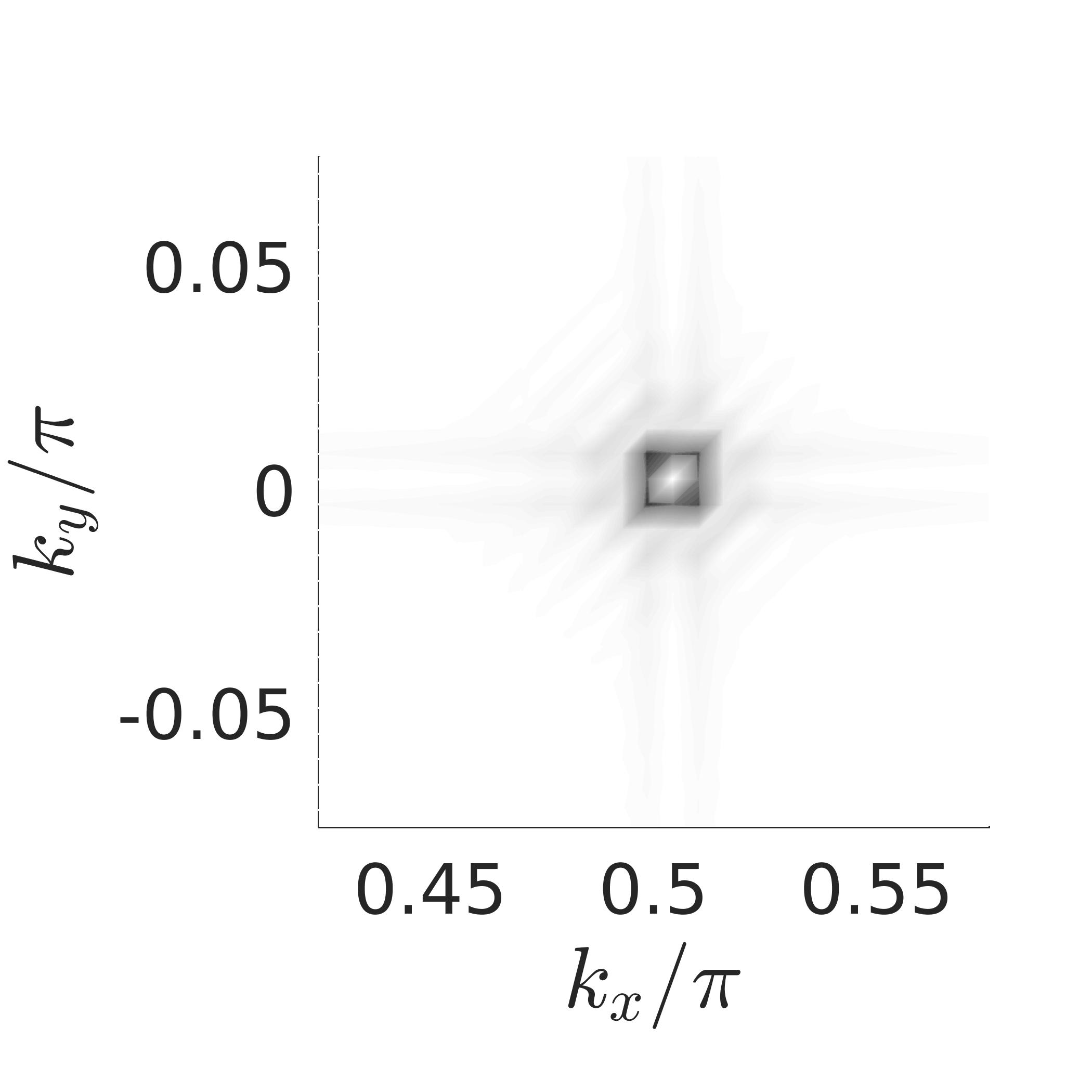}}
\subfloat[\label{figure_6_3}]{\includegraphics[width=0.24\textwidth]{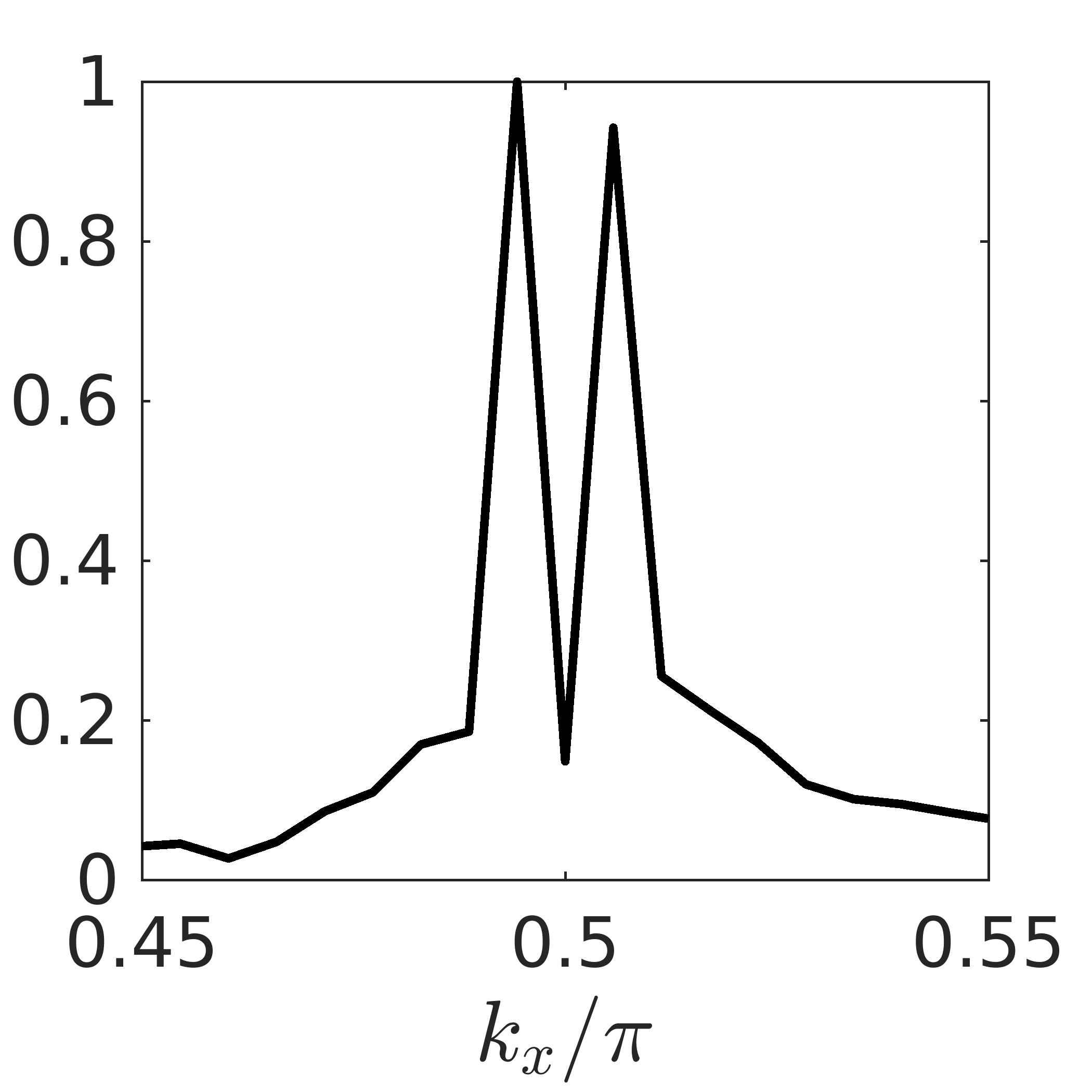}}
\subfloat[\label{figure_6_4}]{\includegraphics[width=0.24\textwidth]{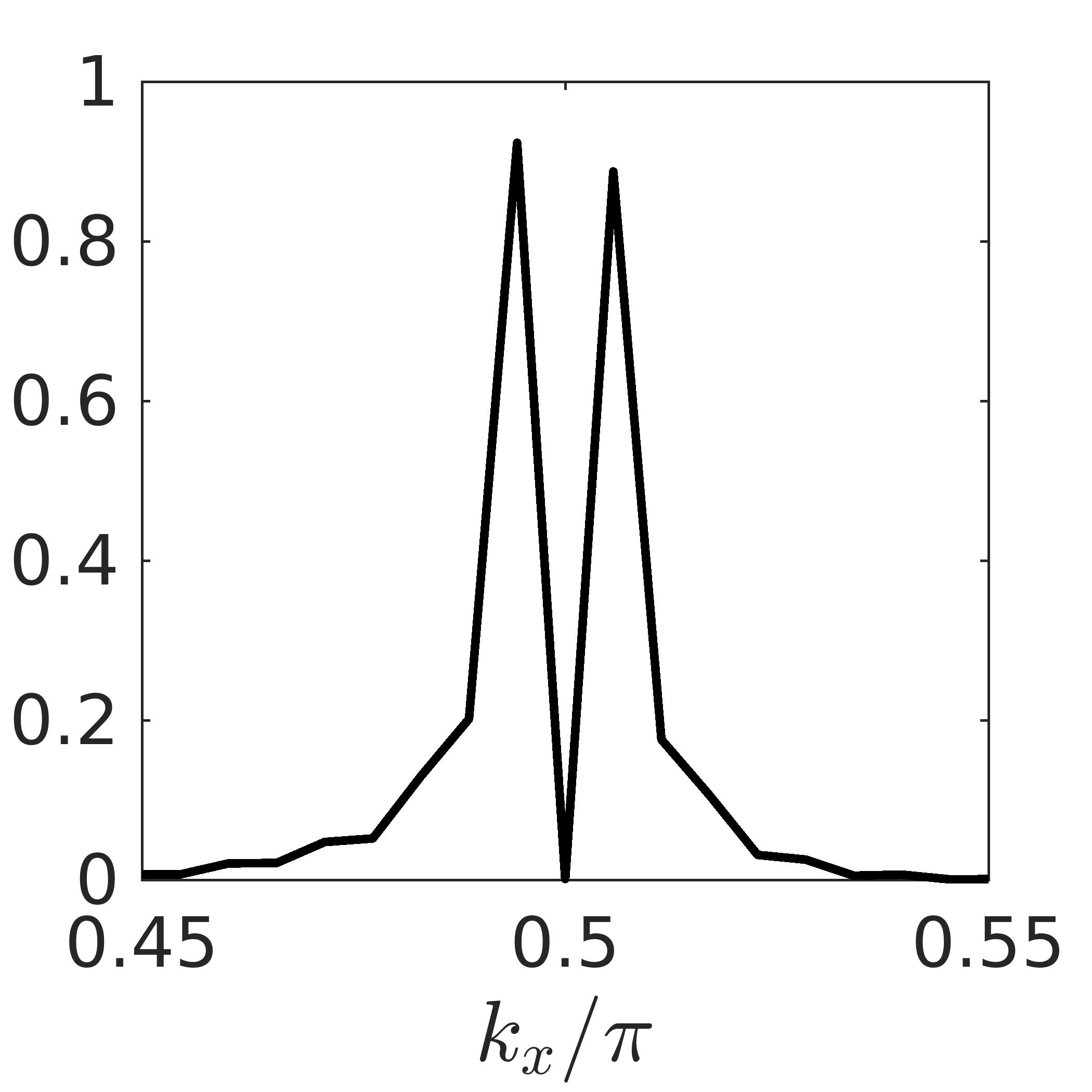}}
\hspace{0pt}
\caption{Fourier transforms of the LDOS for our defects around the dominant Fourier component: $2Q\mathbf{e}_x$ (zoomed in near the CDW ordering wave vector) in the background of a half-vortex of the PDW and a double dislocation of the CDW. 
The density of $k$-points is determined by the lattice size, here $N=400$. Top row: FT-LDOS for (a) the half-vortex and (b) the double dislocation. Notice the 
Fourier harmonic around the ordering wave vector vanishes/are suppressed. The split peaks are seen more clearly by plotting its amplitude along cuts in $k$-space. Here we have the absolute value of the FT along the line $k_y=0$ shown for the 
half-vortex in (c) and for the double dislocation in (d).}
\label{figure_6}
\end{figure*}

On the other hand, the half-vortex in \figref{figure_4_1} does not posses the QPI patterns seen for the double dislocation and the Abrikosov vortex. This is because 
in the case of the half-vortex one component of the PDW is always nonzero, which results in a gap for the states within the half-vortex core where the 
PDW becomes effectively a fully gapped FF  state. In contrast, in the cases of the Abrikosov vortex and of the double dislocation {\it both} components of 
the PDW order parameter, $\Delta_{\pm \bm Q}$, vanish at the core, and the BdG states become gapless at there.
Other details associated with LDOS of our defects can be found in Appendix \ref{LDOS_appendix}, where a zero bias probing voltages is 
considered.

\subsection{Patterns of the FT of the topological defects}
\label{split-peaks}
We now analyze in detail the effects that the half-vortex and the double dislocation have on the induced CDW order. 
Recall that inside the core of the half-vortex the SC order parameter is mostly FF type, since the amplitude of one of the two PDW order parameters must vanish at 
the location of the half-vortex, i.e., the origin (see Sec. \ref{NLSM}). Since the FF state breaks inversion symmetry in the $x$-direction, the corresponding CDW 
pattern inherits this broken symmetry. Also, the edge dislocations of the CDW order parameter break inversion symmetry in the $y$-direction. Note that, the parity 
operator, in the $x$-direction, changes the location of the branch cuts and the signs of the winding numbers when it acts on the order parameters, which changes the
 sign of the dislocation charge. As a result, the corresponding charge density patterns are flipped on their head under this operation. Since the location of the branch 
 cuts have no physically significant effects on the charge density, the full vortex is invariant under this operation.

The ordering wave vector $2\mathbf{Q}$ has many features made more apparent in the Fourier transforms of the LDOS, which are shown in 
Figs. \ref{figure_6_1} and \ref{figure_6_2}. Notice that the Fourier transforms of the LDOS for half-vortex and the double dislocation feature split peaks at the 
$2\mathbf{Q}$ ordering wave vector where the amplitude of the FT-LDOS is zero there. In contrast, in the case of the  Abrikosov vortex, the Fourier transform 
of the LDOS is just the transform of $\cos(2\mathbf{Q}x)$ which has  a single peak at the ordering wave vector. Cuts of the FT-LDOS along $k_y=0$ are given 
in Figs. \ref{figure_6_3} and \ref{figure_6_4} to more clearly illustrate these split peaks.
\begin{figure}[t]
\subfloat[\label{figure_7_1}]{\includegraphics[width=.5\textwidth]{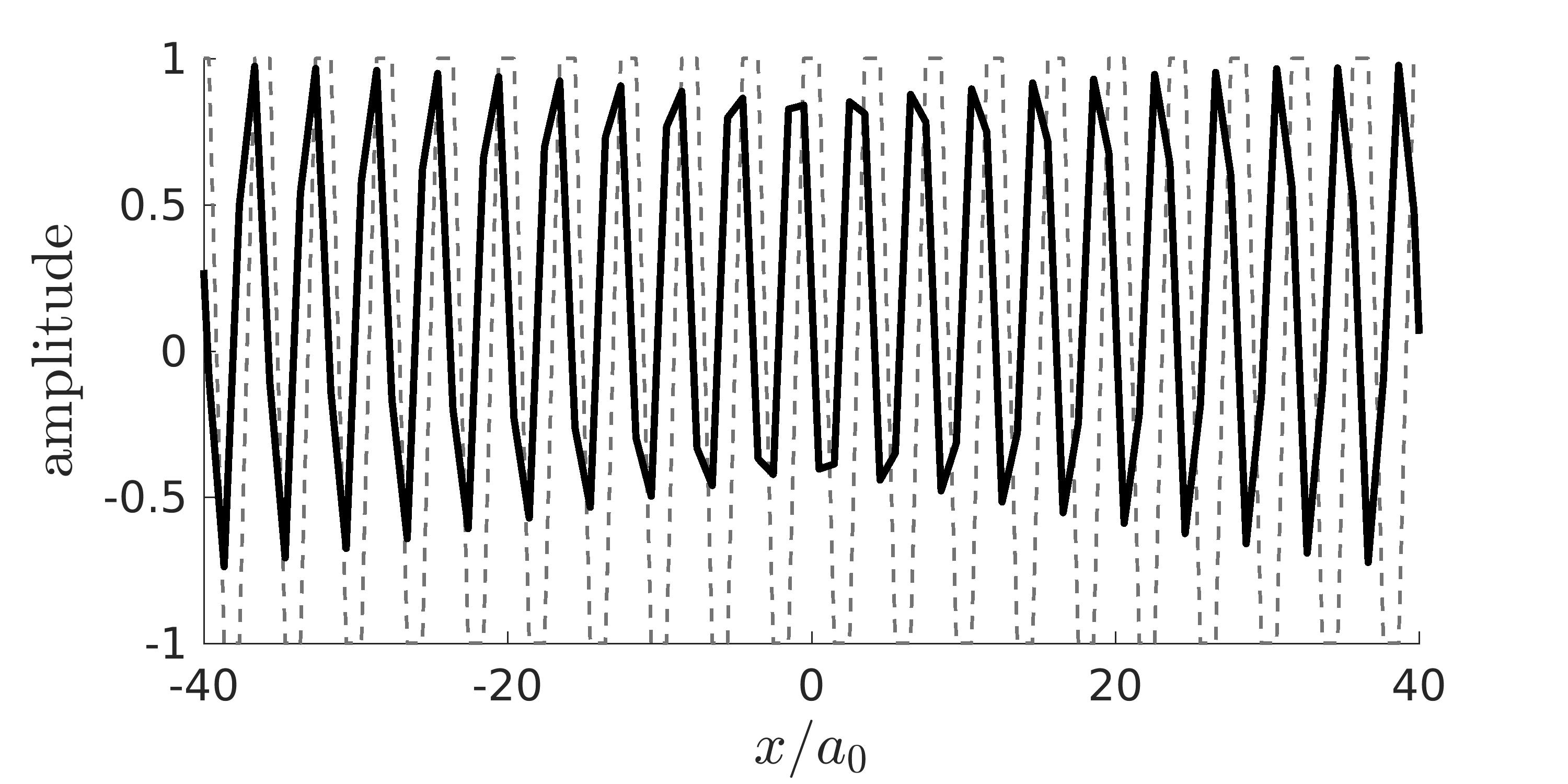}}
\\
\subfloat[\label{figure_7_2}]{\includegraphics[width=.5\textwidth]{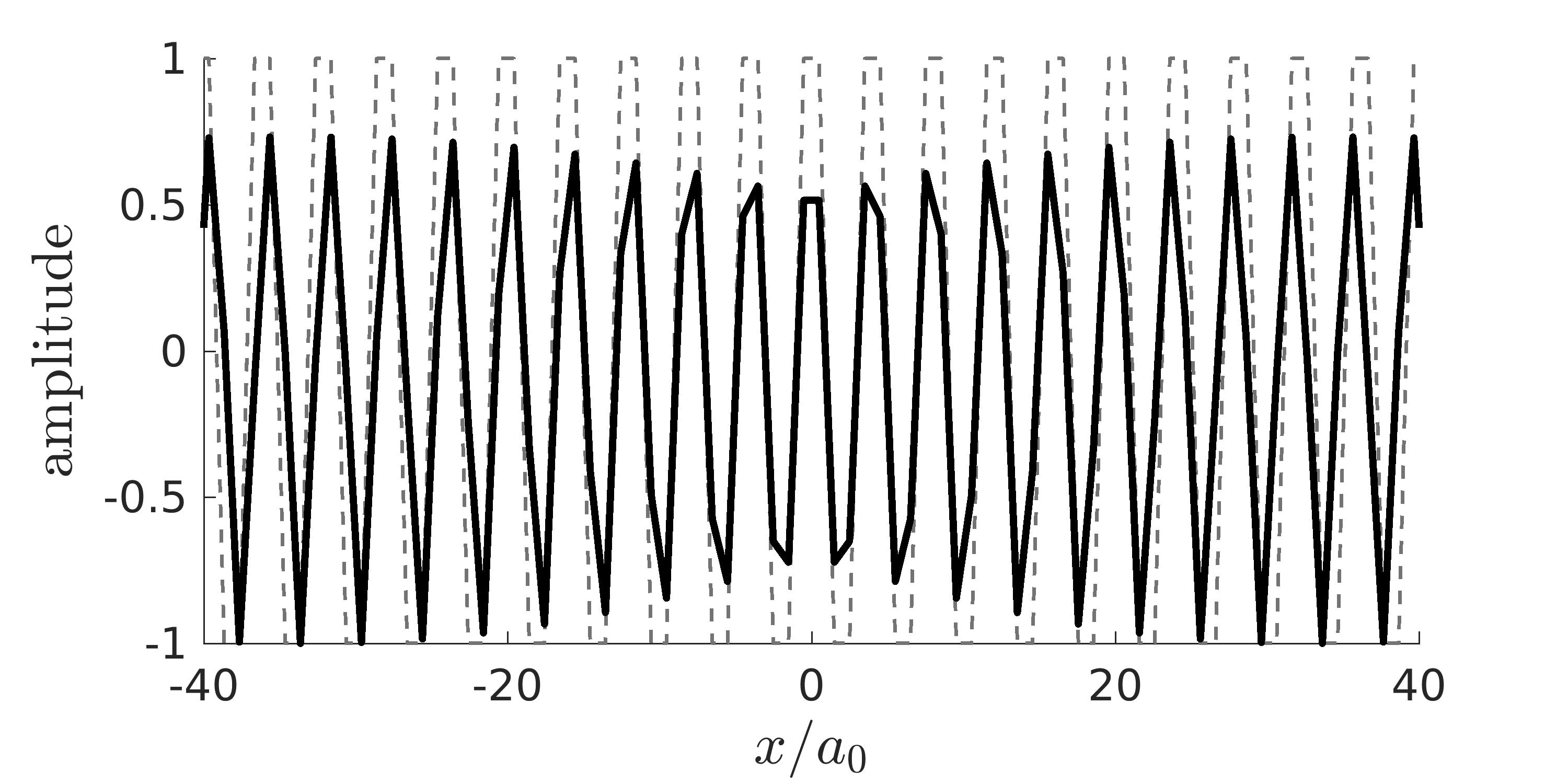}}
\hspace{0pt}
\caption{Real space plots of the effects on the CDW order parameter along the $x-$axis due to the (a) half-vortex and the (b) double dislocation. 
We plot these distorted CDWs on top of a CDW of a defect-free PDW state for comparison. The real space patterns show a jump in the CDW phase around the vortex 
core giving another physical realization of the split peaks seen in the FT-LDOS.}
\label{figure_7}
\end{figure}

The split peaks are signature of the defects of the PDW phase associated with jumps in the phase $\theta_-$, defined in \equref{theta-pm},  
across the core of the topological defect. For example, across the half-vortex the phase jumps by $\pm\pi/2$ since it winds by $\pi$ around the half-vortex. 
This implies that the $2Q$ Fourier component is equal to itself times $i$  across the core of the defect, suggesting that this Fourier component must be zero. 
We can explicitly verify this prediction by examining the  FT of $\rho_{2\mathbf{Q}}(\mathbf{r})$. It is apparent that a nonzero CDW winding number is responsible 
for the vanishing of $\rho_{2\mathbf{Q}}(\mathbf{r})$ at the center of the defect. 
In other words, the phase shift that causes this destructive interference is a measurement of the Burgers vector, which is the topological charge of the dislocation. 
Similar interference patterns  are well known to exist in electron diffraction in crystals of semiconductors with dislocations.

Similar split peaks in the Fourier transform of the tunneling LDOS were also predicted to exist at the PDW halo of an 
Abrikosov vortex of a superconductor 
 in Refs. \cite{Wang-2018,Lee-2018}, but their  physical origin is very different. Indeed, in the case of the vortex halo there is a phase shift in the 
 $\rho_{\mathbf{Q}}(\mathbf{r})$ (instead of $\rho_{2\mathbf{Q}}(\mathbf{r})$) Fourier component of the local charge density caused the Abrikosov vortex of the 
 uniform component of the superconductor.
\begin{figure}[htb]
{\includegraphics[width=.35\textwidth]{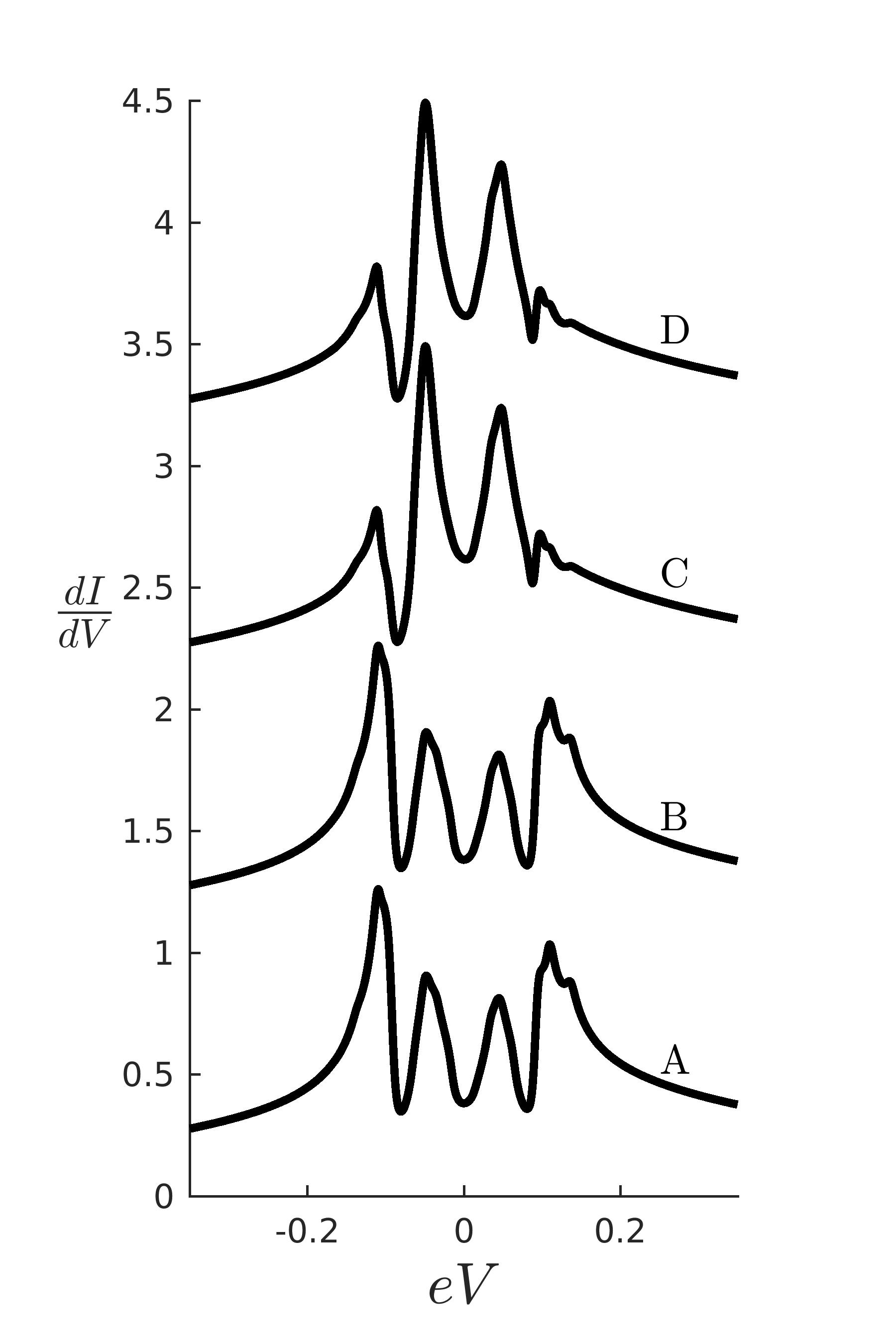}}
\hspace{0pt}
\caption{Comparison of $\frac{dI}{dV}$ curves related to the electron tunneling DOS for a period $8$ PDW (LO state). The electron tunneling DOS is sensitive to the periodicity of the associated period $4$ CDW. The labeled $A-D$ are the tunneling DOS traces at the four inequivalent sites of the CDW. Each consecutive curve is offset by 1 for clarity. Notice that in the PDW traces particle-hole symmetry is present only at low bias (low energies).}
\label{figure_8}
\end{figure}

\begin{figure}[htb]
{\includegraphics[width=.35\textwidth]{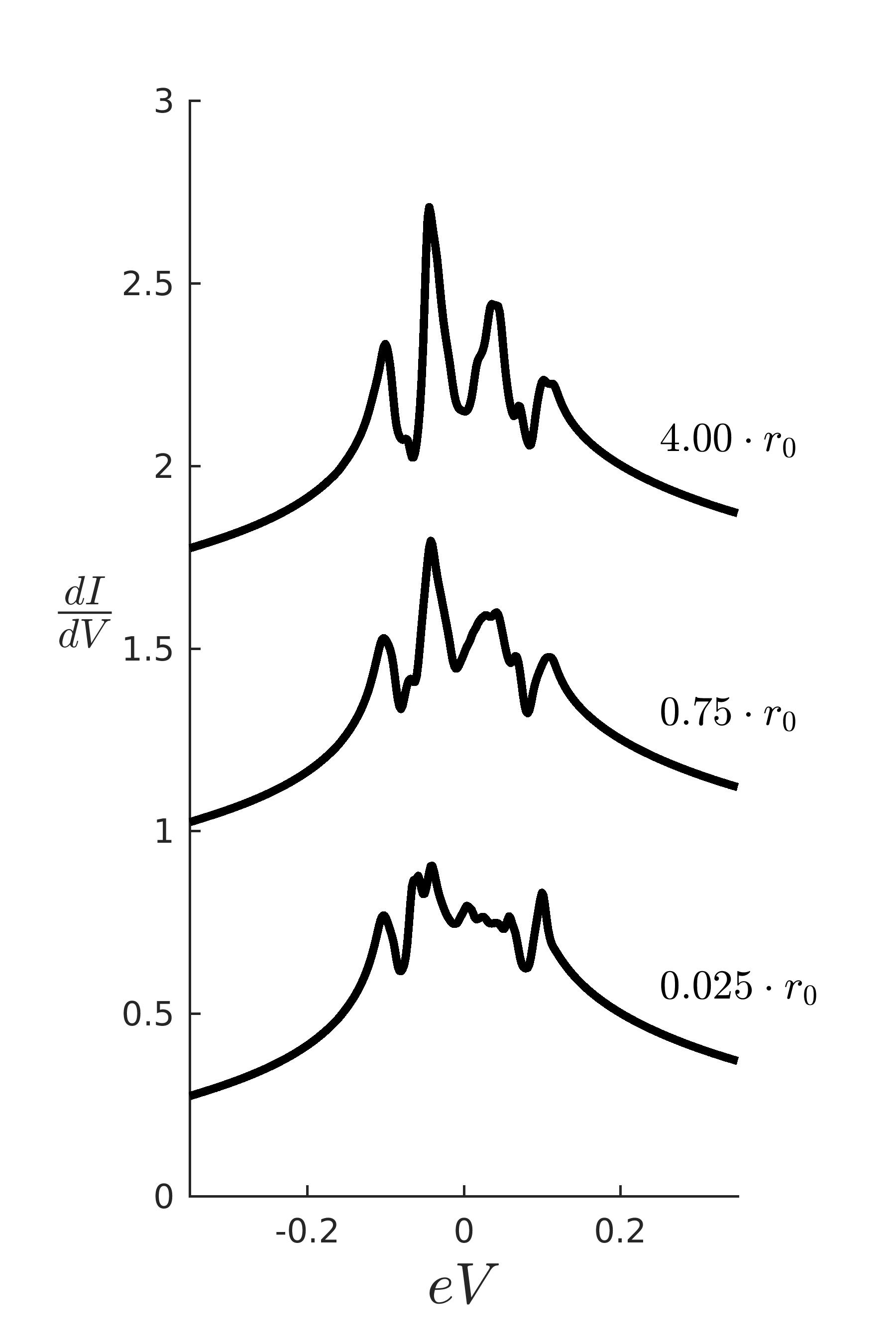}}
\hspace{0pt}
\caption{A set of $\frac{dI}{dV}$ curves for the half-vortex belonging to different lattice sites. The labels indicate the distance the lattice site in question is from the vortex core in fractions of the halo radius, $r_{0}=24a_0$. The top two curves are shifted up by $0.75$ and $1.5$ units in respect to zero. Close to the center of the core the curves resemble a squeezed in FF state more so than an LO state, but as we move outward the coherence peaks grow, and the half-vortex behaves more like an LO state. In appendix \ref{DOS_Sup}, \figref{figure_14} additional curves corresponding to a half-vortex are provided which indicates a 4 lattice site periodicity, like the LO state, but there is also a shift in tunneling spectra associated with the jump in phase of $\theta_+$ across the vortex core. This, along with reduced coherence peaks and additional satellites, distinguished the half-vortex from the pure PDW even far outside the vortex core.}
\label{figure9}
\end{figure}

\begin{figure*}[hbt]
\subfloat[\label{figure_10_1}
]
{\includegraphics[width=.45\textwidth]{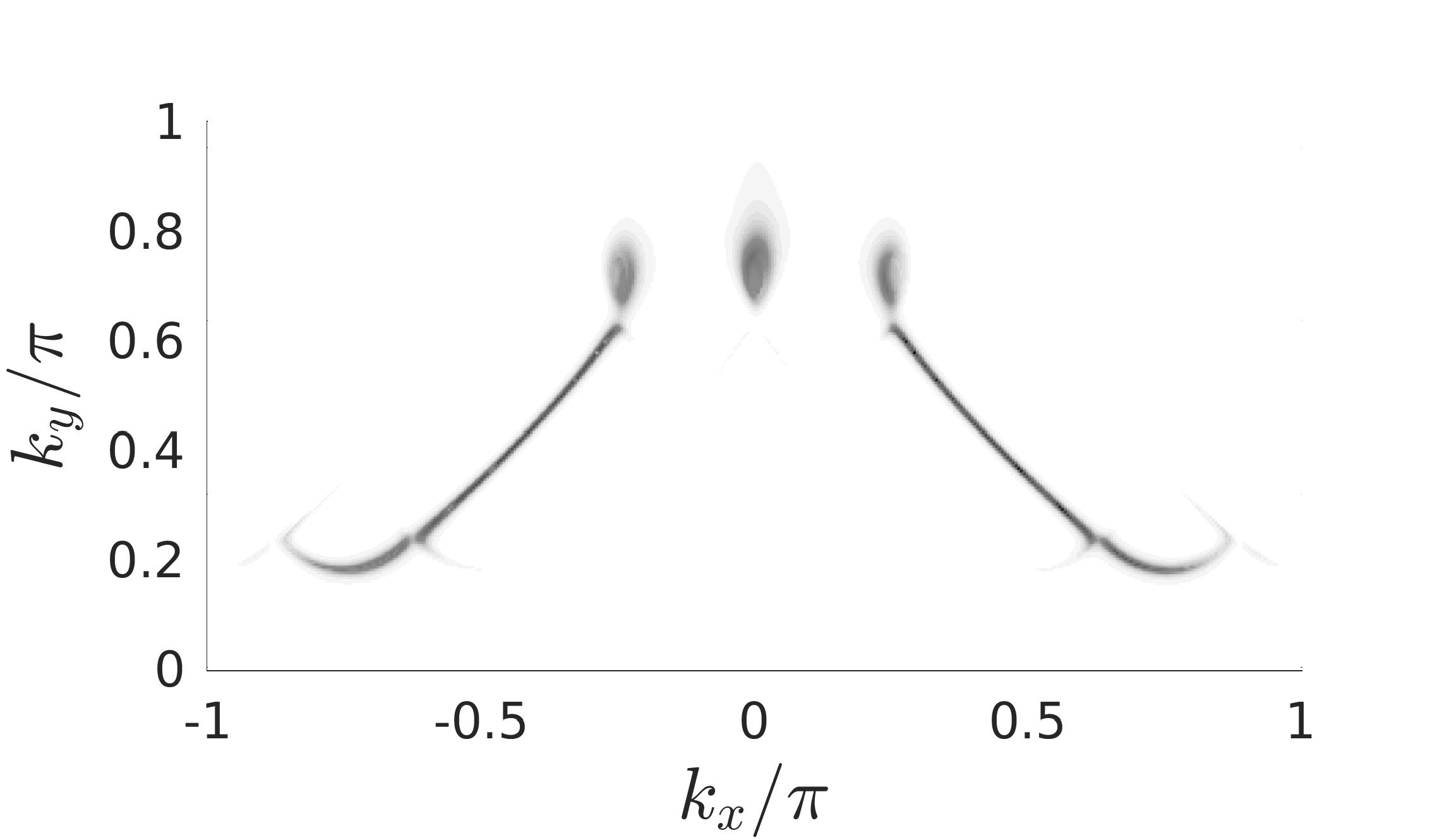}}
\subfloat[\label{figure_10_2}
]
{\includegraphics[width=.45\textwidth]{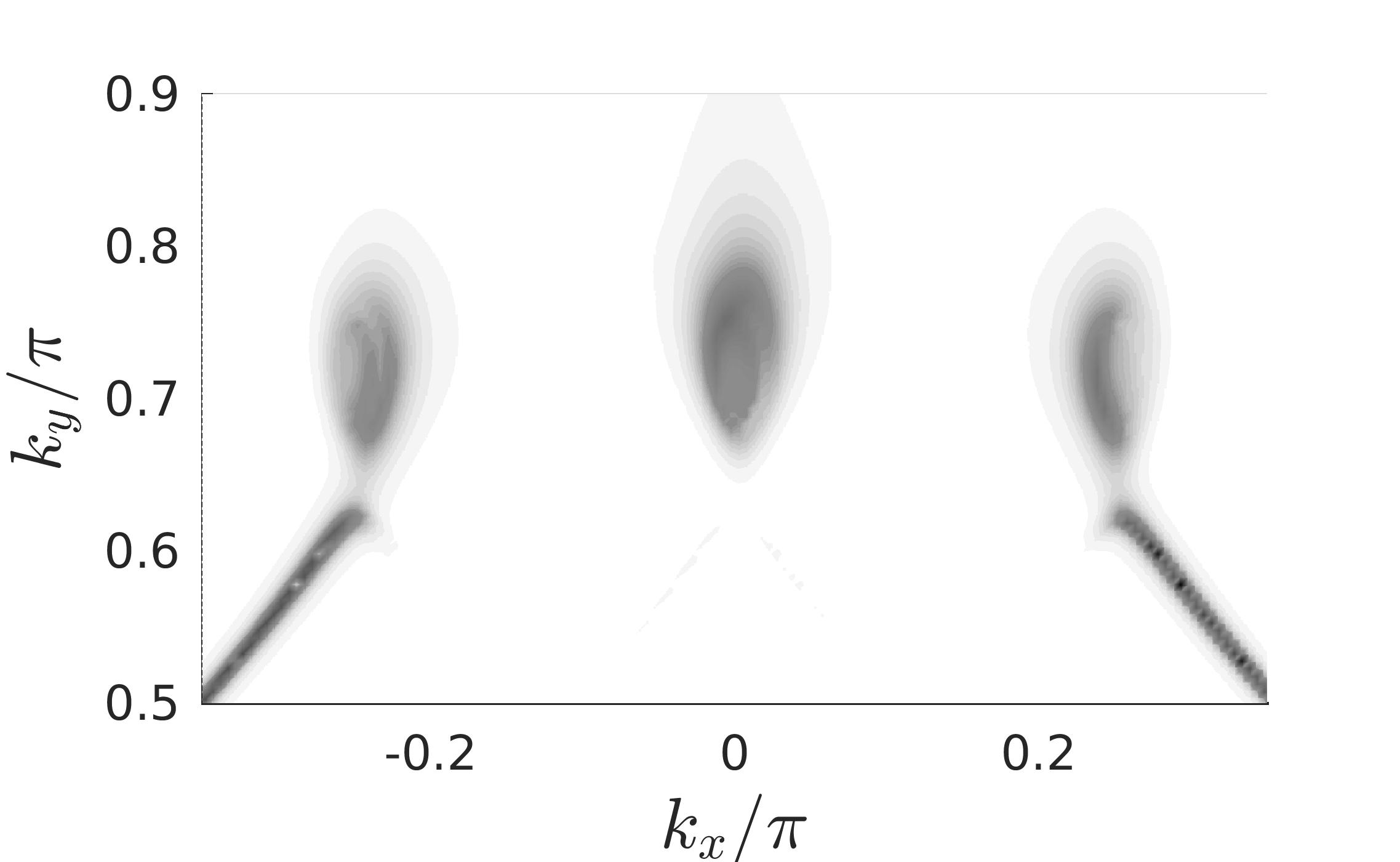}}
\\
\subfloat[\label{figure_10_3}]
{\includegraphics[width=.45\textwidth]{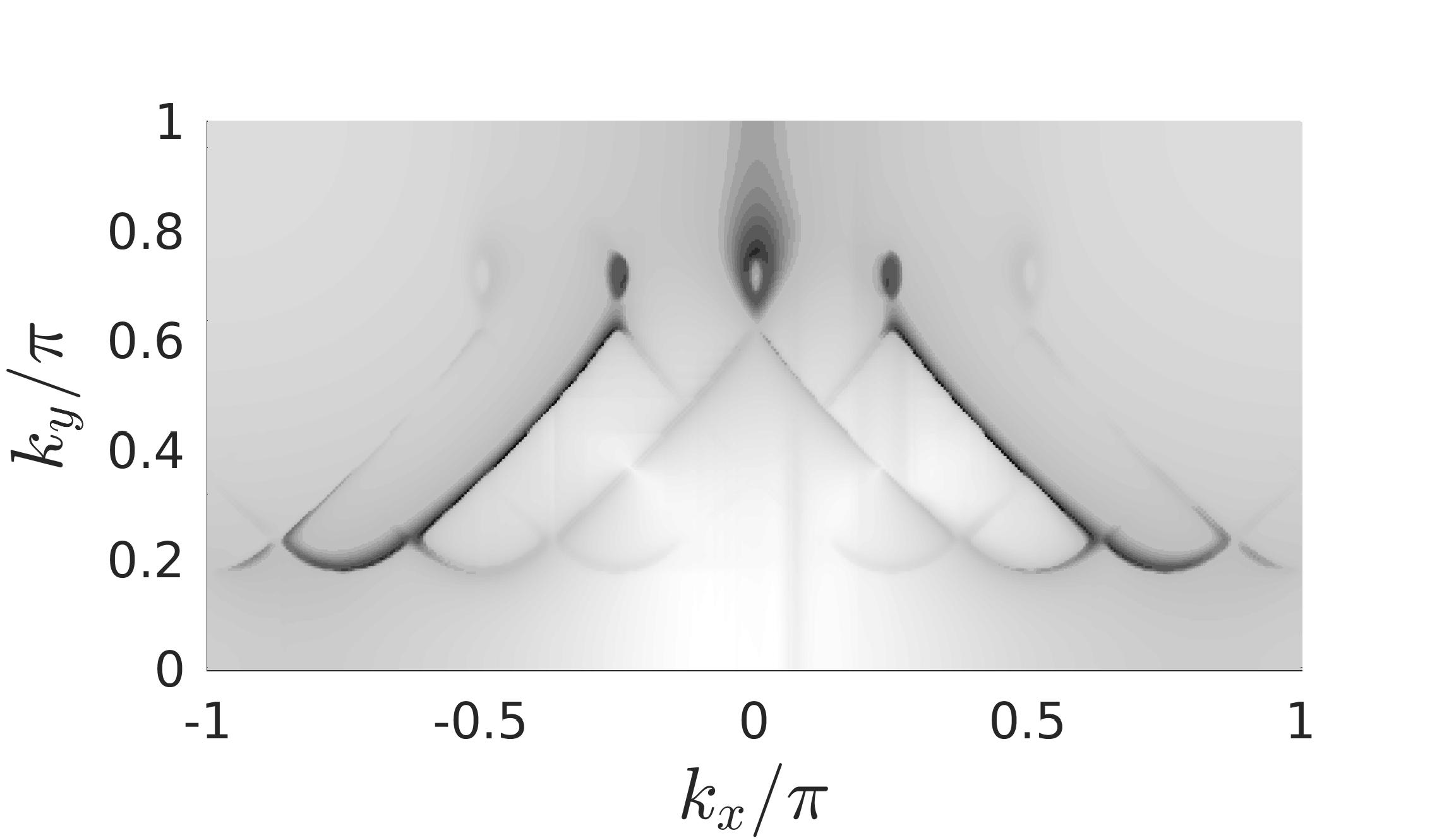}}
\subfloat[\label{figure_10_4}]
{\includegraphics[width=.45\textwidth]{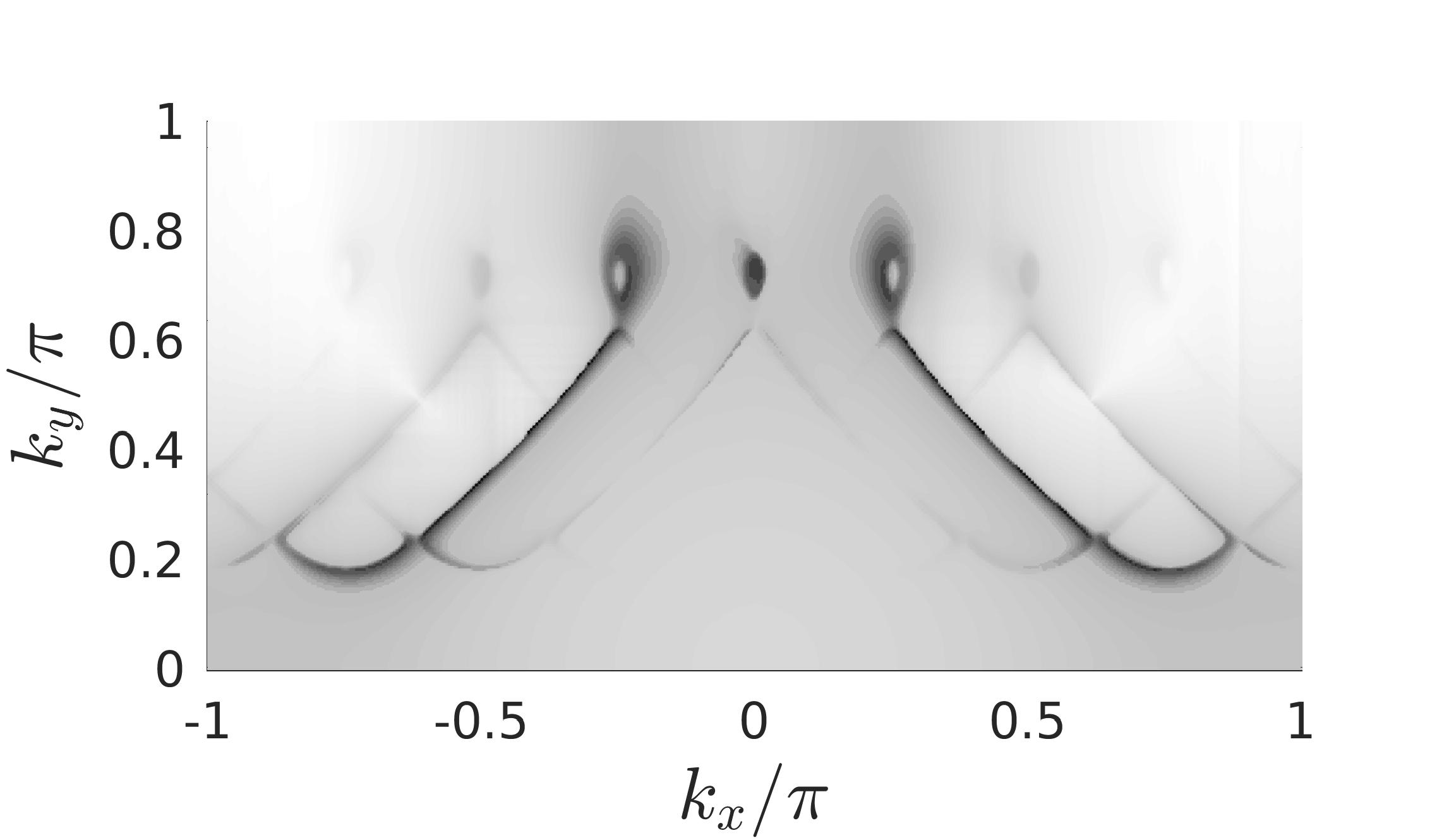}}
\caption{The Bogoliubov Fermi surface, $A(\mathbf{k},\omega=0)$, for a PDW state with a half-vortex of radius $r_0=24a_0$ and a $d$-wave form factor. In (a) we plot the upper portion of the Bogoliubov Fermi surface, which indicates a redistribution of spectral weight in respect to the defect-free PDW (see Appendix \ref{spec_appendix} for plots). In (b) we zoom in on these loops, which indicate inversion symmetry breaking. We also partition the Fermi surface of the half-vortex into (c) the particlelike portion and (d) the holelike portion. Notice the intensity of the spectral weight belonging to the Fermi arcs tend to be either particlelike or holelike depending on the side you are on.}
\label{figure_10}
\end{figure*}

Alternatively, when there is a winding number in both PDW components, we can picture the phase jump as occurring in both winding numbers 
independently. Recalling the form of the induced $2Q$-CDW from \equref{2Q}, we see that complex conjugation flips the phase-winding of one of the PDW 
components. Thus, when we have a vortex in both components $\Delta_{\pm \bm Q}$ with the same winding, i.e., the Abrikosov vortex, the phase differences cancel 
each other, and there is no net phase jump across the core of the vortex. Hence, in the case of the Abrikosov vortex there is not a split peak. 
Equivalently, the Abrikosov vortex does not have any dislocation  charge associated with it to cause a split peak to exist. 
We can contrast this with the double dislocation where the phase jump adds up to $\pi$, and the amplitude at the CDW ordering wave vector should vanish by an 
identical argument to that of the half-vortex. We could again perform the FT to verify these results explicitly or, alternatively,  argue that it should hold  by way of the amount of dislocation charge 
associated with a given defect.

We finish this section by analyzing cuts of the real space patterns of the double dislocation and the half-vortex across the vortex cores, seen in 
\figref{figure_7}. The split peaks have a clear signature in the real space patterns. Since they arise due to jumps in phase of $\theta_-$ across the vortex core, 
the CDW pattern in real space  showcases these phase jumps. To demonstrate this we place a waveform of the defect-free PDW in the background of the half-vortex and the 
double dislocation (dotted line).

Starting with the double dislocation we can see the $\pi$ phase shift which occurs across the vortex core. 
The associated CDW pattern of the double dislocation has a ``sawtooth'' pattern, seen in Fig.  \ref{figure_7_2}, 
 to the left and to the right of the vortex core, which is even in $x$ being inversion symmetric. Comparison with the daughter $2Q$-CDW of the defect-free PDW 
 gives us a subtle indication of the $\pi$ phase shift. A given sawtooth pattern lies within one of 
the  waveforms of the defect-free PDW. Sufficiently far from the core of the vortex the teeth of the saws are odd in respect to the underlying  waveform, 
peaking on the right side of the  wave on the l.h.s. and vice versa for the r.h.s. of the vortex. This is indicative of a $\pi$ phase shift because the locations of the
 maximums and minimums of the double dislocation's CDW change their relative orientation within the square wave, and is indeed needed to maintain inversion 
 symmetry along the $x$-direction. 

Instead, for the half-vortex [\figref{figure_7_1}] the induced CDW on the l.h.s. of the vortex is (basically) in phase with the background CDW.
  Again it is a sawtooth pattern, 
but the minima and maxima of the two waves coincide. Within the core of the vortex the two patterns slightly dephase from each other, and the half-vortex's 
CDW is zero when the defect-free PDW amplitude is at a maximum, meaning there was a $\pi/2$ phase shift. This was to be expected from the above discussion. 
Unlike in the case of the double dislocation the half-vortex's CDW is asymmetric about $x=0$. This too is to be expected from the breaking of inversion symmetry.

\subsection{Tunneling DOS Spectra of the Topological Defects}\label{tunneling-DOS}

We now compare and contrast experimental signatures associated with $dI/dV$ curves belonging to various superconducting order parameters. These curves illustrate that the half-vortex can be thought of as an interpolation between an FF-like state (in the core) to a LO-like state (at infinity). Also, evidence of inversion symmetry breaking in the tunneling data for the half-vortex state is discussed. We also changed the energy resolution in this section to $0.01$ eV to smooth out the curves and make the low bias particle-hole symmetry more apparent.

We begin with a comparison plot between the inequivalent sites for a defect-free LO state shown in \figref{figure_8}. These plots showcase coherence peaks, like a uniform superconductor or low-bias particle-hole symmetry as well as additional satellite peaks and a zero-bias electron DOS. Additionally, the {\it electron} tunneling data for the PDW state is periodic with half the period of the PDW since the data sense the associated CDW (for a detailed analysis see Ref. \cite{Kivelson-2003}), which is established by varying the $x$ coordinate. Thus the curves have a periodicity of four lattice spacings here, and not eight like our PDW order parameter, so we label the four representative curves with letters $A-D$. Alternatively, the periodicity is 4, and not 8, because the remaining four lattice sites have a SC gap which is $\pi$-phase shifted in respect to the first four, and a normal metal STM tip is blind to this effect. The full periodicity of the PDW state can be detected with {\it pair} (Jospehson) tunneling. This will be discussed elsewhere. The defect-free FF state (not shown) possesses a constant tunneling-DOS across the entire plane, so only one representative curve is needed. Clearly then the FF state possesses data which are rotationally invariant (by $\pi/2$ in the CuO$_2$ planes) unlike the above LO state. 
\par We now compare the LO states to a set of curves corresponding to the half-vortex (\figref{figure9}). The labels on each curve indicates how far we are from the vortex core in fractions of the halo radius, $r_0=24a_0$. We first note that near the core of the half-vortex the $\frac{dI}{dV}$ curve is not that of a free particle. In fact, it resembles a squeezed in FF state, even possessing the discrete rotational symmetry of the lattice (not shown). As we travel out to the edge of the vortex, the half-vortex begins to become more LO-like than FF-like. This is the interpolation from a FF to a LO state that the half-vortex undergoes, apparent from our boundary conditions in \equref{bc-alpha-t}. 
\par The presence of a half-flux quanta will further distinguish the half-vortex from the defect-free PDW far outside the vortex core. In Appendix \ref{DOS_Sup} we provide supplementary $dI/dV$ plots taken outside the core of the vortex, which indicated there is a relative shift in the tunneling DOS curves. Indeed, \figref{figure_14} in Appendix \ref{DOS_Sup} compares the half-vortex tunneling spectrum to the right and to the left of the vortex core [Figs. \ref{figure_14_1} and \ref{figure_14_2}, respectively]. Again there is a periodicity present (outside the vortex core) in these plots, and just like an LO state, the $dI/dV$ curves repeat every four lattice spacings. It should also be noted that the curves presented in the appendix have a defect which is placed on the CuO$_2$ bonds opposed to at the center of the plaquette. This changes the appearance of the $dI/dV$ curves, but the spectral weight associated with the quasiparticles remains the same. That is, we shifted the period $4$ CDW by half a lattice spacing, which costs us no, or very little, energy to do so.

The half-vortex becomes more LO-like outside the core of the half-vortex, but since this topological defect breaks inversion symmetry, its tunneling DOS must reflect this, unlike the LO state. A smoking gun signature of inversion symmetry breaking is present by comparing the $dI/dV$ curves to the left and to the right of the vortex core [figures \ref{figure_14_1} and \ref{figure_14_2}, respectively]. It is apparent the data on the left are two lattice spacings behind the right (or vice versa), which we attribute to the jump in $\theta_+$ by $\pi/2$ across the vortex core. 
Indeed, the accumulated phase belonging to a Cooper pair with nonzero COM momentum: $\mathbf{Q}\cdot\mathbf{r}=\pi/2$ if $\mathbf{r}=2a_0\mathbf{e}_x$. We will see another example of inversion symmetry breaking in the next 
section when we discuss the spectral functions for the half-vortex. This jump in phase distinguishes the half-vortex from a defect-free LO state, even outside the half-vortex core.

Note the full vortex also possesses a phase jump of $\Delta\theta_+=\pi$ across the vortex core, but this gives a relative shift of four lattice spacings when comparing electron tunneling DOS on the left and right hand sides of the full vortex. This means there is no analogous signature belonging to the full vortex as there was for the half-vortex when using a normal metal tip. This is simply because the LO state has the same periodicity as this shift. On the other hand, in the case of a superconducting tip there will be a difference in the pair tunneling DOS on the right- and left-hand sides of the full vortex. Finally, we note that the presence of a gap in the core of the half-vortex distinguishes it from the other two topological defects since the latter two have a vanishing gap here. This results in a free particle tunneling DOS in the core of the defect for the double dislocation and the full vortex opposed to a squeezed in FF state seen in the core of the half-vortex [see \figref{figure_15_1} in Appendix \ref{DOS_Sup}]. Finally, we can see in \figref{figure_15_2} that the half-vortex possesses additional satellites peaks past these other two defects, and so the full vortex and the double dislocation more closely resemble the pure PDW outside the core. The inequivalent gaps corresponding to the distinct Fourier components of the half-vortex is responsible for these additional satellites.

\subsection{Spectral Functions\label{spec-section}}

\begin{figure*}[hbt]
\subfloat[\label{figure_11_1}]
{\includegraphics[width=.33\textwidth]
{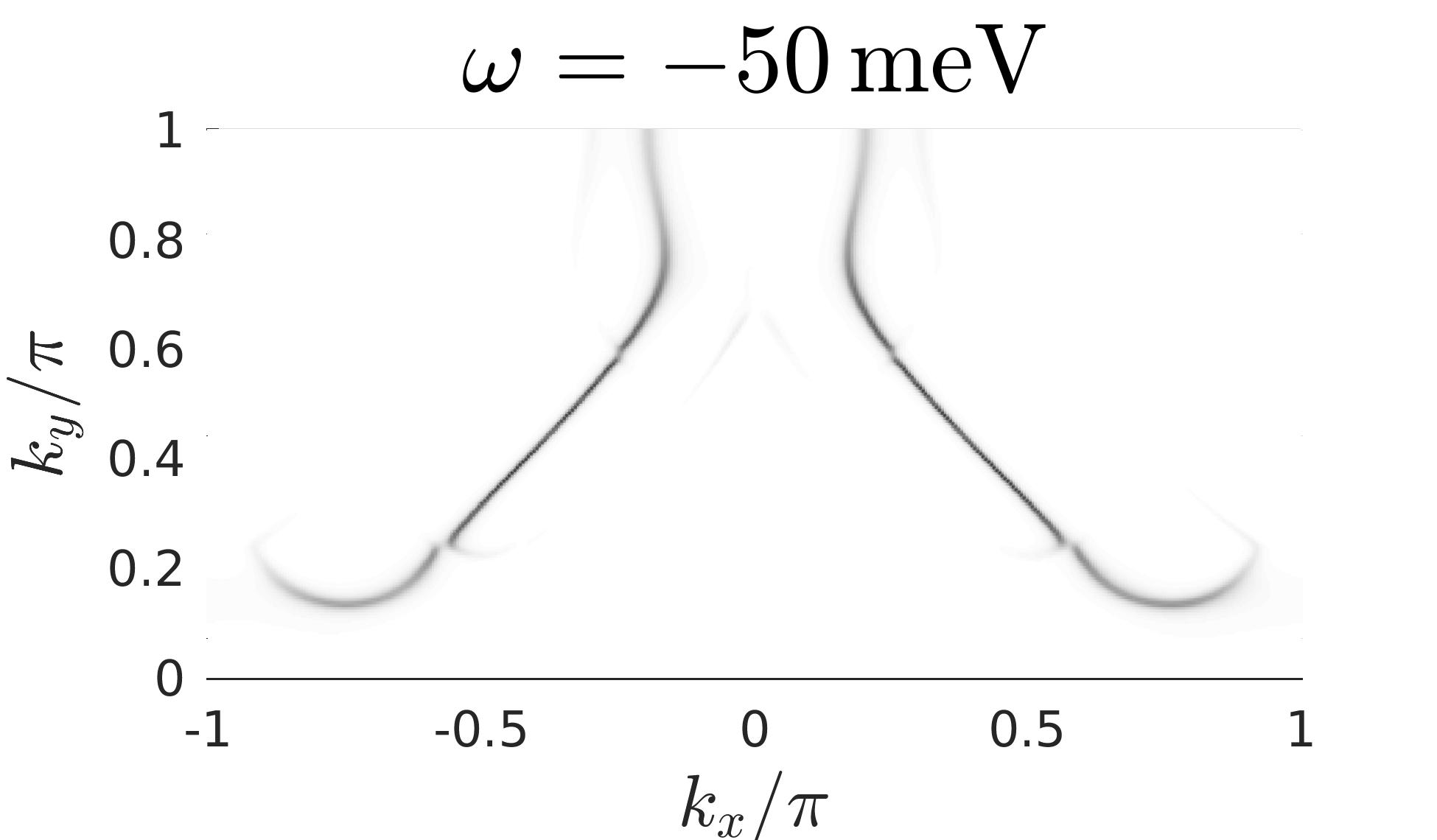}}
\subfloat[\label{figure_11_2}]
{\includegraphics[width=.33\textwidth]
{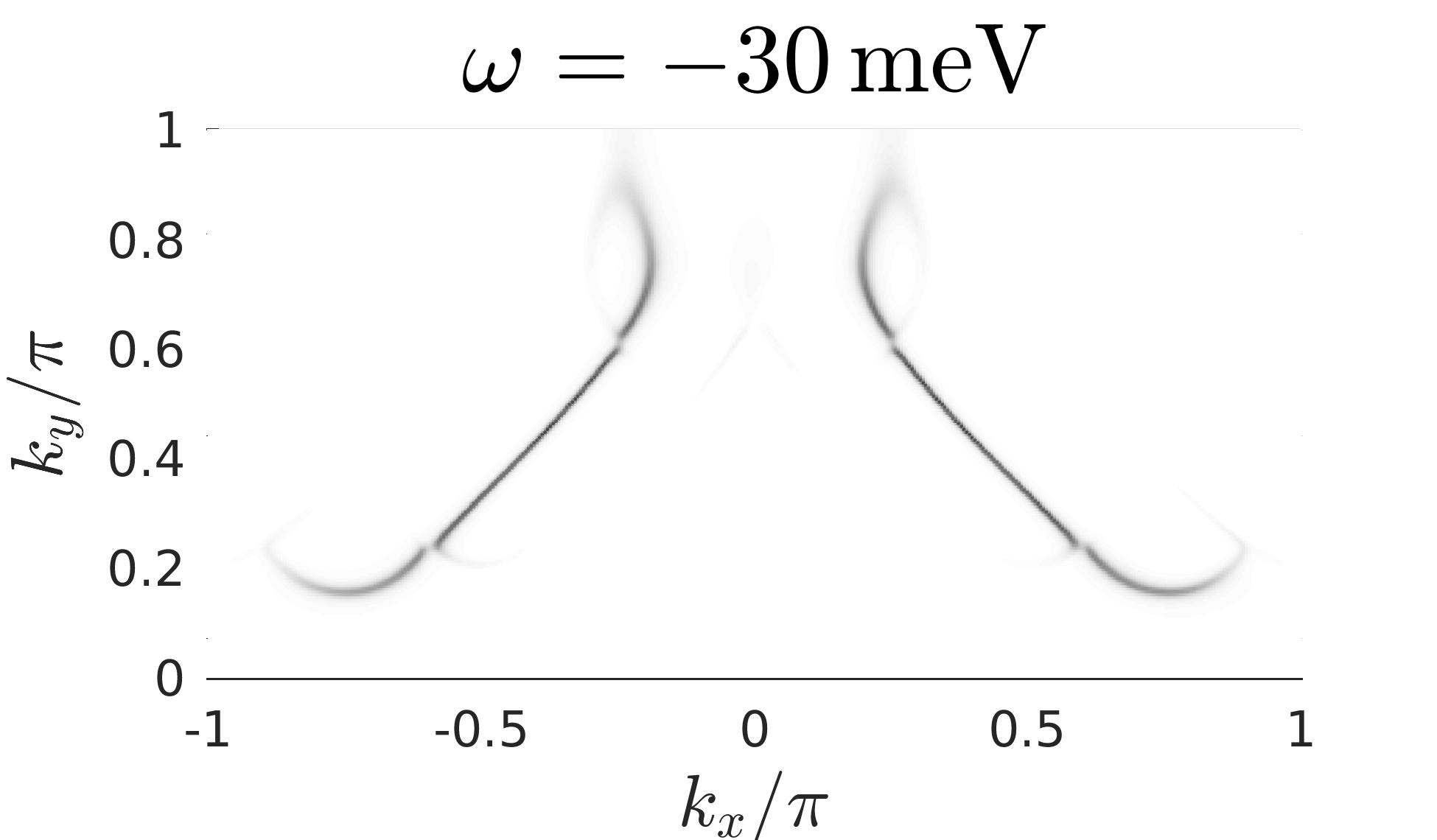}}
\subfloat[\label{figure_11_3}]
{\includegraphics[width=.33\textwidth]
{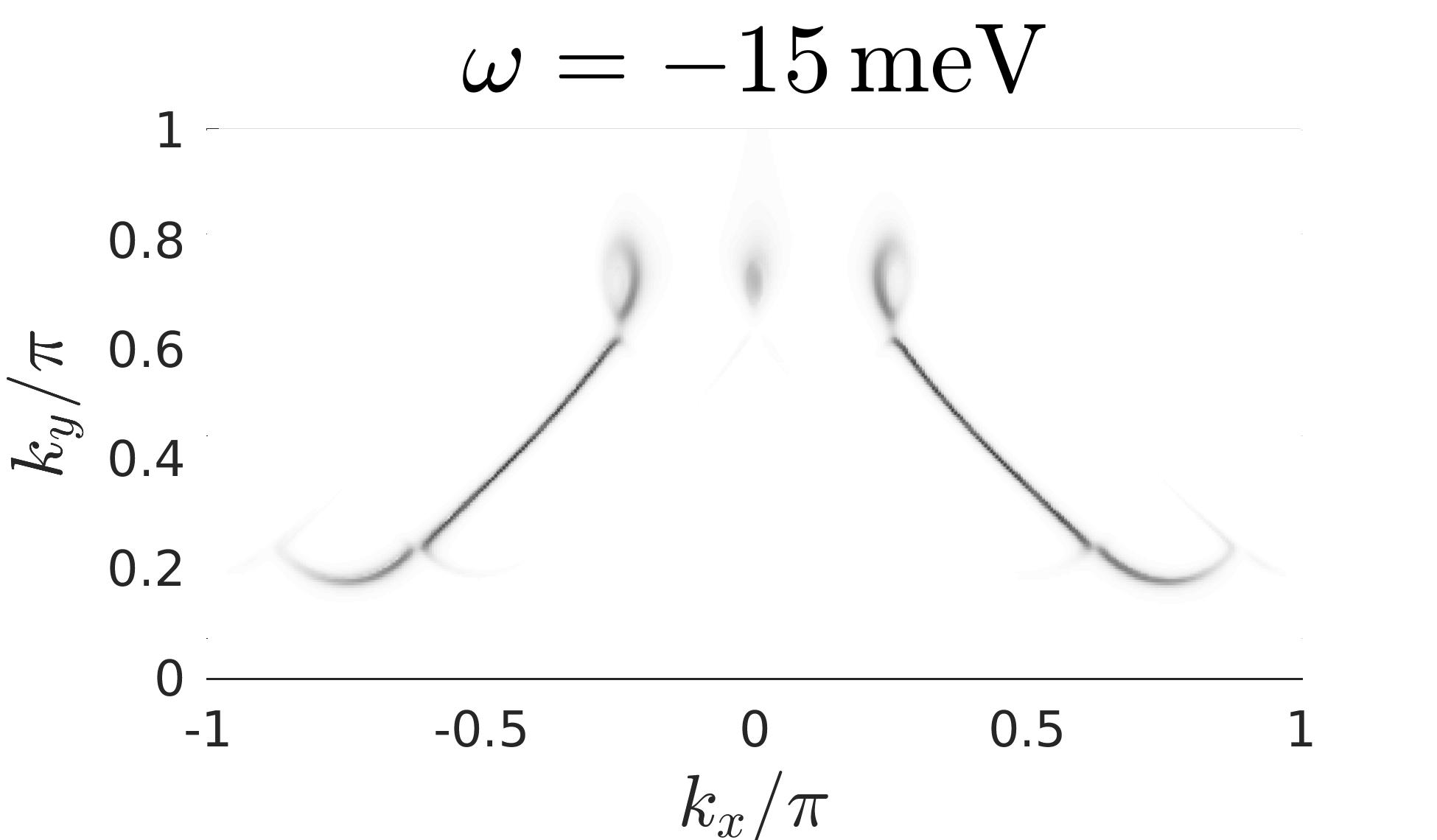}}
\\
\subfloat[\label{figure_11_4}]
{\includegraphics[width=.33\textwidth]{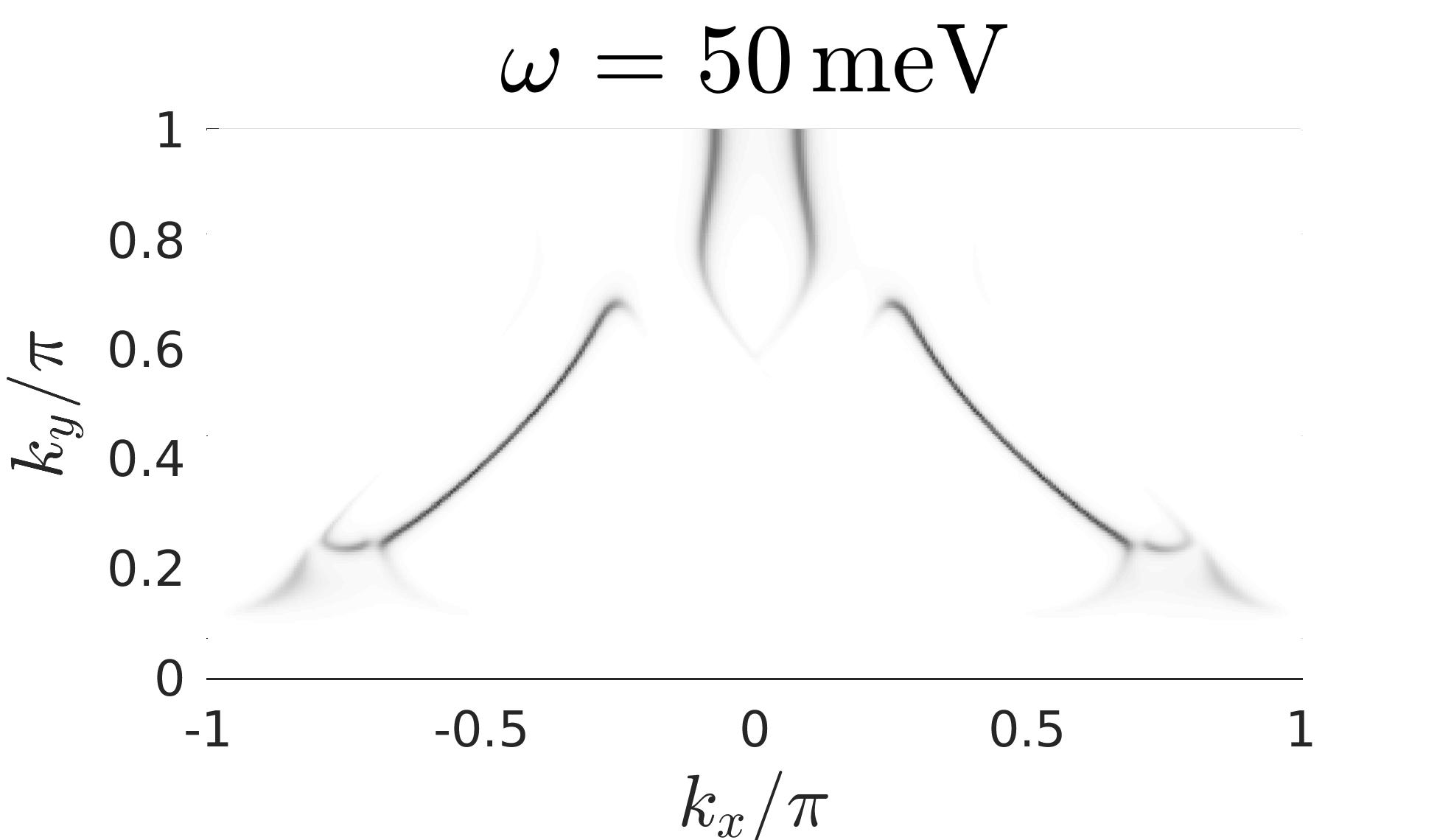}}
\subfloat[\label{figure_11_5}]
{\includegraphics[width=.33\textwidth]{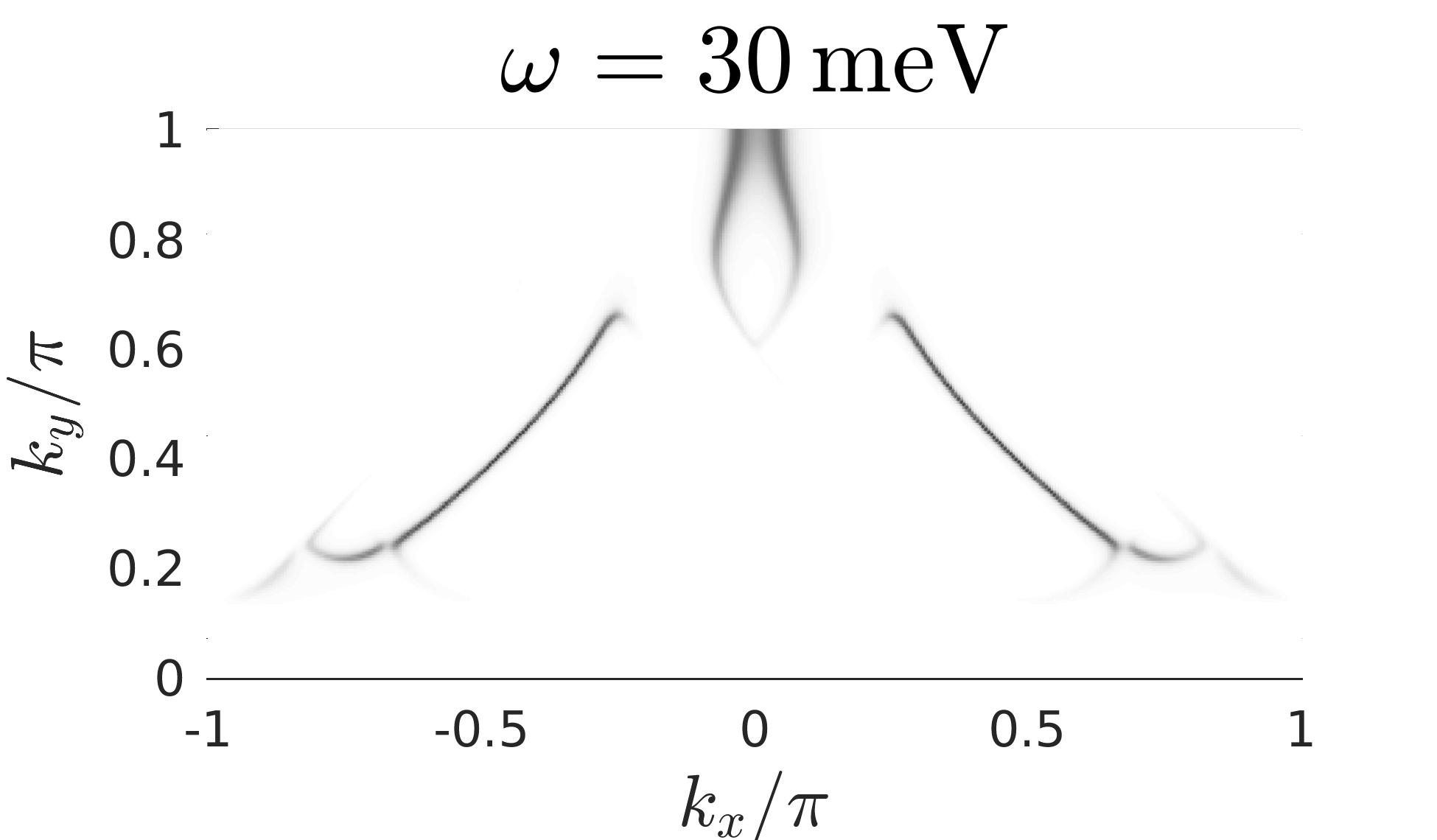}}
\subfloat[\label{figure_11_6}]
{\includegraphics[width=.33\textwidth]{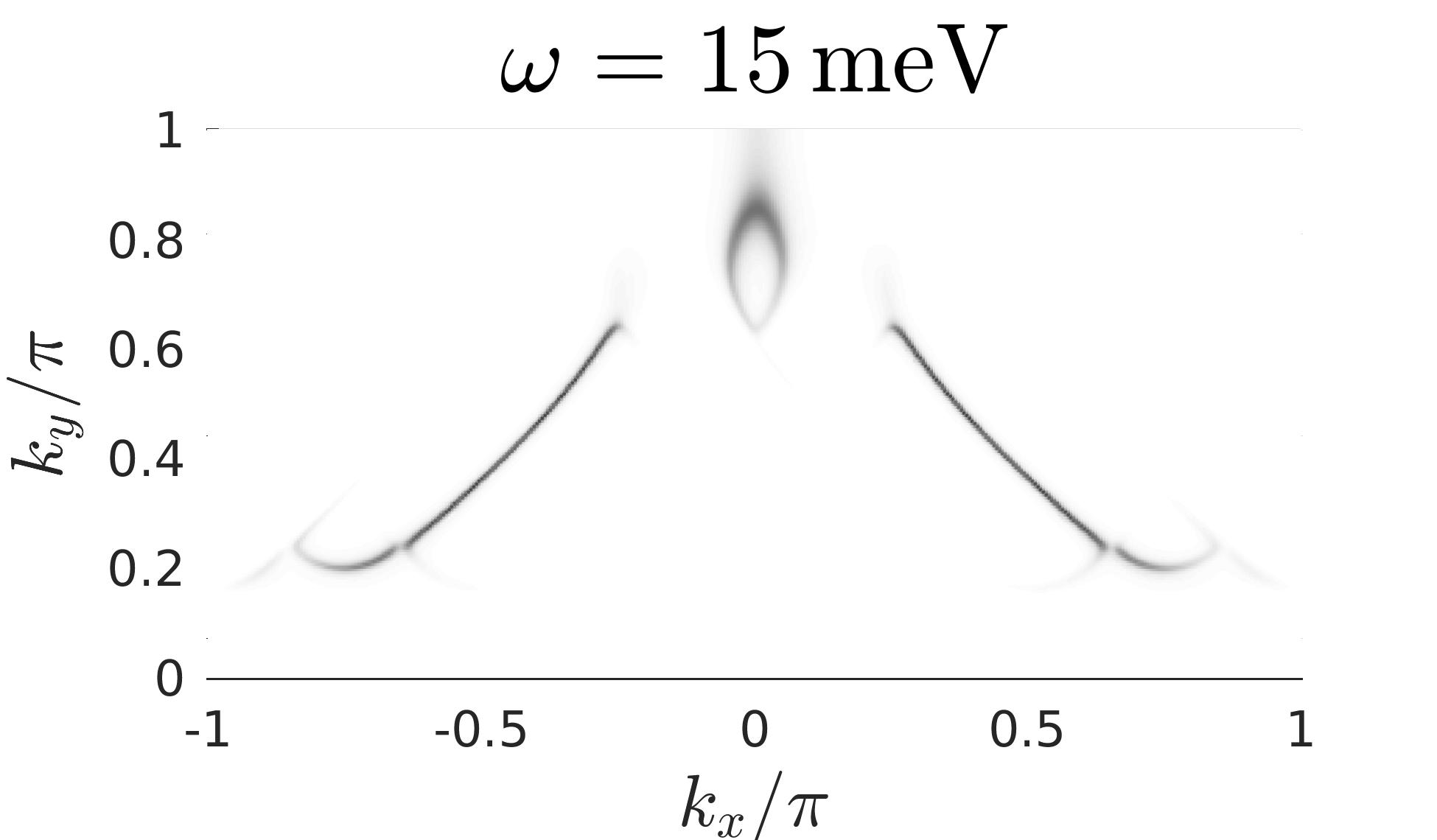}}
\caption{The spectral function, $A(\mathbf{k},\omega)$, of the PDW with a half-vortex ($r_0=24a_0$) evaluated at various energies: $\omega$. 
From plots (a)-(f) we can see the 
holelike character grows with a negative bias and the particle-like portions with positive bias. This redistribution of spectral weight can be used to map out the 
dispersion.}
\label{figure_11}
\end{figure*}
In this section we analyze the spectral functions of the  PDW state in the presence of topological defects, again, paying special attention to the half-
vortex. We will also be particularly interested in the plots of the spectral function $A(\mathbf{k},\omega = 0)$, defined in \equref{Spectral}, which counts how 
many quasiparticle states are connected to the ground state within our energy resolution, $\epsilon$. In a metal the spectral function at zero frequency yields the 
locus of points corresponding to the Fermi surface in the Brillouin zone. In the case of a PDW the spectral function at zero frequency reveals the locus of the Fermi surface of the Bogoliubov 
quasiparticle.
 Here we will use the term \textit{Bogoliubov Fermi surface} to represent the locus of points on the Brillouin zone where there are pockets of Bogoliubov quasiparticle. 
 Since the Bogoliubov quasiparticles are admixtures of electrons and holes different portions of the Bogoliubov Fermi surfaces have electron or holelike character.
Still we expect some  resemblance between the Fermi surface of the normal state and the Bogoliubov  Fermi surfaces of the PDW states; we give more on this below.

The Bogoliubov Fermi surfaces of the PDW state have been examined in Refs. \cite{Berg-2009,Baruch-2008} and in a quasi-1D model of the PDW in 
Ref. \cite{Soto-Garrido-2015}. In these references it was shown that in a time-reversal invariant superconductor, such as the PDW with wave vector $\bm Q$,  the pockets 
are separated by gaps in $k$-space, where the SC gap is nonzero,  where the condition $\xi_{\mathbf{k}}=\xi_{\mathbf{k}\pm\mathbf{Q}}$ is satisfied; 
here $\xi_{\mathbf{k}}$ is the quasiparticle dispersion in the non superconducting state. These gaps appear where the Fermi surface of the normal state is perfectly 
nested and we pair electrons with their time reversed partner.


In this subsection we are interested in the effects of the topological defects on the spectral functions of a PDW. A plot of a few Fermi surfaces of Bogoliubov quasiparticle of the PDW in the presence of a half-vortex can be found in \figref{figure_10}. Here we consider a half-vortex of $r_0=24a_0$ possessing a $d$-wave form factor. Spectral plots of a pristine PDW with both an $s$-wave and $d$-wave form factor as well as log plots of the Abrikosov vortex and the double dislocation (both with $d$-wave form factors) can be found in Appendix \ref{spec_appendix}, \figref{figure_16}. The latter two plots greatly resemble the spectral function of the pure PDW, small differences only becoming apparent on taking a logarithm. As in the spectral functions of the defect-free PDW of Ref.\cite{Berg-2009} the portion of the Bogoliubov Fermi surfaces closely resemble ``arcs'' along the ungapped parts of the normal Fermi surface (\textit{i.e.}, in the absence of the PDW state).

We first observe that the normal state dispersion is partially retained for the presence of the half-vortex. This also holds true for the other defects and is illustrated in Appendix \ref{spec_appendix}, \figref{figure_16}, where we also overlaid a copy of the normal state Fermi surface with that of the pure PDW [\figref{figure_16_2}]. These gapless regions retain the normal state character \cite{Berg-2008-QPI}. Utilizing the weak coupling argument above we realize the modified nesting condition suggests an $s$-wave form factor would also possess these Fermi arcs (see \figref{figure_16_1}).

The most striking feature of the spectral function in a PDW with a half-vortex is the redistribution of spectral weight to regions above the arcs forming discernible loops seen in \subfigref{figure_10_1}. In \figref{figure_10_2} we zoom in on these loops to indicate a degree of inversion symmetry breaking, seen in the distribution of spectral weight in these loops and along the arcs. Inversion symmetry is broken in the core of the half-vortex where the SC state becomes close to that of an FF state. 
Note that the formation of these loops does not occur so dramatically for the other two topological defects, but it still happens to some degree. The half-vortex is special in the sense that it couples to both the CDW and the SC degrees-of-freedom, unlike the other two topological defects. The asymmetry in the charge density induces a significant reshuffling of the spectral weight for the half-vortex according to these plots.

A point worth mentioning at this stage is the apparent coexistence of the ``Fermi-Arcs'', just mentioned, and electron/hole pockets \cite{Meng-09}. An arc usually refers to a large section of the Fermi surface which is seemingly open-ended. The pockets on the other hand are small closed surfaces. An experimental probe known as angle-resolved photo emission spectroscopy (ARPES) can help determine the Fermi surface, but the holelike regions are invisible to ARPES \cite{Chakravarty-03}. We can demonstrate this by looking at log plots of the particle-like Fermi surface [\subfigref{figure_10_3}] and the holelike Fermi surface [\subfigref{figure_10_4}]. Here it can be seen that the front and backsides of these arcs have primarily particle-like or holelike character, respectively. Note that this is the case for all the other PDW order parameters as well.

We can further examine the holelike and particlelike character of certain regions of the Bogoliubov Fermi surface in the presence of the half-vortex by looking at the spectral function for various probing voltages [\subfigrangref{figure_11_1}{figure_11_6}]. We see the spectral weight shifts around from one set of loops to another depending on the sign of the bias. Indeed, \figref{figure_11} demonstrates a negative bias will grow the holelike loops of the Fermi surface, while positive biases the electron-like portions. This redistribution of spectral weight pertains to the fact holes are at a negative energy in respect to the Fermi energy and vice versa for the particles.


\section{Discussion and Conclusions} 
\label{summary}

Evidence for the existence of pair density wave superconducting phases (or, at least a PDW component) has continued to grow. In addition to the panoply  
of evidence in the cuprate superconductors \cite{Agterberg-2020} new evidence for PDW order has now been found in other materials such as 
the heavy fermion superconductor $UTe_2$  
\cite{Aishwarya-2023,Gu-2023},  in a monolayer iron superconductor Fe(Te,Se) \cite{Liu-2023}, in  EuRbFe$_4$As$_4$ \cite{Zhao-2023}, and in the kagome 
superconductor CsV$_3$Sb$_5$ \cite{Chen-2021}. This growing body of evidence of the existence of PDW superconducting states makes the characterization of 
these phases an important problem.

In this paper we investigated the electronic structure of the BdG Hamiltonian of a unidirectional PDW in two dimensions in the presence of its three 
topological defects: the half-vortex, the Abrikosov vortex, and the double dislocation. In essence we showed that the topological defects of a PDW generically have 
``halos'' which provide evidence for the nature of this superconducting state. However, it is important to distinguish the halos of the PDW topological defects, which 
occur \textit{in the absence} of a magnetic field, to the PDW halo of a superconducting vortex which requires \textit{the presence} of a magnetic field 
\cite{Edkins-2018,Wang-2018,Lee-2018}. In contrast, the half-vortex and double dislocation topological defects of the PDW can be created only by impurities.

This work was partly motivated by evidence for half vortices found by Du and collaborators \cite{Du-2020} in STM experiments 
in the superconducting state of the high $T_c$ superconductor 
{\BSCCO}. The results of our work  will also be useful for investigating in the PDW superconducting state the heavy fermion superconductor $UTe_2$. 
Recent  STM experiments in this 
material have revealed that in its vortices the associated CDW has a dislocation-anti-dislocation  dipole structure \cite{Aishwarya-2024}. 
Our results provide alternative ways to investigate the nature of the PDW superconductors by investigating the structure of its interesting topological defects.

For practical reasons the  PDW was taken to be commensurate with the lattice 
spacing of the CuO$_2$ planes  with a periodicity of $8a_0$. The restriction to a commensurate PDW was needed for our numerics. However, in a truly commensurate 
PDW topological defects such as the half-vortex  and the double dislocation have a linearly divergent energy instead 
of a logarithmic divergent energy for an incommensurate PDW. Also,  in the presence of a uniform component of the $d$-wave superconducting order the half-vortex also 
has a linearly divergent energy. We have not discussed this case here. Nevertheless, in both cases half vortices can appear in the vicinity of static impurities. At any 
rate, a nearly commensurate PDW looks like a locally commensurate state with \textit{discommensurations} to account for the incommensurate character. The same 
physics is know to occur in conventional CDW states \cite{McMillan-1976}. This is also what is seen in  STM experiments in the cuprate superconductor {\BSCCO} 
where the observed CDW order is  locally commensurate   \cite{Mesaros-2016}.

The PDW state and its topological defects was treated using the  Landau-Ginzburg theory of the Refs. \cite{Berg-2009,Berg-2009a,Fradkin2015}, 
which describes the PDW and its CDW as intertwined orders. 
 In this approach, the CDW order parameter is a composite operator of the two independent PDW order parameter fields. 
 The static configuration of the half-vortex was derived  using a non-linear sigma model approximation valid deep in the PDW phase. 
 The BdG Hamiltonian of the PDW was then adapted to include the changes in the PDW order parameter in the presence 
of the topological defects. The calculation of the electronic states described by the BdG Hamiltonian is not self-consistent. 
Using the resulting Green functions of our effective theory, obtained numerically, we investigated the $4a_0$ CDW of the PDW phase as well 
as the effects of the topological defects on the electronic states. 
We should note that this approach is not self-consistent since the PDW order parameter (with or without defects) is fixed. 
The lack of self-consistency require some caveats on our results that are discussed below.

The half-vortex of the PDW is particularly interesting as it is essentially a dislocation of the CDW order parameter pinned to a half-flux 
quantum of the superconductor \cite{Agterberg-2020,Agterberg-2008,Berg-2009,Berg-2009b}. 
In this paper we investigated several aspects of the core of the half-vortex. We showed that the half-vortex of the PDW, which is an  LO type state, 
has a ``halo'' of an FF state. This FF state causes inversion symmetry to be broken at the core. 
Another interesting effect that arises in the presence of a half-vortex is  the splitting of the  peaks of the associated CDW at the ordering wave vector $2\mathbf{Q}$. 
The split peak  arises from a $\pi/2$-phase shift across the half-vortex core. We verified this explicitly  via examination of the Fourier transform of  the  LDOS. 
On the other hand, the double dislocation is also shown to  exhibit a split peak, which is due to a $\pi$-phase shift across its core. 
As expected, we found that there is no such split peak seen in the Abrikosov vortex consistent with the fact that this topological defect  
does not involve a dislocation of the CDW order of any type.

We  analyzed in detail the quasiparticle spectral function of a PDW with a half-vortex defect. In addition to the ``arc-like'' structure  at the Fermi 
surface of the Bogoliubov quasiparticle states which are seen in the defect-free PDW state \cite{Berg-2009,Baruch-2008}, we found that the half-vortex  induces  
asymmetric ``loop-like'' structures above the ``arcs''. We attributed the existence of these loops  to  the breaking of inversion symmetry at the core of the half-vortex. 
While much of the quasiparticle spectral function is very similar to that of the defect-free PDW, the presence of the half-vortex, and its inversion symmetry breaking, 
has a clear imprint in the spectral function. We also analyzed the real space position and the voltage dependence of the local differential conductance across a half 
vortex core. This dependence gives additional evidence for the existence of an FF component of the PDW in the core of the half-vortex.

Since this is not a self-consistent theory, both the spectral functions and the differential tunneling conductance results should be reliable at low energies 
but cannot be trusted at energies (voltages) substantially higher than the superconducting gap. As a matter of principle we expect that at energies well above the gap 
the superconducting order parameter should be progressively suppressed and the Bogoliubov quasiparticle effectively should become ``normal'' electrons. This also 
implies that the composite order parameters such as the CDW should also be progressively  suppressed well above the gap. This is not what  happens in our numerics 
which computed the BdG spectrum in a  fixed background of the superconducting order. To have a fully self-consistent theory requires a viable physical mechanism for 
a PDW which cannot be obtained by a weak coupling BCS-type theory. This is an open problem and a matter of current research.

In most systems in which PDW has been observed it happens in, at best, coexistence with a uniform superconducting state. This happens even in the 
case of {\LBCO}, which has, so far, provided the best evidence for PDW order (see Refs. \cite{Agterberg-2020,Tranquada-2023} and references therein). Thus it is 
important to understand what changes are brought about to our results when a PDW coexists with uniform superconducting order parameter. We plan to address this 
problem in a separate publication.

Contrary to the case of vortices in a  superconductor, whose number and separation are controlled by an external magnetic field, impurities are  needed 
to create the half-vortices and double dislocations of a PDW. Here we considered the problem of  single isolated topological defects. In practice this will require very 
clean systems so that the impurities are separated over large distances, larger than the size of the halos. On the other hand, a finite density of disorder has large 
qualitative effects in states such as the PDW, including the destruction of long-range CDW order \cite{Imry-1975,Efetov-1977}, and most intriguingly a possible 
charge-4$e$ superconducting state \cite{Berg-2009b} by proliferation of double dislocations as proposed in Ref. \cite{Mross-2015}. 
These important open problems  are beyond the scope of this paper.

\begin{acknowledgments}
We thank Ryan Levy, Raman Sohal, and Yuxuan Wang for useful discussions pertaining to the setup of the numerical simulation and Vidya Madhavan for addressing issues with the low bias particle-hole symmetry in the STM plots. Furthermore, we thank Peiyan Wu for his help in parallelizing the eigen-solvers, and for correcting the orthogonalization algorithms in these canned routines. M.R. thanks Mike Stone with his help in clarifying certain aspects of our solution process. This work was supported in part by the US National Science Foundation through Grant No. NSF DMR 2225920 at the University of Illinois (M.R. and E.F.) and by the (2017) Sloan Foundation (M.R.).
\end{acknowledgments}

\newpage
%
\clearpage

\begin{appendix}
\section{Bogoliubov-Valatin transformation\label{BV}}
Here we outline the solution process of the following Bogoliubov-de Gennes (BdG) Hamiltonian
\begin{equation}
        \begin{aligned}
    \hat{H}&= -\sum_{i,j,\sigma} t_{ij}
    \hat{c}^\dagger_{i\sigma}
    \hat{c}_{j\sigma}
    +\sum_{i,j}\Big(
    \tilde{\Delta}_{ij}
    \hat{c}^\dagger_{i\uparrow}
    \hat{c}^\dagger_{j\downarrow}
    +\text{H.c.}    
    \Big).
    \end{aligned}
\end{equation}
We associate each Latin index with position in this appendix. It is standard to diagonalize this operator using a Bogoliubov-Valatin (BV) transformation, but we will opt for a Nambu formalism instead. In the end we obtain the same BV transformation defining the same quasiparticles; the alternative route of defining our quasiparticle operators before diagonalizing can be found in Ref. \cite{Lieb-1961}. 
\par Let's define the following operator: $\mathbf{\tilde{\psi}}^T=[\mathbf{c}_\uparrow,\mathbf{c}_\downarrow]$, where the above components are vectors that consist of electron operators, $\big[\mathbf{c}_\sigma\big]_{i}=\hat{c}_{\mathbf{r}_i\sigma}$. Defining our Nambu spinor as $\tilde{\Psi}^T =[\tilde{\psi},\tilde{\psi}^\dagger]$ we can express our above Hamiltonian as a matrix product:
\begin{align}
    \hat{H}&=\tilde{\Psi}^\dagger \mathbf{\tilde{H}} \tilde{\Psi}=
    \tilde{\Psi}^\dagger \Big(\mathbf{\tilde{V}} 
    \mathbf{E} \mathbf{\tilde{V}}^{-1}\Big) \tilde{\Psi}
    \equiv\tilde{\gamma}^\dagger \mathbf{E}
    \tilde{\gamma}.
    \label{BdG_transform_eq}
\end{align}
The explicit definitions of the terms presented in \equref{BdG_transform_eq} will be covered in the next few paragraphs, but essentially this is just a similarity transformation. The matrix $\mathbf{\tilde{H}}$ takes the following generic form:
\begin{equation}
    \begin{aligned}
        \mathbf{\tilde{H}}
        =
        \frac{1}{2}\left[  
        \begin{array}{c c c c}  
        \mathbf{T}&
        \mathbf{\tilde{\Delta}}
        \\
        \mathbf{\tilde{\Delta}}^\dagger&  
        -\mathbf{T}^T
\end{array}  
\right]. 
    \end{aligned}
\end{equation}
Defining $[\mathbf{t}]_{ij}=-t_{ij}$ and $\mathbf{\Delta}_{\uparrow\uparrow}$/$\mathbf{\Delta}_{\uparrow\downarrow}$ (\textit{etc}.) for triplet/singlet superconductivity we have the following forms for our sub matrices:
\begin{equation}
    \begin{aligned}
        &\mathbf{T}
        =
        \left[  
        \begin{array}{c c c c}  
        \mathbf{t}& \mathbf{0}
        \\  
        \mathbf{0}& \mathbf{t}
        \end{array}
        \right]
        ,\;\;
        \mathbf{\tilde{\Delta}}
        =
        \left[
        \begin{array}{c c c c}  
        \mathbf{\Delta}_{\uparrow\uparrow}
        &\mathbf{\Delta}_{\uparrow\downarrow}
        \\
        \mathbf{\Delta}_{\downarrow\uparrow}& \mathbf{\Delta}_{\downarrow\downarrow}
        \end{array}  
        \right] 
        =-\mathbf{\tilde{\Delta}}^T.
    \end{aligned}
\end{equation}
This last condition on the SC matrix is a consequence of the anti-commutation relations, and the formation of this matrix should be chosen such that the above product reproduces the original Hamiltonian. 
\par We next define the quasiparticle operators, $\tilde{\gamma}=[\mathbf{b}^\dagger,\mathbf{b}]^T$, which can be expressed in terms of the electron creation and annihilation operators with the unitary matrix inducing the similarity transform, $\mathbf{\tilde{V}}$. Our similarity transform takes a very simple form because there are no zero modes in our spectrum due to the finite size of the system (needed for numerical diagonlization). This reveals we have $\pm$ energy pairs, and they are related via complex conjugation of the eigenvalue equation and swapping the top blocks of the BdG equations with the bottom. That is, given (sorted) eigenenergies $E_l>0$ ($l \in \{1,\dots ,2N^2\}$) we have the corresponding negative energy solutions
\begin{equation}
\begin{aligned}
\mathbf{\tilde{H}}
\begin{bmatrix}
\mathbf{\tilde{u}}_l\\
\mathbf{\tilde{v}}_l
\end{bmatrix}
=
E_l
\begin{bmatrix}
\mathbf{\tilde{u}}_l\\
\mathbf{\tilde{v}}_l
\end{bmatrix}
\quad
\Rightarrow
\quad
\mathbf{\tilde{H}}
\begin{bmatrix}
\mathbf{\tilde{v}}^*_l\\
\mathbf{\tilde{u}}^*_l
\end{bmatrix}
=
-
E_l
\begin{bmatrix}
\mathbf{\tilde{v}}^*_l\\
\mathbf{\tilde{u}}^*_l
\end{bmatrix}.
    \end{aligned}
    \label{ph_E}
\end{equation}
 We now define the following $2N^2\times 2N^2$ matrix $\mathbf{\tilde{u}}$ by setting it's $l^{th}$ column equal to $\mathbf{\tilde{u}}_l$, and similarly for $\mathbf{\tilde{v}}$, so the row index corresponds to the lattice site and the matrix $\big[\mathbf{E}\big]_{lk} = E_l\delta_{lk}$. The unitary matrix inducing the similarity transform takes the following form:
\begin{align}
    \mathbf{\tilde{V}}=\begin{bmatrix}
    \mathbf{\tilde{v}}^*
    &
    \mathbf{\tilde{u}}\\
    \mathbf{\tilde{u}}^*
    &
    \mathbf{\tilde{v}}
    \end{bmatrix}.
\end{align}
The block matrices, $\mathbf{\tilde{u}}$ and $\mathbf{\tilde{v}}$, contain the real space coherence factors, and provide us with the following electron operators in terms of our quasiparticle operators:
\begin{equation}
    \begin{aligned}
    &\hat{c}_{i\uparrow }= \tilde{v}^*_{il}\hat{b}_{l}^\dagger
    +\tilde{u}_{il}\hat{b}_{l}
    ,\;\;\;\;\;\;
    \hat{c}_{i\uparrow }^\dagger= \tilde{v}_{il}\hat{b}_{l}
    +\tilde{u}^*_{il}\hat{b}^\dagger_{l},
    \\
    &\hat{c}^\dagger_{i\downarrow }= \tilde{u}^*_{il}\hat{b}_{l}^\dagger
    +\tilde{v}_{il}\hat{b}_{l}
    ,\;\;\;\;\;\;
    \hat{c}_{i\downarrow }= 
    \tilde{u}_{il}\hat{b}_{l}
    +\tilde{v}^*_{il}\hat{b}_{l}^\dagger.
    \label{full_config}
\end{aligned}
\end{equation}
We can substitute these into our Hamiltonian above, and we find
\begin{align}
    \hat{H}=\sum_{E_l>0} E_l \hat{b}_l^\dagger \hat{b}_l+E_G.
\end{align}
The explicit form for $E_G$ and $\ket{G}$ will not matter for what follows. The ground state is taken to satisfy
\begin{align}
    \hat{b}_l\ket{G}=0.
\end{align}
Defining $S_{ij}=\tilde{v}^*_{il}(\tilde{u}^*)^{-1}_{lj}$, it can be shown the ground state is a coherent state of Cooper-pairs; that is, it is related to the vacuum state, $\ket{0}$, in the following way
\begin{equation}
    \begin{aligned}
        \ket{G} = 
        \mathcal{N}
        \exp\Big(
        \frac{1}{2}
        \hat{c}^\dagger_{i\uparrow}
        S_{ij}
        \hat{c}^\dagger_{j\downarrow}
        \Big)
        \ket{0}
    \end{aligned}.
\end{equation}
What we'll need below is the single-particle excited states
\begin{align}
    \hat{H}\hat{b}_l^\dagger\ket{G}=(E_G+E_l)\hat{b}_l^\dagger\ket{G}
\end{align}
These above relations are the building blocks of our Green functions, found in Appendix \ref{zero-temp-Green}. Before deriving those expression, it is advantageous to reconfigure our Hamiltonian for the case of a singlet SC because this will be computationally more efficient.

\section{Numerical set up for the Diagonalization of the Bogoliubov-de Gennes Equations
\label{numerical-diagonalization}}
This appendix outlines the set up for numerical diagonalization of the BdG Hamiltonian introduced in the last appendix. When we are considering a singlet SC, we can simplify the superconducting matrix to the following form:
\begin{equation}
    \begin{aligned}
        \mathbf{\tilde{\Delta}}
        =
        \left[
        \begin{array}{c c c c}  
        \mathbf{0}
        &\mathbf{\Delta}
        \\
        -\mathbf{\Delta}& \mathbf{0}
        \end{array}  
        \right] 
        ,\;\;
        \mathbf{\Delta}^T=\mathbf{\Delta}.
    \end{aligned}
\end{equation}
It pays to change the basis of our Nambu spinor, so we can work with an effective Hamiltonian of half the dimension of both our column and row space. We define
\begin{equation}
    \begin{aligned}
        \Psi
        =
        \begin{bmatrix}
            \mathbf{c}_{\uparrow}&
            \mathbf{c}^\dagger_{\downarrow}&
            \mathbf{c}_{\uparrow}&
            \mathbf{c}^\dagger_{\uparrow}&
        \end{bmatrix}^T
        =\mathbf{O}\tilde{\Psi}
        \Rightarrow&\hat{H}
        =\Psi^\dagger
        \mathbf{V}
        \mathbf{E}
        \mathbf{V}^\dagger
        \Psi.
        \label{transform_nambu}
    \end{aligned}
\end{equation}
The orthogonal transformation in question takes the following simple form and gives an explicit relation between the similarity transforms in question:
\begin{equation}
    \mathbf{O}
    =
    \begin{bmatrix}
        \mathbf{I}&\mathbf{0}&\mathbf{0}&\mathbf{0}\\
        \mathbf{0}&\mathbf{0}&\mathbf{0}&\mathbf{I}\\
        \mathbf{0}&\mathbf{I}&\mathbf{0}&\mathbf{0}\\
        \mathbf{0}&\mathbf{0}&\mathbf{I}&\mathbf{0}
    \end{bmatrix},\;
    \mathbf{\tilde{V}}=\mathbf{O}^T\mathbf{V}.
\end{equation}
Defining the following matrices:
\begin{equation}
     \begin{aligned}
    \mathbf{H}_\pm=
    \begin{bmatrix}
    \mathbf{t}
    &
    \pm\mathbf{\Delta}
    \\
    \pm \mathbf{\Delta}^\dagger
    &
    -\mathbf{t}
    \label{Hpm}
    \end{bmatrix},
\end{aligned}
\end{equation}
we find that our transformed Hamiltonian takes the following form [$\psi_\pm$ are defined inline according to \equref{transform_nambu}]:
\begin{equation}
    \begin{aligned}
        \mathbf{H}
        =
        \mathbf{O}
        \mathbf{\tilde{H}}
        \mathbf{O}^T
        =
        \begin{bmatrix}
            \mathbf{H}_+&\mathbf{0}
            \\
            \mathbf{0}&\mathbf{H}_-
        \end{bmatrix}
    \end{aligned}
    ,\;\;
    \Psi = \begin{bmatrix}
        \psi_+\\
        \psi_-
    \end{bmatrix}.
\end{equation}
The matrix is block diagonal, meaning we can diagonalize the two sub-Hamiltonians independently. The eigenvectors of the lower block are related to those of the upper via the same transformation shown in \equref{ph_E}. In addition to this, the energies of each block come in $\pm$ pairs. This can be seen by complex conjugating the eigenvalue equations defined by \equref{Hpm}, then applying the following orthogonal matrix: $      \mathbf{O}'=
        \begin{bmatrix}
            \mathbf{0}&\mathbf{I}
            \\
            -\mathbf{I}& \mathbf{0}
        \end{bmatrix}$.

Solving one of these sub-blocks is enough. After doing so, and organizing our eigenvectors from least to greatest (in energy), we arrive at the following similarity transforms:
\begin{equation}
    \begin{aligned}
        \mathbf{V}_+
        &=
        \begin{bmatrix}
            \mathbf{v}^*&
            \mathbf{u}
            \\
            -\mathbf{u}^*
            &
            \mathbf{v}
        \end{bmatrix}
        ,\;\;
        \mathbf{V}_-
        =
        \begin{bmatrix}
            \mathbf{v}^*&
            -\mathbf{u}
            \\
            \mathbf{u}^*
            &
            \mathbf{v}
        \end{bmatrix}
        \;\;
        \Rightarrow
        \;\;
        \mathbf{V}
        =
        \begin{bmatrix}
        \mathbf{V}_+&\mathbf{0}
        \\
        \mathbf{0}&\mathbf{V}_-
        \end{bmatrix}.   \label{singlet_config}
    \end{aligned}
\end{equation}
Thus, we can simply diagonalize $\mathbf{H}_+$, then use our above string of relations to find the coherence factors given in the previous appendix if needed. We work with these coherence factors defined above in the following appendix. 
\section{Retarded Green Function at zero Temperature\label{zero-temp-Green}}
Using the results laid out in the previous appendixes we can find the retarded zero temperature Green functions. We start with the real space retarded Green function at zero temperature
\begin{align*}
G_\sigma(\mathbf{r}_i,\mathbf{r}_j,t)
=&-i\theta(t)
\big\langle \big\{\hat{c} _{ \mathbf{r}_i\sigma}(t),\hat{c}^\dagger_{ \mathbf{r}_j\sigma}
\big\}\big\rangle  \\
=&-i\theta(t)
\big\langle \big\{e^{i\hat{H}t}\hat{c}  _{\mathbf{r}_ i\sigma}e^{-i\hat{H}t},\hat{c}^\dagger_{ \mathbf{r}_j\sigma}
\big\}\big\rangle.
\end{align*}
 Using the action of our quasiparticle operators on our ground state, and relation (\equref{full_config} or \equref{singlet_config}), we can evaluate these terms. Recall that we are in the singlet configuration, so we may work with a single spin, say, $\sigma=\uparrow$, and drop the spin label:
 \begin{align*}
G(\mathbf{r}_i,\mathbf{r}_j,t)
&=-i\theta(t)\sum_{E_l> 0}\Big(
      v_{il}v_{jl}^*e^{iE_lt}
        +  u^*_{il}u_{jl}e^{-iE_lt}
        \Big).
 \end{align*}
 
 We are interested in the Fourier transform, which takes us from the time domain to the frequency domain. This integral requires a dampening factor, $\epsilon$, for convergence, which represents the energy resolution, taken to be 2.5 meV. The Fourier transform gives us the Green function in a familiar form:
\begin{widetext}
    
\begin{equation}
\begin{aligned}
G(\mathbf{r}_i,\mathbf{r}_j,\omega)
\equiv
\int_{-\infty}^{\infty}
G(\mathbf{r}_i,\mathbf{r}_j,t)
e^{i(\omega+i\epsilon)t}
dt
&=
\sum_{E_l\geq 0}
\Big(
\frac{u_{il}u_{jl}^*}
{\omega- E_l+i\epsilon}
+  
\frac{v^*_{il}v_{jl}}
{\omega+ E_l+i\epsilon}
\Big).
\end{aligned}
\label{Green-rw}
\end{equation}
We'll also need to Fourier transform, to $k$-space to obtain the spectral function
\begin{equation}
    \begin{aligned}
    G(\mathbf{k_a},\mathbf{k_b},\omega)
   &=\frac{1}{N}
   \sum_{i,j}
      G(\mathbf{r}_i,\mathbf{r}_j,\omega)
      e^{i\mathbf{k_a}\cdot\mathbf{r_i}}
     e^{-i\mathbf{k_b}\cdot\mathbf{r_j}}=
     \sum_{E_l\geq 0}\Bigg(
      \frac{ \tilde{u}_{l}(\mathbf{k_a}) \tilde{u}^*_{l}(\mathbf{k_b})}{\omega- E_l+i\epsilon}
        +  \frac{\tilde{v}_{l}(\mathbf{k_b}) \tilde{v}^*_{l}(\mathbf{k_a})}{\omega+ E_l+i\epsilon}
        \Bigg).
        \label{Green-kw}
    \end{aligned}
\end{equation} 
With the same definition as in the text for the coherence factors
\begin{align*}
    \tilde{u}_{l}(\mathbf{k})
   &=\frac{1}{N}
   \sum_{i} u_{il}      e^{i\mathbf{k}\cdot\mathbf{r_i}},
   \;\;\;
   \tilde{v}_{l}(\mathbf{k})
   =\frac{1}{N}
   \sum_{i} v_{il}      e^{i\mathbf{k}\cdot\mathbf{r_i}}.
\end{align*}

\end{widetext}

\begin{figure*}
\vspace{0pt}
\hspace{0pt}
\subfloat[\label{figure_12_1}]
{\includegraphics[width=0.33\textwidth]{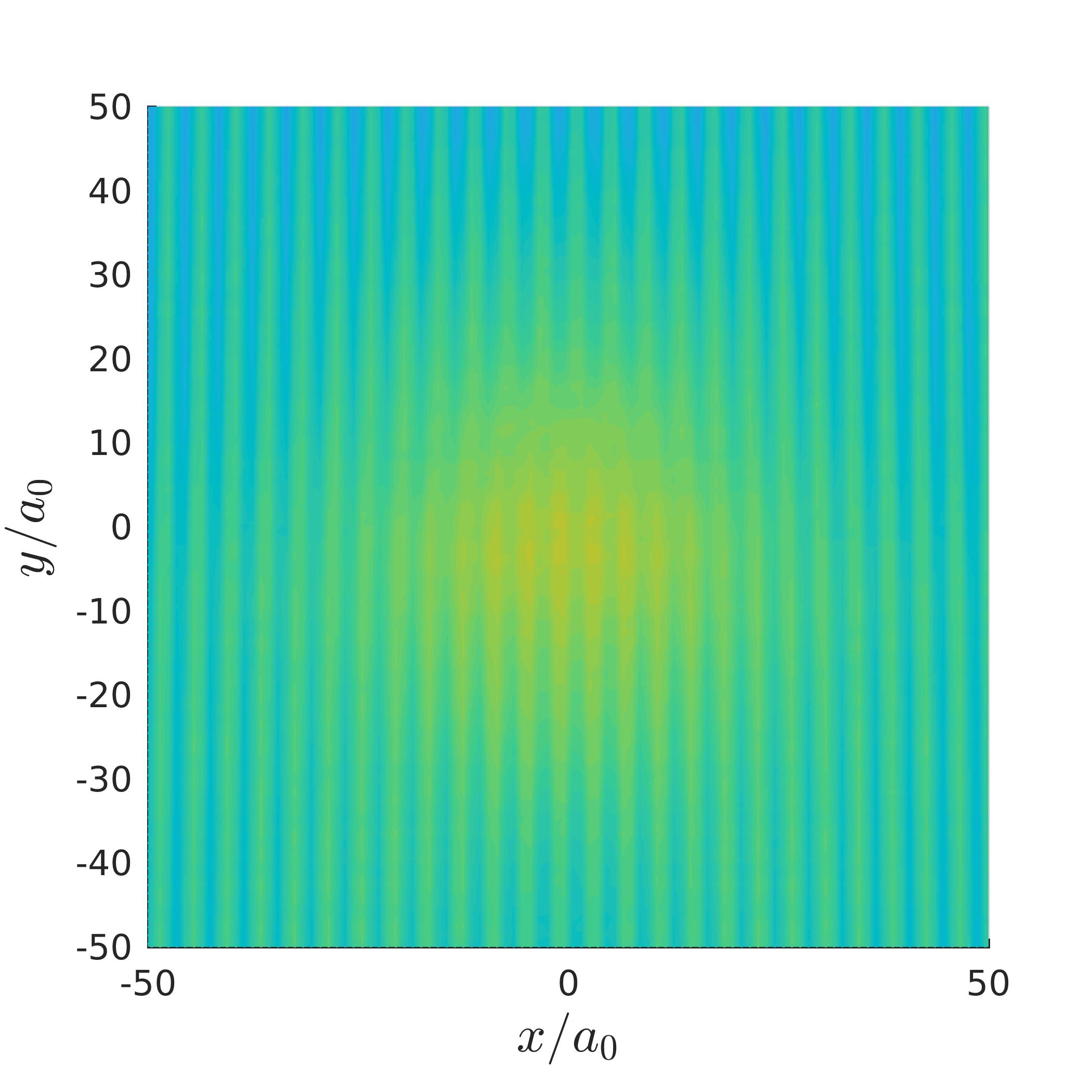}}
\subfloat
[\label{figure_12_2}]
{\includegraphics[width=0.33\textwidth]{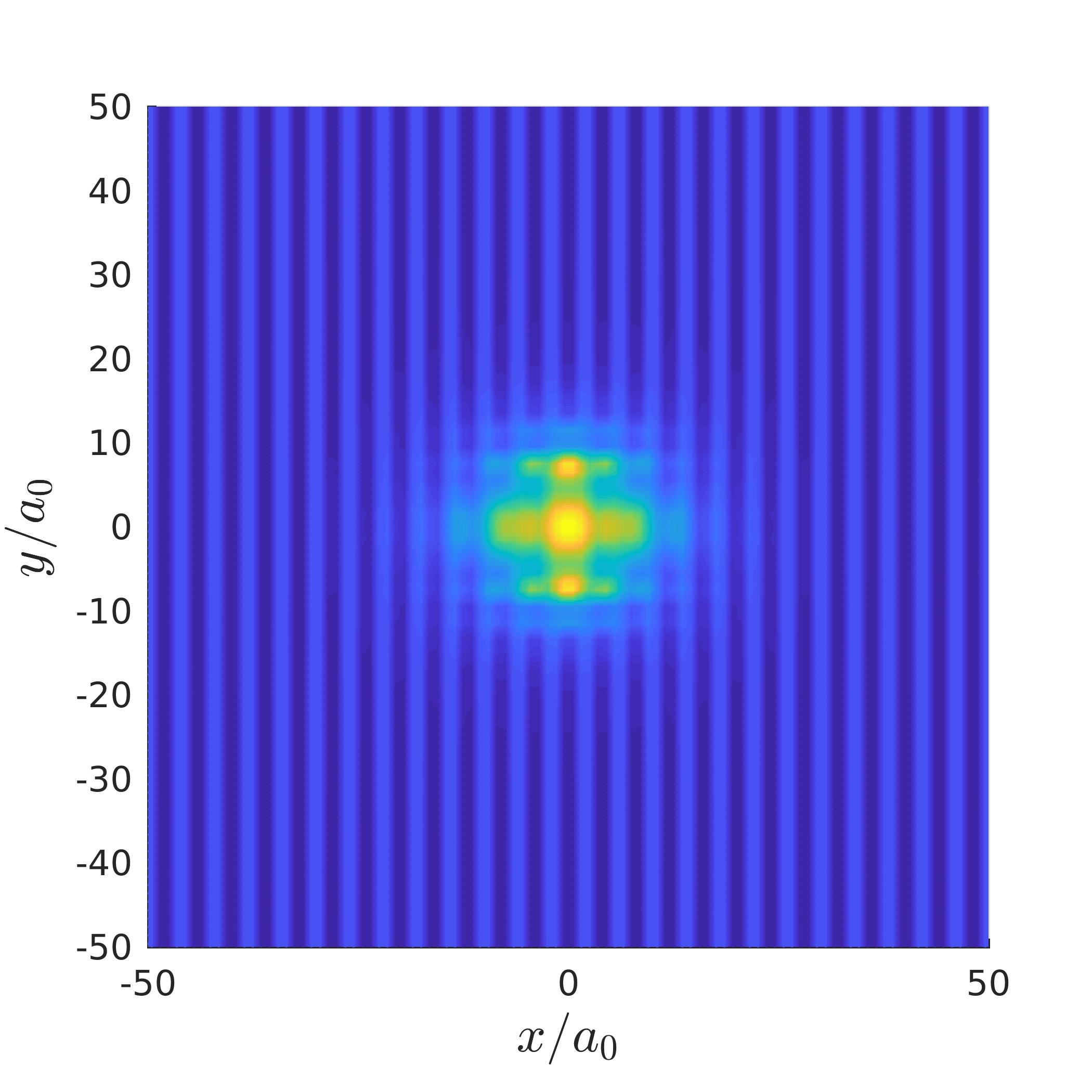}}
\subfloat[\label{figure_12_3}]
{\includegraphics[width=0.33\textwidth]{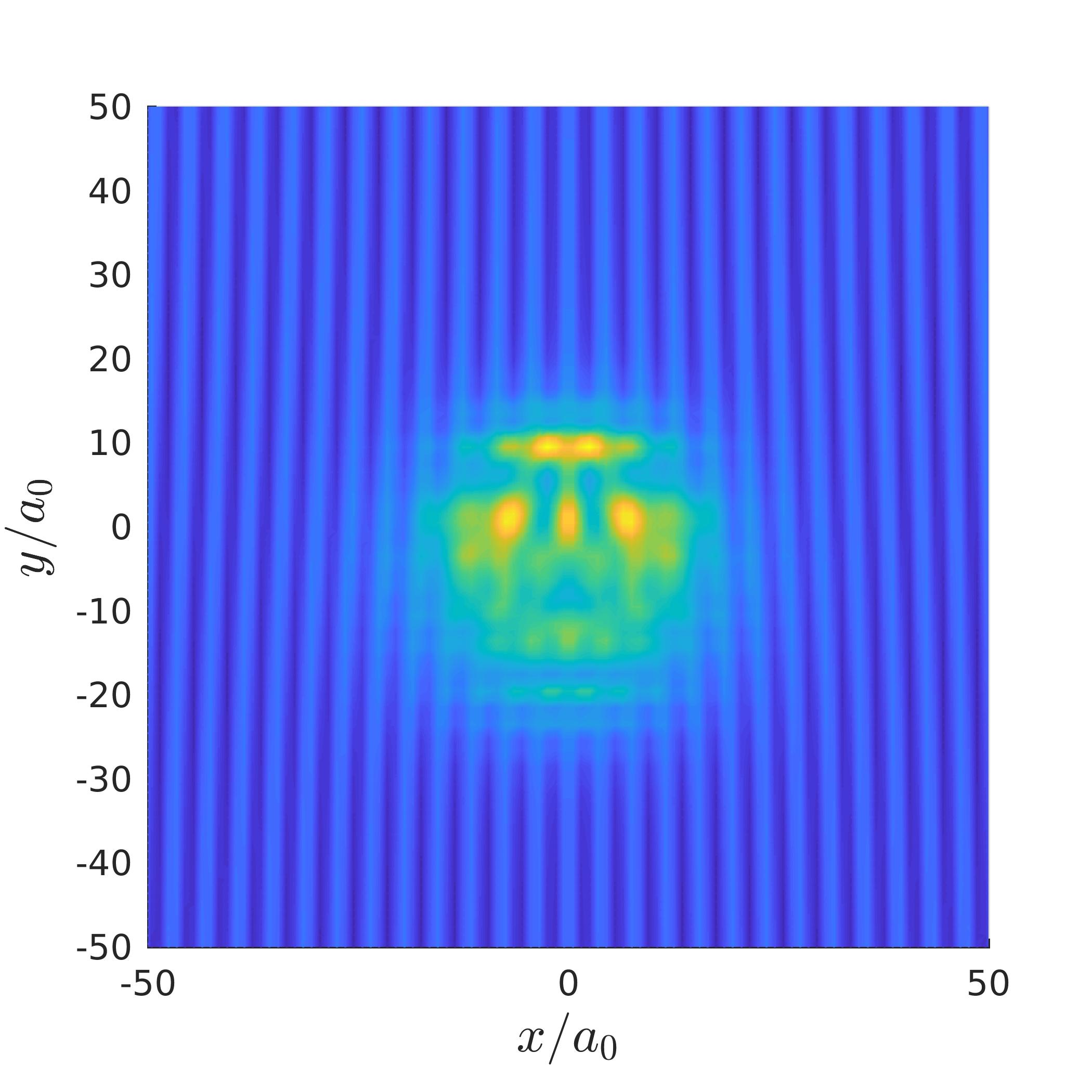}} 
\\
\subfloat[\label{figure_12_4}]
{\includegraphics[width=0.33\textwidth]{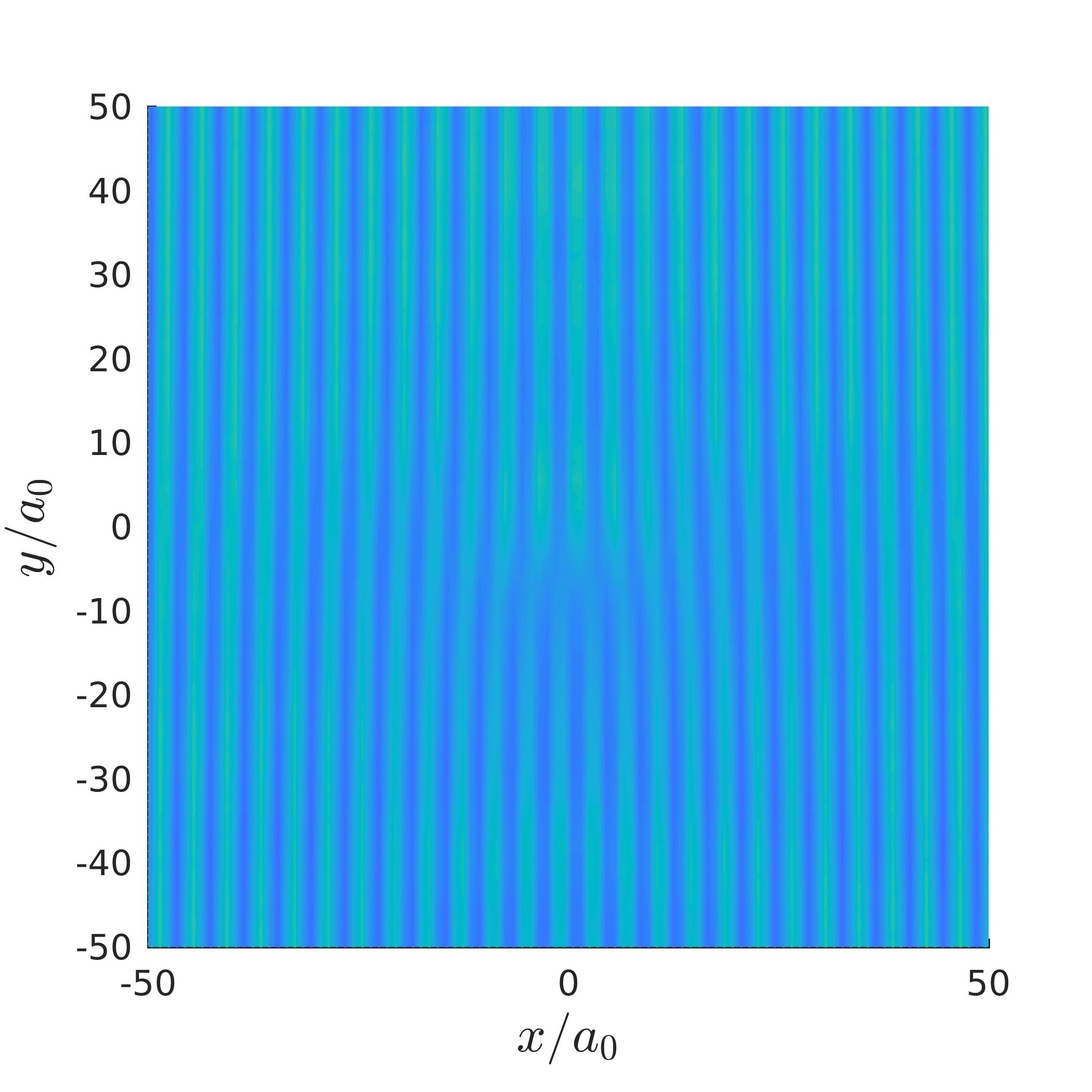}}
\subfloat
[\label{figure_12_5}]
{\includegraphics[width=0.33\textwidth]{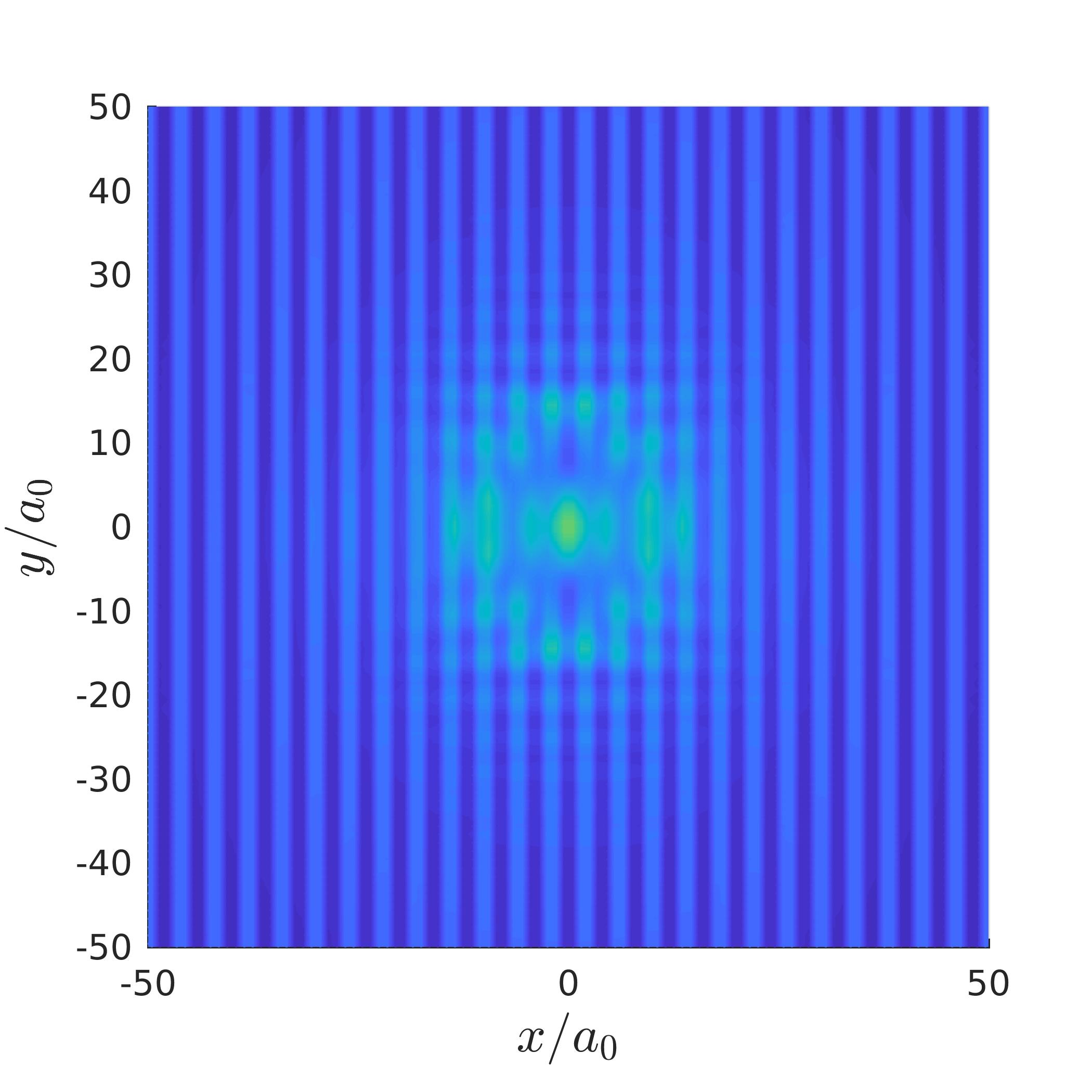}}
\subfloat[\label{figure_12_6}]
{\includegraphics[width=0.33\textwidth]{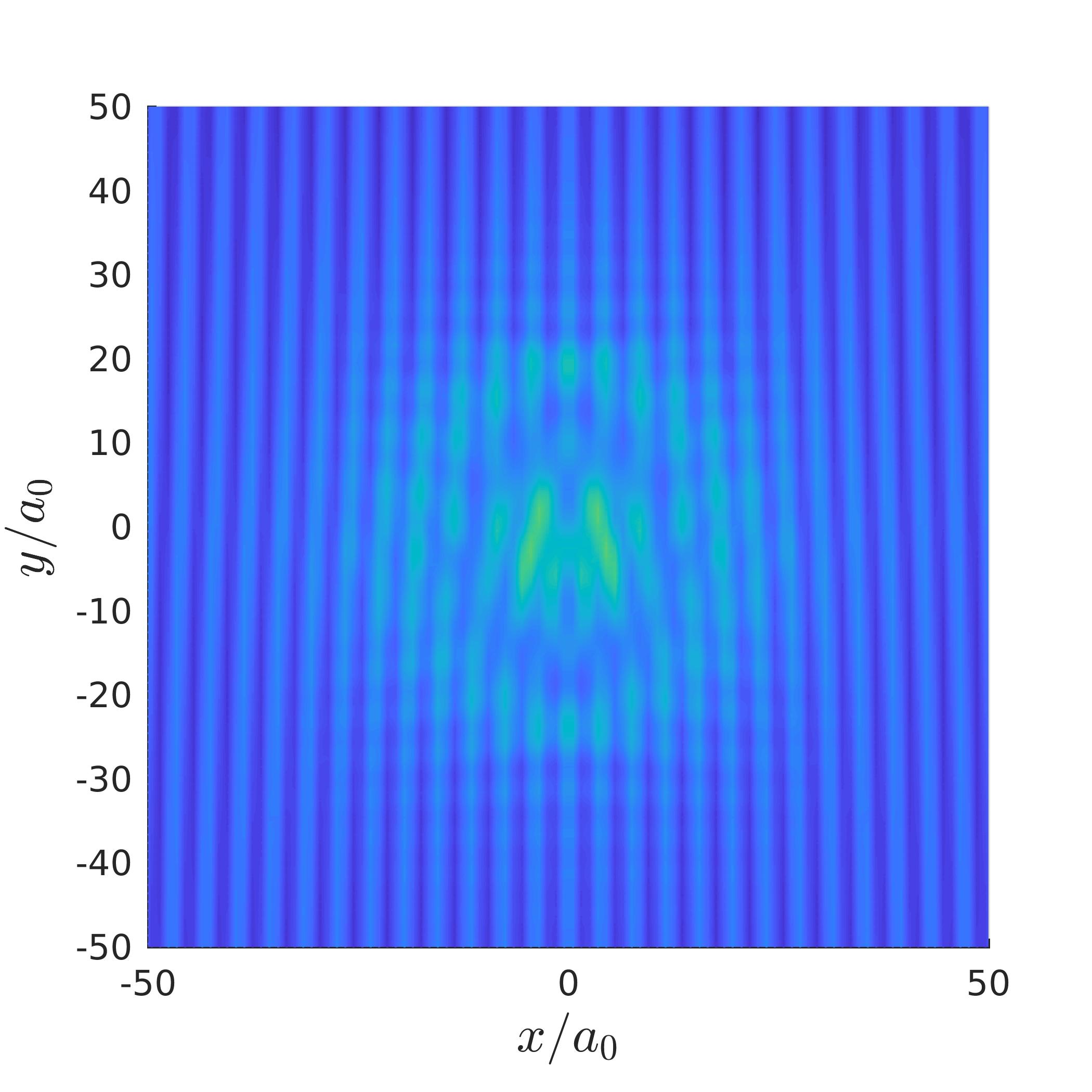}} 
\hspace{0pt}
\caption{(Online Color) Comparison plots of the LDOS for a PDW order SC with the three topological defects [described in \equref{order-parameter-configs}]. First, we consider the zero bias LDOS given in the first row for (a) the half-vortex, (b) the Abrikosov vortex and (c) the double dislocation. In the second row we include plots of the LDOS at probing voltages of $0.75\bar{\Delta}$ for (d) the half-vortex, (e) the Abrikosov vortex, and (f) the double dislocation. Each plot is normalized to the scale of the defect as in the main text.}
\label{figure_12}
\end{figure*}

\begin{figure}
\vspace{0pt}
\hspace{0pt}
{\includegraphics[width=.5\textwidth]{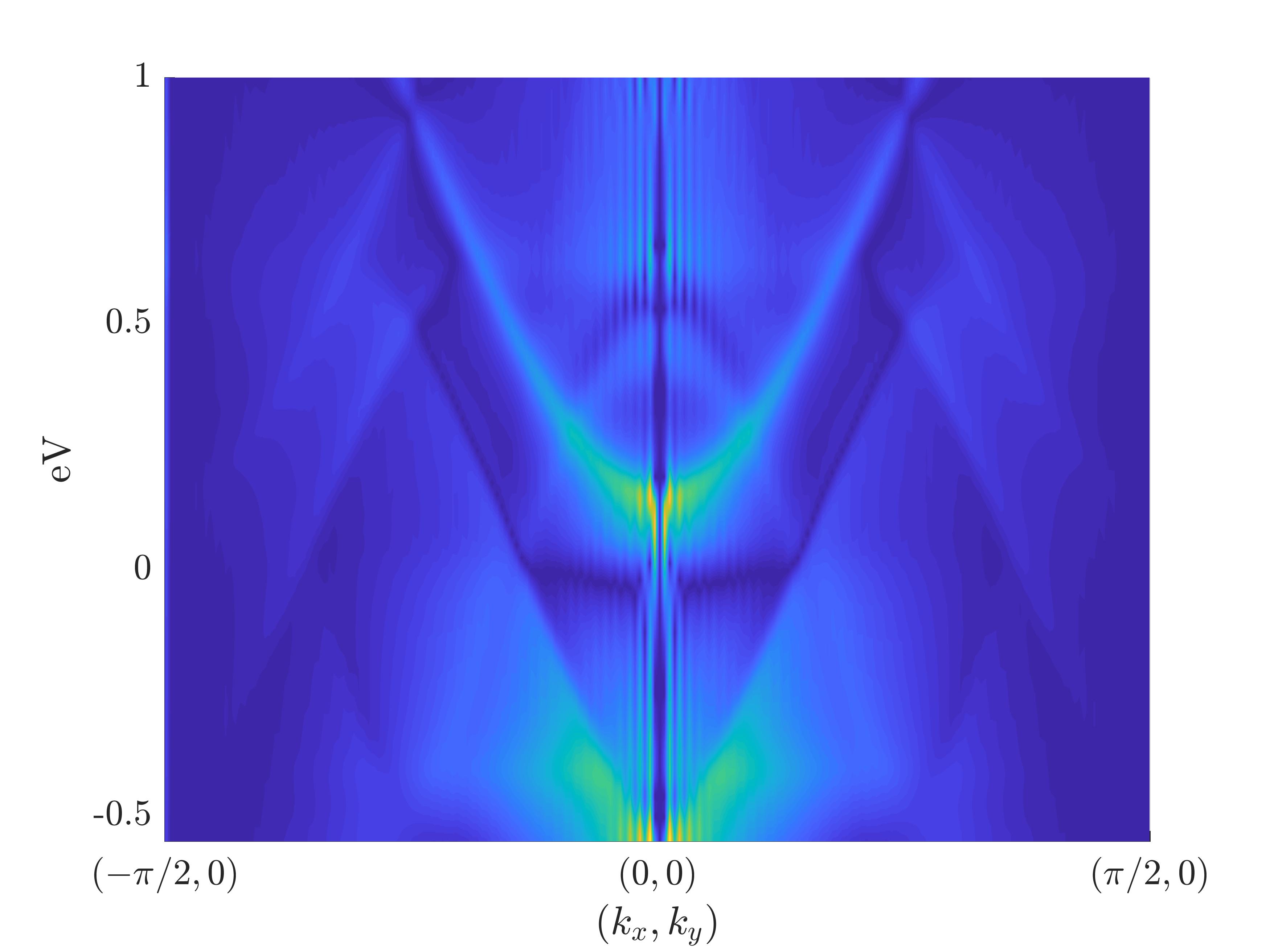}}
\hspace{0pt}
\caption{(Online Color) The Fourier transform of the LDOS for an Abrikosov vortex of the PDW state for a range of probing voltages. We take cuts along the line $k_y=0$ and plot the absolute value of the FT vs probing voltage. This figure showcases the QPI for the Abrikosov vortex, for which we note the dispersing quasiparticles; hence, these patterns do not represent a bound state. We suppressed the intensity around the $\Gamma$-point in this plot. A similar pattern of the QPI is present in the case of the double dislocation, suggesting it too does not possess bound quasiparticles.}
\label{figure_13}
\end{figure} 
%
\section{Plots of \texorpdfstring{$\rho_{2\mathbf{Q}}$}{TEXT} and the LDOS} \label{LDOS_appendix}
This appendix provides comparison plots between the induced CDW for our three defects using \equref{2Q} directly. We still see the signatures of the defects described in the text (\textit{e.g.}, a edge dislocation). Representative plots of the $\rho_{2\bm{Q}}$ are shown in \figref{figure_3}. Comparison plots to the LDOS at zero bias are shown in \subfigrangref{figure_12_1}{figure_12_3}. Note that the core of the vortex is noticeably different in \figref{figure_3} than in \figref{figure_12} for the subplots which correspond to the full vortex and the double dislocation. This is due to the fact the LDOS calculation possesses information regarding the quasiparticles, and the patterns are a result of quasiparticle interference. The half-vortex looks similar in both these figures because its core contains a fully gapped FF state. On the other hand, the other two defects, the full vortex and the double dislocation, possess order parameters with a vanishing superconducting gap at the core. The integrated LDOS (the charge density) for the full vortex and the double dislocation also possesses a nonzero weight within the vortex core due to the fact the vortex can accommodated quasiparticles, but the dynamic features are integrated out. In \subfigrangref{figure_12_4}{figure_12_6} we also provide a set of LDOS at a bias of $0.75\bar{\Delta}$. Notice the depletion of states that occurs in the core of the vortex even in the case of a half-vortex where there is a partial gap present.
\par Note these quasiparticles do not reside solely in the vortex core and are thus not bound states. We can illustrate this by taking a FT of the LDOS and plotting cuts in momentum space over a range of energies to observe a dispersion relation. The nonzero Fourier harmonics at each energy, $\omega$, corresponds to the scattering wave vectors connecting different regions of the surface determined by the spectral function evaluated at the same $\omega$. The regions of large joint DOS on this surface provides us with the dominate Fourier harmonics, and if these regions disperse, we see it in the FT of the LDOS. In \figref{figure_13} the QPI of the Abrikosov vortex is provided for cuts in $k$-space along the nodal direction, and it indicates we have dispersing quasiparticles by the change in Fourier harmonics.
\begin{figure}
\subfloat[\label{figure_14_1}]{\includegraphics[width=.32\textwidth]{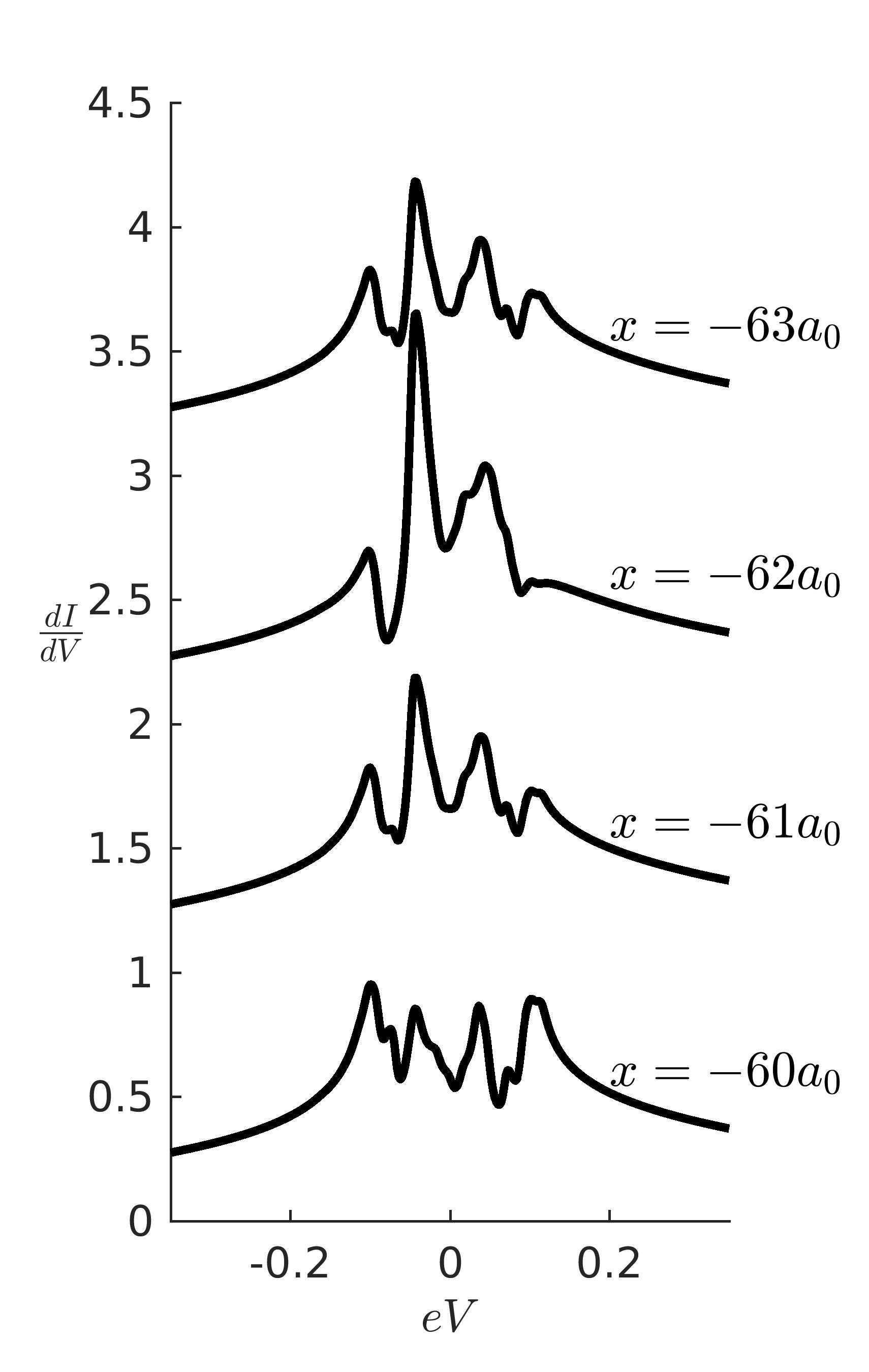}}
\\
\subfloat[\label{figure_14_2}]{\includegraphics[width=.32\textwidth]{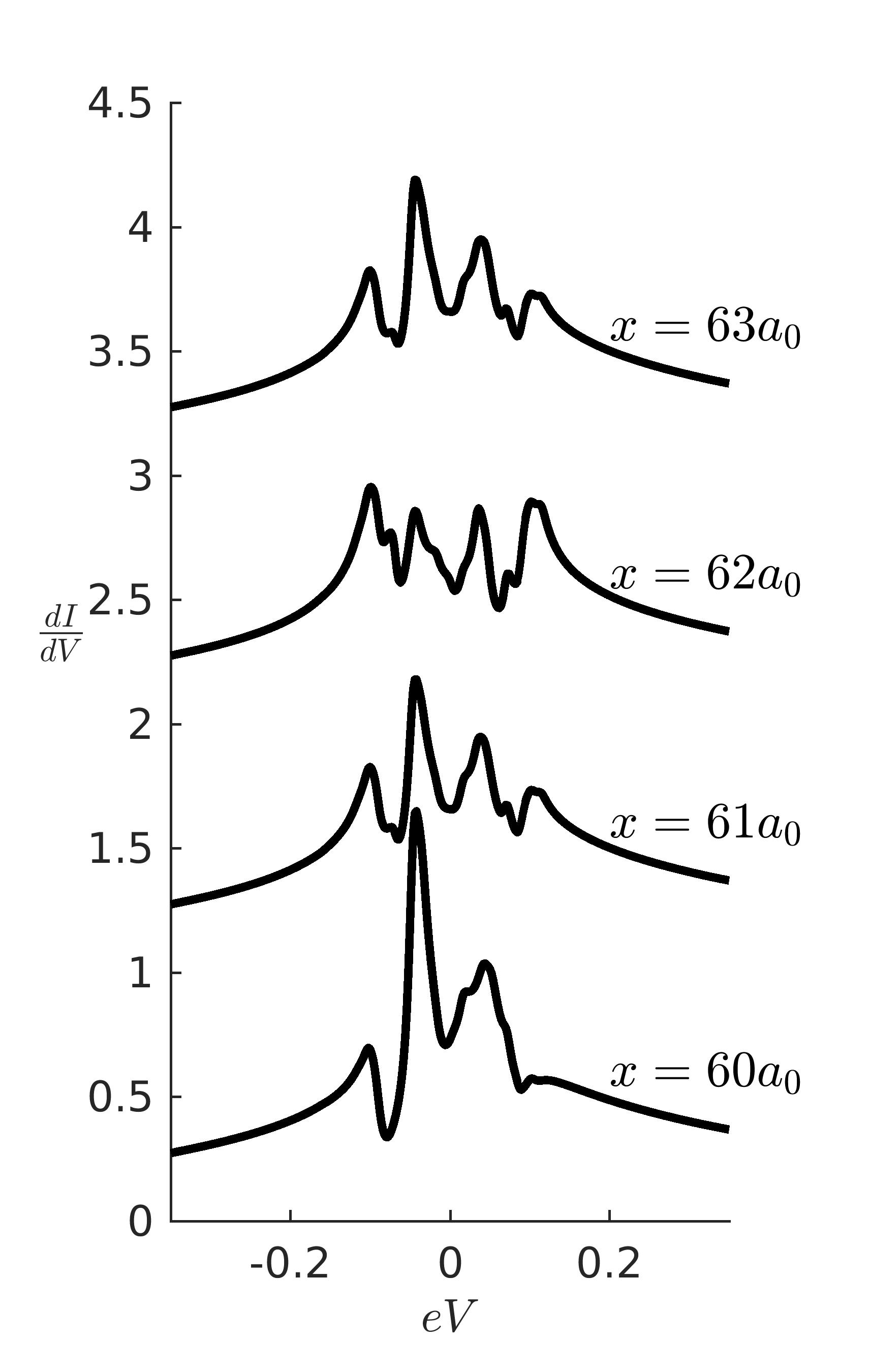}}
\hspace{0pt}
\caption{Comparison of $\frac{dI}{dV}$ curves related to the tunneling DOS for the half-vortex (taken to live on the CuO bonds and not the center of the plaquette) at different lattice sites, indicated with the $x$ position relative to the vortex core (with a relative shift of 1 between curves). We look at two sets of curves taken outside the vortex core of radius $r_0=24a_0$, (a) one to the left (negative $x$) and (b) one to the right (positive $x$). Both sets of curves repeat every four lattice spacings (in $x$), like an LO state. This indicates a relative shift between the curves in the two columns by two lattice spacings. The shift seen in these plots are a result of the jump in $\theta_+$, defined in \equref{theta-pm}, occurring across the core of the vortex.}
\label{figure_14}
\end{figure}

\begin{figure*}
\subfloat[\label{figure_15_1}]{\includegraphics[width=.45\textwidth]{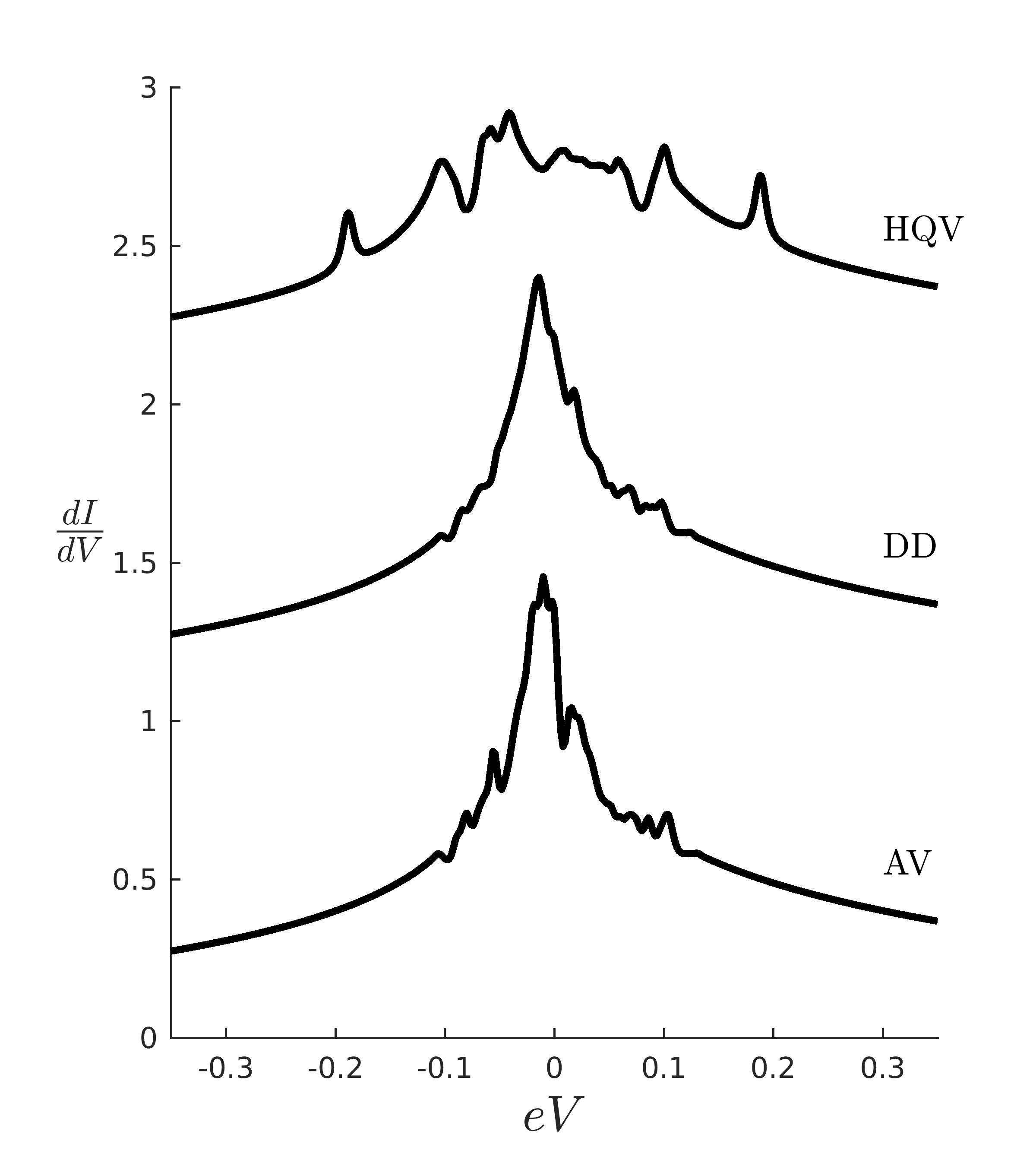}}
\subfloat[\label{figure_15_2}]{\includegraphics[width=.45\textwidth]{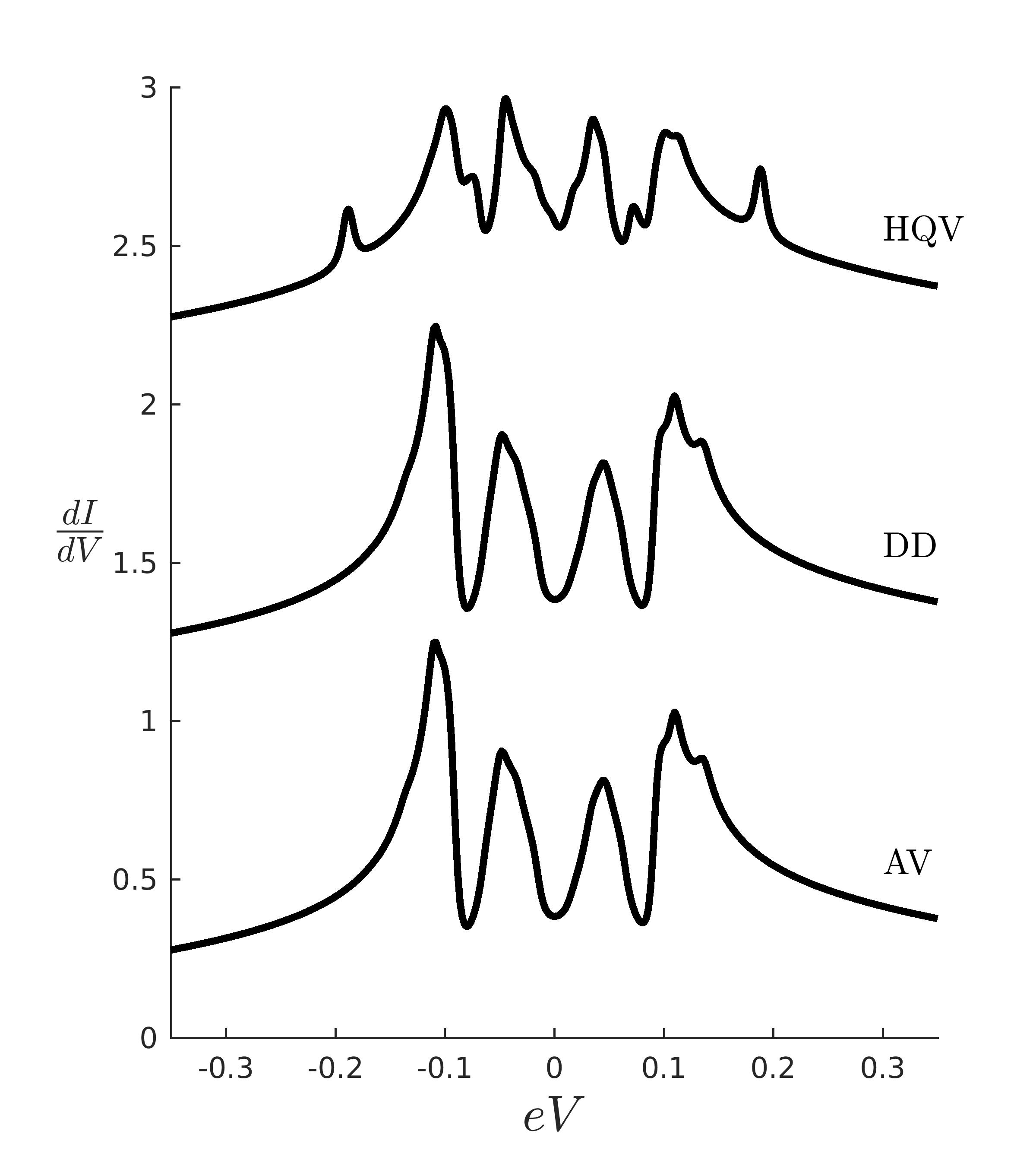}}
\hspace{0pt}
\caption{Comparison of $\frac{dI}{dV}$ curves related to the tunneling DOS for the half-vortex, the double dislocation and the Abrikosov vortex at two different lattice sites. The amplitudes are expressed in arbitrary units, each being normalized to the scale of the respective defect. In terms of fractions of the halo radius, $r_0=24a_0$, the plots are taken at a distance of (a) $0.025\;r_0$ and (b) $4\; r_0$ in respect to the vortex cores. Inside the core (a), the double dislocation and the Abrikosov vortex resemble a free particle dispersion since both the PDW components are small here. However, in the case of the half-vortex the core state is still superconducting, resembling a squeezed in FF state. Outside the core (b) all the curves resemble the pure PDW, but the half-vortex possesses additional satellite peaks due to the fact it possesses inequivalent PDW components.}
\label{figure_15}
\end{figure*}
\section{Tunneling DOS Comparison}\label{DOS_Sup}
Here we compare the plots corresponding to the tunneling DOS for the half-vortex outside the vortex core on its l.h.s. and r.h.s. [Figs. \ref{figure_14_1} and \ref{figure_14_2} respectively]. The tunneling DOS are labeled with only $x$ coordinates to indicate how far we are in respect to the half-vortex core ($x$ being measured relative to its center). This indicates we are looking at consecutive $x$ positions in a given sub figures as well as we are comparing curves to the left (negative sign) and right (positive sign) of the vortex. The particular positions used for the representative curves are somewhat arbitrary in the sense we want only positions outside the core. Also, note that the defect is placed on the CuO bonds here opposed to the center of the plaquette as was done in the main text for the LDOS. The spectral functions corresponding to these two situations are equivalent, meaning the energetics are very similar, but the shape of the $dI/dV$ curves will certainly change character since the gap amplitude takes on different values on the lattice sites.
\par Both sets of curves, to the right and to the left of the vortex core, have a periodicity of four lattice spacings, just like that of the pure PDW in the text (the same conclusion holds for the placement of the defect at the center of the plaquette). However, unlike a pure LO state the data appears to have a relative shift of two lattice spacings when comparing the curves from the right to the left (again this holds true for the half-vortex placed at the center of the plaquette). As was discussed in the text, this is due to the jump in $\theta_+$ by $\pi/2$ across the half-vortex's core and can be explained as an accumulation in phase due to the non-zero COM-momentum of our Cooper pairs; that is, $\mathbf{Q}\cdot\mathbf{r}=\pi/2$ if $\mathbf{r}=2a_0\mathbf{e}_x$. This is a smoking gun signature of the half-vortex since a bona-fide LO state would not break inversion symmetry like so. 
\par Now we compare the tunneling DOS of our topological defects inside and outside the core of the vortex [Figs. \ref{figure_15_1} and \ref{figure_15_2}, respectively]. The Abrikosov vortex and the double dislocation both resemble a free particle tunneling DOS, albeit some additional wiggles. This is simply because we are not right at the center of the core. Also, notice that the curves for these two defects are not in perfect agreement here, due to the form of the topological defect. Indeed, the electron DOS has to be distinct in the core of each vortex due to the difference in the respective defect. The half-vortex still possesses signatures of superconductivity in the core due to the fact it has FF character there. In fact, it resembles a squeezed in FF state.
\par Outside the core the double dislocation and the Abrikosov vortex look almost identical to one another and also the pure PDW (LO state). This is because the topological defect has less of an influence on the electronic states far away from the core. The half-vortex has additional wiggles in its tunneling DOS corresponding to satellite peaks. These additional peaks arise here and not for the other two defects because of the half-vortex's inequivalent Fourier components.
\begin{figure*}
\subfloat[\label{figure_16_1}]{\includegraphics[width = .33\textwidth]{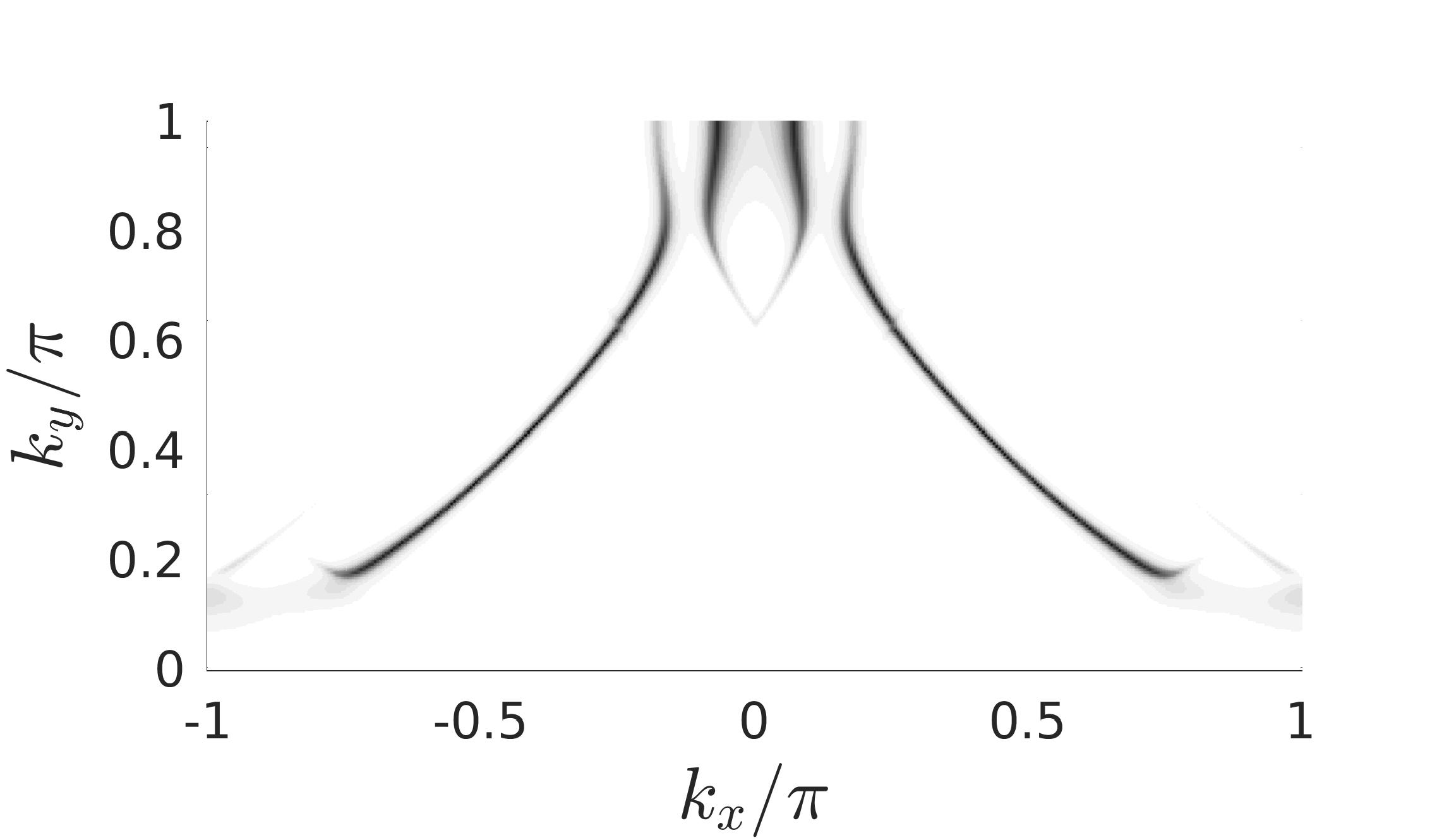}
} 
\subfloat[\label{figure_16_2}]{\includegraphics[width = .33\textwidth]{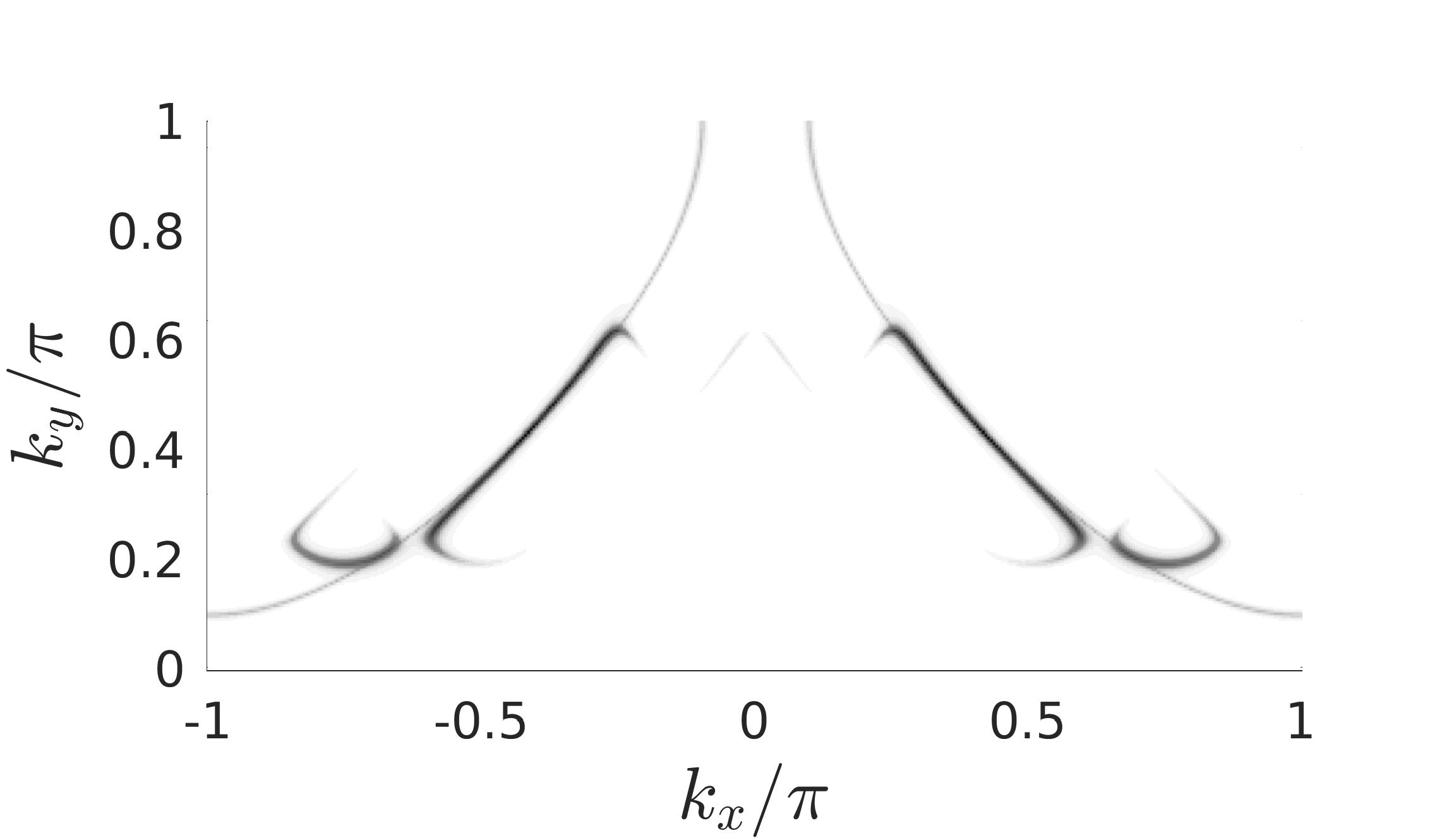}
} 
\subfloat[\label{figure_16_3}]{\includegraphics[width = .33\textwidth]{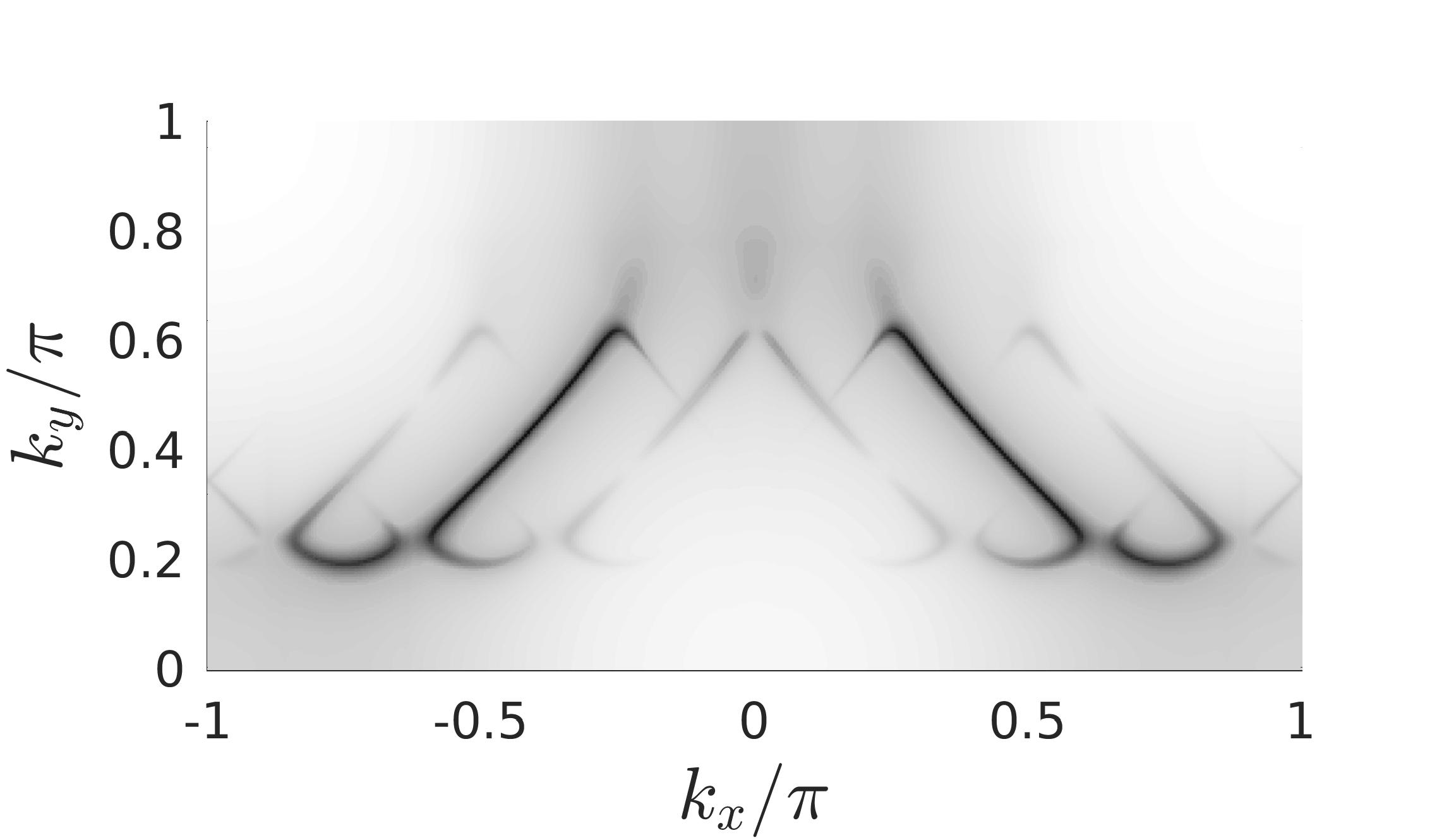}
} 
\caption{Comparison plots of various Fermi surfaces. First, we provide a plot for (a) a defect-free PDW with an $s$-wave form factor compared with (b) a defect-free PDW with a $d$-wave form factor superimposed with the normal state dispersion (shown in light gray). The main arcs seen in both plots correspond to the normal state dispersion. In (c) we have a log plot of the Abrikosov vortex with a $d$-wave form factor, which greatly resembles the double dislocation and the pure PDW. Both defects prompt a slight redistribution of spectral weight, but nothing too dramatic like the half-vortex. Weaker features are uncovered with the log plots where the additional arcs correspond to the normal state dispersion shifted by $\pm\mathbf{Q}$.}
\label{figure_16}
\end{figure*}   
\section{Comparison of Spectral functions for $s$-wave and $d$-wave form factors and additional order parameters}\label{spec_appendix}
In this appendix we compare some additional spectral functions not included in the main text. First, let's focus on the spectral functions corresponding to defect-free PDW states, one with an $s$-wave form factor and the other $d$-wave [Figs. \ref{figure_16_1} and \ref{figure_16_2}, respectively]. The normal state dispersion is retained in both cases along the so-called Fermi arcs, so certain features are robust. The LDOS looks more or less the same for both form factors (not shown). Figures \ref{figure_16_3} is the natural log of the spectral function corresponding to the Abrikosov vortex, which closely resembles the double dislocation (not shown). Slight redistributions of the spectral weights, occur for both these defects, but nothing as dramatic as that seen in the case of the half-vortex.
\end{appendix}
\end{document}